\documentclass[prb,aps,twocolumn]{revtex4}

\usepackage{graphicx}
\usepackage{dcolumn}
\usepackage{bm}
\usepackage{color}
\usepackage{ulem}

\begin{document}
\newcommand{\avrg}[1]{\left\langle #1 \right\rangle}
\newcommand{\eqsa}[1]{\begin{eqnarray} #1 \end{eqnarray}}
\newcommand{\eqwd}[1]{\begin{widetext}\begin{eqnarray} #1 \end{eqnarray}\end{widetext}}
\newcommand{\hatd}[2]{\hat{ #1 }^{\dagger}_{ #2 }}
\newcommand{\hatn}[2]{\hat{ #1 }^{\ }_{ #2 }}
\newcommand{\wdtd}[2]{\widetilde{ #1 }^{\dagger}_{ #2 }}
\newcommand{\wdtn}[2]{\widetilde{ #1 }^{\ }_{ #2 }}
\newcommand{\cond}[1]{\overline{ #1 }_{0}}
\newcommand{\conp}[2]{\overline{ #1 }_{0#2}}
\newcommand{\nn}{\nonumber\\}
\newcommand{\cdt}{$\cdot$}
\newcommand{\bra}[1]{\langle#1|}
\newcommand{\ket}[1]{|#1\rangle}
\newcommand{\braket}[2]{\langle #1 | #2 \rangle}
\newcommand{\bvec}[1]{\mbox{\boldmath$#1$}}
\newcommand{\blue}[1]{{#1}}
\newcommand{\bl}[1]{{#1}}
\newcommand{\bn}[1]{\textcolor{blue}{#1}}
\newcommand{\rr}[1]{{#1}}
\newcommand{\bu}[1]{{#1}}
\newcommand{\red}[1]{{#1}}
\newcommand{\fj}[1]{{#1}}
\newcommand{\green}[1]{{#1}}
\newcommand{\gr}[1]{\textcolor{green}{#1}}
\definecolor{green}{rgb}{0,0.5,0.1}
\definecolor{blue}{rgb}{0,0,0.8}
\preprint{APS/123-QED}

\title{
\fj{
Cofermion Theory for Pseudogap Phenomena
and Superconducting Mechanism
of Underdoped Cuprate Superconductors
}
}

\author{Youhei Yamaji}
\altaffiliation[Present address: ]{
Department of Physics,
Rutgers University, Piscataway, New Jersey 08854, USA.}
\author{Masatoshi Imada}
\affiliation{Department of Applied Physics, University of Tokyo, Hongo, Bunkyo-ku, Tokyo, 113-8656, Japan}%


\date{\today}

\begin{abstract}
We study 
{pseudogap phenomena and Fermi-arc formation
experimentally observed in typical two dimensional doped Mott insulators,
namely,
underdoped cuprate superconductors.
To develop a physically unequivocal theory,
we
start from the slave-boson mean-field theory for the Hubbard model
on a square lattice.}
Our crucial step is to further take into account the charge dynamics
and fluctuations.
The extra charge fluctuations seriously modify
{low-energy single-particle spectra} of doped Mott insulators
near
the Fermi level:
An electron added around an empty site (or a hole added around a doubly occupied site)
constitutes {\it composite fermion} (cofermion),
called {\it holo-electron} (or {\it doublo-hole})
at low energy in distinction from the normal quasiparticles.
These unexplored composite fermions substantiate the extra charge fluctuation.
We show that the quasiparticles hybridize with the holo-electrons and doublo-holes.
The resultant hybridization gap is identified as
the pseudogap observed in the underdoped region of the high-$T_{{\rm c}}$ cuprates.
Because the Fermi level crosses the top (bottom)
of the low-energy band formed just below (above) the hybridization gap 
in the hole-doped (electron-doped) case,
it causes a Fermi-surface reconstruction, namely,
a {\it topological change in the Fermi surface} forced by the penetration of zeros
of the quasiparticle Green function.
This reconstruction signals the emergence of a non-Fermi-liquid phase.
The pseudogap, and the resultant formation of pocket or arc of the Fermi surface 
reproduce the experimental results for the cuprate superconductors in the underdoped region.
The pairing channel opens not only between two quasiparticles but also
between a quasiparticle and a cofermion.
This pairing solves the puzzle \bu{o}f the dichotomy between the $d$-wave superconductivity and the
precursors of the the insulating gap in the antinodal region.
We propose and analyze them as the mechanism of the  high-temperature superconductivity for the cuprates.
\end{abstract}

\maketitle
%
\section{Introduction}\label{Sec.I}
The discovery of cuprate superconductors
has triggered extensive studies on
the nature of low-energy electronic excitations evolving
in doped Mott insulators.
\bu{T}he \bu{extensive} interest
\bu{on the doped Mott insulators exists}
\bu{because}
it must be directly related
with the origin of the high temperature
superconductivity itself.

From the early stage of the studies, 
{experimental observations
\bu{have}
indicated that quasiparticle states
of normal phases of cuprates
change qualitatively
\bu{with the increase of}
the doping,
together with changes in superconducting
transition temperatures.}
It has been shown\cite{Ong87,Takagi89} that the Hall coefficient 
$R_H$ changes its sign near the so-called optimal doping
for the highest superconducting
transition temperatures, followed by a steep increase of 
its amplitude $|R_H|$ with lowering doping concentration, 
typically indicating a drastic change in 
low-energy quasiparticles of the normal state between
the underdoped and overdoped regions.
Such a drastic change is also but differently suggested from observations of
pseudogap phenomena\cite{Yasuoka89,Rossat-Mignod,Loram93,Homes93,Nishikawa93,Marshall96},
where the spin and charge excitations are unexpectedly suppressed in the underdoped region.
Various types of non-Fermi-liquid properties are accompanied in this region.

Recent improve\bu{d}
experimental tools ha\bu{ve}
enabled 
resolving low-energy single-particle spectra
near the Fermi level.
In particular,
strongly momentum-dependent quasiparticle states in the
hole-underdoped cuprates\cite{Damascelli_RMP,Yoshida06}
observed by angle-resolved photoemission spectroscopy (ARPES) studies
have renewed the interest in the low-energy spectrum of the cuprate superconductors.
In contrast to the overdoped region, where a large Fermi surface
crossing the region around the so-called antinodal points
$(\pm\pi,0)$ and $(0,\pm\pi)$ in the 2D Brillouin zone
for the CuO$_2$ plane is clearly observed,
low-energy quasiparticle states around the antinodal points
are missing in the underdoped cuprates.
It
\bu{emerges as}
a truncation of the large
Fermi surface observed in the overdoped region\bu{.}
\bu{The resultant truncated structure is called}
the ``Fermi arc".

\begin{figure}[h]
\begin{center}
\includegraphics[width=6cm]{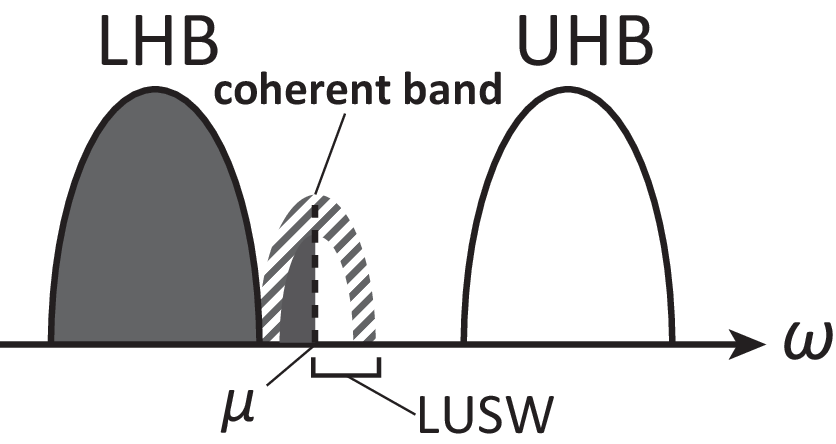}
\end{center}
\caption{
Schematic
global density of states of hole-doped Mott insulators.
The coherent band around the Fermi level $\mu$,
the upper Hubbard band (UHB), and
the lower Hubbard band (LHB) are schematically illustrated.
The low-energy unoccupied spectral weight
(LUSW) is also indicated.
\label{SW_HB}}
\end{figure}
From more fundamental point of view,
the normal state of the cuprates offers
a challenge of condensed matter physics 
as a typical open issue of ``Mott physics", 
namely, nature of strongly correlated metals in
the proximity to the Mott insulator\cite{Peierls,Mott37,Imada_RMP}.
The experimentally observed arc-like Fermi surface
in the underdoped cuprates
is a hallmark of the proximity to the Mott insulators.
As we illustrate in Fig.\ref{SW_HB}, global energy spectra of
doped Mott insulators are known to
consist of essentially three energy ``bands";
a coherent band around the Fermi level $\mu$ and two incoherent bands, 
namely, the upper Hubbard band (UHB) located above $\mu$, and the lower Hubbard band (LHB)
located below $\mu$.
For the hole-doped (electron-doped) systems, the coherent band is formed
around the top of the LHB 
(the bottom of the UHB).
{The spectral weight formed just above (below) $\mu$ within the coherent band
is often called the {\it low-energy unoccupied spectral weight} (LUSW)
in the hole-doped (electron-doped)
Mott insulators.
}

Such a global structure of spectra can be roughly described by a 
simple picture given by the dynamical mean-field theory\cite{Georges_RMP}. 
It has unified two scenarios of earlier studies by Hubbard\cite{Hubbard1}, and
Brinkman and Rice\cite{Brinkman} based on the Gutzwiller approximation\cite{Gutzwiller}.
The Hubbard approximation captures the formation of UHB and LHB. 
In the Brinkman-Rice picture, Mott insulators appear as a consequence of 
homogeneous vanishing of
\bu{the} quasiparticle
weight uniformly on the Fermi surface.
Therefore, this approximation draw\bu{s}
a picture that
quasiparticles are renormalized in a momentum-independent fashion
and the effective mass diverges independently of the momentum position
on the Fermi surface on the verge of the
Mott transition.

In infinite-dimensional systems, such a route of the Mott transition
has turned out to be correct\cite{Metzner89, Muller-Hartmann_a, Georges_RMP}.
However, in finite dimensions, this picture has turned out to be too simple
to understand {the low-energy spectra}
and the nature of the Mott transition:
Realistic theory has to capture significant momentum-dependent
quasiparticle renormalization, as has been experimentally
observed as Fermi-arc formation in the underdoped cuprates,
and has also been revealed  
from various numerical attempts\cite{Imada_Onoda,Onoda_Imada,Senechal04,Stanescu06,Zhang07,Sakai09,Sakai10}.
It has now become increasingly clear that a conceptually new idea for
describing Mott physics including an emergence of the new non-Fermi-liquid phase in the 
underdoped region is desired. 

In this paper, we extend \bu{our} previous attempt\cite{Yamaji11} for
describing the momentum-dependent Mott
physics including the mechanism of Mott transitions at the level of LUSW
and for understanding the experimentally
supported dramatic change in the electronic states of the cuprates.
\bu{We give a detailed comparison with experiments and a novel superconducting mechanism
based on the previous attempt\cite{Yamaji11}, together with the derivation of our theoretical
description in details.}
We particularly pay attention to the possible reconstruction of the Fermi surface.
If the reconstruction emerges, the doped Mott insulator
becomes topologically inequivalent to the conventional Fermi liquid and
hence offers an unexplored avenue for long-sort issue of unconventional metals.

In the present work, we propose a simple and physically transparent
theory that 
accounts for \bu{the} strongly momentum-dependent renormalization 
experimentally
suggested in the cuprate superconductors.
We focus on charge dynamics involved in the low-energy
spectra of the doped Mott insulators.
A key idea is that, near the Mott insulator,
an electron (a hole) added to an empty (a doubly occupied) site costs 
much smaller energy than an electron (a hole) added to singly occupied sites
and behaves as a component of a band separated from the main quasiparticle.
Then near the Mott insulator, this separated excitation
contributes to the LUSW in the hole-doped (electron-doped) Mott insulators.
This is in contrast with the weakly correlated regime, in which
an added electron (hole) constitutes a uniform excitation irrespective of 
the added site, because the added electron becomes uniformly and spatially extended 
with a good wavenumber of the momentum. 
In our theory,
such an electron (a hole) added to an empty site (a doubly occupied site)
forms a composite fermionic excitation
\bu{(or cofermion)},
which we call {\it holo-electron} ({\it doublo-hole}).
\bu{The conventional quasiparticles may
be scattered by the doubly occupied site
(empty site) and are transformed to the cofermions,
which introduces finite life time into the quasiparticles.}
\bu{This scattering and transformation may
alternatively be formulated as the hybridization between the conventional
quasiparticles and the holo-electrons and/or doublo-holes.}
Th\bu{is} hybridization
naturally
generates a hybridization gap 
and the resultant band splitting causes the Fermi-surface reconstructions,
namely, topological changes in the Fermi surface.
The hybridization gap and the topological change account for the experimentally
suggested pseudogap and Fermi-arc formation, respectively, in the cuprates.
Our unique prediction is that the pseudogap is independent of the precursor of the Mott
gap itself, while the pseudogap has the structure of the $s$-wave type symmetry rather than 
the $d$-wave symmetry. In addition, the pseudogap identified here is 
in clear distinction from the superconducting gap as well.  
We propose several experimental tests to prove or disprove
our theoretical consequences.

A way to understand topological changes in the Fermi surface
has recently been proposed
to originate from 
the emergence of zeros
of single particle Green's functions,
which are the points in the $(k,\omega)$-space satisfying ${\rm Re}\ G(k,\omega)=0$
for the momentum $k$ and the frequency
$\omega$.
In other words, the single particle self-energy $\Sigma (k,\omega)$
diverges at the zeros of Green's function.
The idea of the emergence of the zeros has a root in the work by
Dzyaloshinskii, who examined a possible extension of the Luttinger theorem 
into Mott insulating states\cite{Dzyaloshinskii03}.
As is reviewed in Ref.\onlinecite{Rice_PTP},
this idea is applied to the doped Mott insulators to explain
\bu{the}
``Fermi arcs" observed by ARPES.
Recent results of numerical calculations also suggest that the zeros emerging in doped Mott insulators
reconstruct
the Fermi surface and changes its topology\cite{Stanescu06,Sakai09,Sakai10}.
Such \bu{a} topological change is indeed
claimed in a recent ARPES measurement\cite{Meng09}.

The Fermi surface reconstruction itself is not an 
unconventional phenomenon, if it accompanies a
spontaneous symmetry breaking such as an antiferromagnetic order.
In translational-symmetry broken phases with an ordering wavenumber $Q$,
electrons with momentum $k$ hybridize, at least, with electrons at momentum $k+Q$.
Then Fermi surface reconstructions are naturally understood as a consequence of
the hybridization gap as we will discuss later in detail.
It is indeed able to account for the formation of Fermi pockets or arc-like Fermi surfaces
if a hybridization with other Fermionic excitations exist.

However, in the hole-doped cuprates, in spite of recent reports on
time reversal symmetry\cite{Fauque,Xia,Li}
\bu{and rotational symmetry\cite{Daou10} breakings}
in some of the cuprate superconductors,
translational symmetry breakings
have not been universally observed\cite{MacDougall}. On the other hand,
the ``Fermi arch or the truncated Fermi surface have universally 
been experimentally observed in the underdoped cuprates.
To have a unified picture of the Mott physics, 
it is important to understand whether the Fermi surface reconstruction 
occurs as a consequence of the hybridization gap generated by the ``hybridization" 
with some hidden fermionic excitations without assuming a symmetry broken phase.
In the present theory, the hybridization between the conventional quasiparticles
and the emergent fermionic excitations called
\bu{the}
holo-electron and
\bu{the}
doublo-hole naturally
explain such a Fermi-surface reconstruction.

Now we go into some details regarding characteristic low-energy spectra of 
doped Mott insulators.
As is proposed by Meinders, Eskes and Sawatzky\cite{Meinders93}, 
the doping dependence of the LUSW formed just above $\mu$,
has a marked feature for
nearly Mott insulating, correlated electron systems.
As we have mentioned above, an electron added to a singly occupied site 
inevitably costs the on-site Coulomb interaction $U$
and gives an excitation in the UHB,
whereas electrons added to empty sites exclusively contribute to
the low-energy states just above $\mu$.
Here, an empty site can accept an electron irrespective 
of its spin state and creates two unoccupied states, namely up and down spin states. 
Therefore the $N_{{\rm h}}$ holes create  $N_{{\rm h}}$ empty sites in the atomic limit,
where the kinetic energy of electrons $t$ is zero.
Then, they create $2N_{{\rm h}}$ unoccupied states as a consequence.
When the kinetic energy of the electrons $t$ becomes nonzero, 
the number of unoccupied states increases because of the pair creation of 
doubly occupied and empty sites induced by the hopping of electrons
and resultant charge fluctuations. 
Therefore, for hole-doping concentration $x=N_{{\rm h}}/N_{{\rm s}}$ with $N_{{\rm s}}$ 
being the total number of lattice sites,
the LUSW in the doped Mott insulator is always larger than $2x$.
From the early stage of the studies on the cuprates,
the LUSW developing faster than $2x$ has been observed by optical conductivity measurements\cite{Uchida91}.
This quick increase of the spectral weight larger than $2x$ requires
a picture of doped holes very different from the doped semiconductors,
where the LUSW is trivially equal to $x$.
Our scheme presented in this paper naturally explain\bu{s} this 
unconventional feature.

The mechanism of the superconductivity itself is a central open issue of the physics of the cuprate 
superconductors. When the LUSW in the normal state belongs to an unconventional phase with emergent 
excitations, the mechanism has to be understood based on this framework, because the energy scale of the 
superconductivity is even smaller than and governed by the energy scale of LUSW.
Our theory offers an unconventional channnel of the pairing and resultant superconductivity
emerging from the contribution of composite fermions never considered in the literature.
We examine an unexplored type of quasiparticle pairing arising from the pairing potential
generated by pairing fields of cofermions, holo-electrons/doublo-holes and quasiparticles.   
Although the pseudogap formation by the hybridization gap of quasiparticles and holo-electrons/doublo-holes
is destructive to the superconductivity, this unconventional pairing potential serves to creating
superconductors.  This dual character and two sides of the same coin naturally accounts for the 
recent puzzle under debates on the dichotomy
and nature of the gap in the anitinodal region of 
the underdoped cuprates\cite{Tanaka06,Yang08}.

The organization of this paper is the following:
In Sec.~\ref{Sec.II}, we start from the Kotliar-Ruckenstein's mean-field theory 
and review previous extensions for correlated metals for the self-contained description.
In Sec.~\ref{Sec.III} we take into account the charge dynamics, which plays an important role
in formation of the LUSW and
explain how emergent excitations, namely, the holo-electrons and doublo-holes emerge.
The hybridization between quasiparticles and these composite fermionic
excitations naturally causes the Fermi-surface reconstruction,
namely, the topological change in the Fermi surface.
The pseudogap phenomena observed in cuprate
superconductors also emerge because of this hybridization.
\bu{In Sec.~\ref{Sec.III}D, we propose a novel pairing mechanism
evolved from the cofermions as the mechanism of the high temperature superconductivity.}
Our results for single particle spectra, amplitudes of the pseudogap,
Fermi-surface topology, the specific heat coefficient, and the density of states
are presented and compared with experimental results in Sec.~\ref{Sec.IV}.
We \bu{also} estimate the \bu{superconducting} gap amplitude and quantitative aspects of
\bu{the superconducting mechanism}.
Sec.~\ref{Sec.VI} is devoted to discussions and summary. 

%
\section{Previous theories}\label{Sec.II}
\subsection{Hubbard model}
The Hubbard model\cite{Kanamori,Hubbard1,Gutzwiller} defined by the following Hamiltonian,
\eqsa{
	\hat{H}=\sum_{i,j}t_{ij}\hatd{c}{i\sigma}\hatn{c}{j\sigma}+U\sum_{i}\hatn{n}{i\uparrow}\hatn{n}{i\downarrow},\label{Hubbard}
}
is a canonical model, which describes the competition between the kinetic energy $t_{ij}$ and
the on-site Coulomb interaction $U$, where $\hatd{c}{i\sigma}$ ($\hatn{c}{i\sigma}$) is a creation (annihilation)
operator of $\sigma$-spin electron on $i$-th site and  $\hatn{n}{i\sigma}=\hatd{c}{i\sigma}\hatn{c}{i\sigma}$ is
a number operator.
Hereafter we focus on the Hubbard model on the square lattice as a model for the cuprates.
In this chapter, the nearest-neighbor hopping and the next-nearest-neighbor hopping are set as $t_{ij}=-t$ and $t_{ij}=+t'$,
and further neighbor hoppings are ignored.  

The solution of the Hubbard model
on two-dimensional lattices
remains an open problem.
To get an insight into the nature of the correlated metallic phase, there 
exist a variety of
numerical methods, which give accurate results for finite size clusters such as
exact diagonalization\cite{Dagotto94} and
quantum Monte Carlo\cite{Preuss}.
Cluster extensions of the dynamical mean-field theory\cite{Hettler98,Kotliar01},
improved variational Monte Carlo
methods\bu{\cite{Umrigar07,Tahara08}},
Gaussian Monte Carlo method\cite{Aimi07A,Aimi07B}
and the path integral renormalization group\cite{Kashima01A,Kashima01B,Mizusaki06}
are also {available} \bu{in the literature}.
However, they all have some limitations.
For example,
limitations on cluster size
are severe in the exact diagonalization and the quantum Monte Carlo, while
resolution{s} in the momentum space
is severely limited in
the cluster extension of the dynamical mean-field theory.

Without relying on accurate numerical methods,
there exist other ways to extract essential physics
from analytical or conceptually correct limits.
The Landau's Fermi-liquid picture\cite{Landau57}
offers such an example.
Although we have no exact solutions in the thermodynamic limit,
low-energy single-particle spectra of the Hubbard model
are believed to behave following the 
Landau's Fermi-liquid picture for less correlated
systems except for 1D systems.
The original proposal for the Mott insulating states
itself is not originally based on the numerical results, but
proposed by Peierls and Mott from a
gedanken experiment
on metallic crystalline hydrogen-like atoms\cite{Mott1,Mott2,Mott3}.
Such phenomenological theories based on physical intuitions
offer insights into difficult issues in a
wide range of condensed matter physics.
In this paper, we try to make a step towards
constructing such a physically transparent theory
\bu{that} account\bu{s} for the
unconventional properties of doped Mott insulators.

\subsection{Kotliar-Ruckenstein formalism}
One of the simplest picture to describe correlated metals and
metal-insulator transitions at half filling (Mott transitions)
along the line of the original idea by Mott
is
the Brinkman-Rice scenario\cite{Brinkman}
based on the Gutzwiller approximation\cite{Gutzwiller}.
As our starting point,
we {employ} the Kotliar-Ruckenstein's (KR's) slave-boson formalism,
which gives the same results as the Gutzwiller approximation. 
In this subsection, we briefly outline
the KR's slave-boson formalism for
the Hubbard model\cite{KR}, to make the present paper self-contained.
This slave-boson formalism
gives a starting point for
the mean-field description of strongly correlated electron systems by replacing original electrons with four kinds of bosons and one kind of \bu{spinful} fermion.

We start with the local Hilbert space of the Hubbard model, which is expanded by a set of the Fock states;
the empty state (holon) $\ket{0}$, the singly occupied states $\ket{\uparrow}$, $\ket{\downarrow}$, and  the doubly occupied state (doublon) $\ket{\uparrow\downarrow}$ .
Corresponding to each Fock state, one slave boson is introduced as
$\hatn{e}{}$ for $\ket{0}$, $\hatn{p}{\sigma}$ for  $\ket{\sigma}$, and $\hatn{d}{}$ for $\ket{\uparrow\downarrow}$ .
In addition to these bosons {$\hatn{b}{}$ ($b=e, p_{\sigma}$ or $d$)},
a fermion operator $\hatn{f}{\sigma}$ is introduced to stand for the $\sigma$-spin quasiparticle. 
The correspondence
relation to
local basis for the lattice wave functions {is} given as
\eqsa{
	\ket{0}_{i}\doteq\ket{{\rm vac}}_{i}^{{\rm F}}\otimes\hatd{e}{i}\ket{{\rm vac}}_{i}^{{\rm B}},\label{Eq_2}
}
\eqsa{
	\hatd{c}{i\uparrow}\ket{0}_{i}\doteq\hatd{f}{i\uparrow}\ket{{\rm vac}}_{i}^{{\rm F}}\otimes\hatd{p}{i\uparrow}\ket{{\rm vac}}_{i}^{{\rm B}},
}
\eqsa{
	\hatd{c}{i\downarrow}\ket{0}_{i}\doteq\hatd{f}{i\downarrow}\ket{{\rm vac}}_{i}^{{\rm F}}\otimes\hatd{p}{i\downarrow}\ket{{\rm vac}}_{i}^{{\rm B}},
}
\eqsa{
	\hatd{c}{i\uparrow}\hatd{c}{i\downarrow}\ket{0}_{i}\doteq
	\hatd{f}{i\uparrow}\hatd{f}{i\downarrow}\ket{{\rm vac}}_{i}^{{\rm F}}\otimes
	\hatd{d}{i}\ket{{\rm vac}}_{i}^{{\rm B}},\label{Eq_5}
}
where $\ket{{\rm vac}}_{i}^{{\rm F (B)}}$ is a vacuum of {the} $i$-th site for fermionic (bosonic) degrees of freedom.
{Equations (\ref{Eq_2})-(\ref{Eq_5}) represent the mapping between}
wave functions written by the original electrons and the fermions $\hat{f}$
{combined} with
bosons {$\hat{b}$}.
This mapping
\bu{is} derived
when the electron operators are replaced with composite ones as
\eqsa{
	\hatd{c}{i\sigma}\doteq
	\left(\hatd{p}{i\sigma}\hatn{e}{i}+\hatd{d}{i}\hatn{p}{\overline{i\sigma}}\right)\hatd{f}{i\sigma}.\label{re1}
}
We should note that
the replacement given by Eq.(\ref{re1}) is not a unique one.
It {is} known that operators equivalent to the right hand side of Eq.(\ref{re1}) can be given as
\eqsa{
	\hatd{c}{i\sigma}\doteq
	\hatn{z}{i\sigma}
	\hatd{f}{i\sigma},
}
where $\hatn{z}{i\sigma}$ is defined\cite{KR,Raimondi} as
\eqsa{
	\hatn{z}{i\sigma}
	=
	{\hat{g}_{i\sigma}}^{(1)}
	\left(\hatd{p}{i\sigma}\hatn{e}{i}+\hatd{d}{i}\hatn{p}{\overline{i\sigma}}\right)
	{\hat{g}_{i\sigma}}^{(2)},
}
\eqsa{
	{\hat{g}_{i\sigma}}^{(1)}&=
	\left(1-\hatd{p}{i\overline{\sigma}}\hatn{p}{i\overline{\sigma}}-\hatd{e}{i}\hatn{e}{i}\right)^{p_{1}},
	\label{p_1}\\
	{\hat{g}_{i\sigma}}^{(2)}&=\left(1-\hatd{p}{i\sigma}\hatn{p}{i\sigma}-\hatd{d}{i}\hatn{d}{i}\right)^{p_{2}}.
	\label{p_2}
}
The operators $\hat{g}_{i\sigma}^{(1)}$ and $\hat{g}_{i\sigma}^{(2)}$ act as identities  
when these are
operated to
$\left(\hatd{p}{i\sigma}\hatn{e}{i}+\hatd{d}{i}\hatn{p}{\overline{i\sigma}}\right)$, for any
{powers}
$p_{1}$ and $p_{2}$.
This ambiguity of the correspondence has been
utilized before\cite{KR}.

In the expanded Hilbert space,
the Hubbard Hamiltonian Eq.(\ref{Hubbard}) is rewritten as
\eqsa{
	\hat{H}=
	\sum_{i,j}t_{ij}
	\hatn{z}{i\sigma}
	\hatd{f}{i\sigma}
	\hatn{f}{j\sigma}
	\hatd{z}{j\sigma}
	+U\sum_{i}\hatd{d}{i}\hatn{d}{i}.
	\label{KR_Hubbard}
}
The on-site Coulomb interaction is replaced by a ``chemical potential" for doublons $\hatd{d}{i}$. 
Then the correlation among electrons is
now contained as
hopping process of quasiparticles $\hatd{f}{i\sigma}$
disturbed by the associated motion of slave bosons.
The quasiparticle hopping is accompanied
with four kinds of bosonic motions
generated by $\hatn{z}{i\sigma}\hatd{z}{j\sigma}$,
namely, physical process\bu{es} of
hopping of holons $\hatd{e}{i}$ and 
doublons $\hatd{d}{i}$, pair creations and annihilations of holons and doublons.

When we employ the path integral description of the system by making use of the coherent states for bosons and fermions,
we need to introduce a set of
constraints to eliminate unphysical states in the expanded Hilbert space,
which arise
when we introduce slave bosons to describe the local Fock states.
First, only one boson should occupy each local state. There are only four local physical Fock states, and
these four states are exhausted by
four different kinds of slave bosons.
Therefore\bu{,} we need the first constraint
\eqsa{
	\hatd{e}{i}\hatn{e}{i}+\sum_{\sigma}\hatd{p}{i\sigma}\hatn{p}{i\sigma}+\hatd{d}{i}\hatn{d}{i}=1.\label{cnst1}
}
Second, the number operator of \bu{the} $\sigma$-spin quasiparticle is necessarily given as  
\eqsa{
	\hatd{f}{i\sigma}	\hatn{f}{i\sigma}=\hatd{p}{i\sigma}\hatn{p}{i\sigma}+\hatd{d}{i}\hatn{d}{i}.\label{cnst2}
}

In the path integral form, the partition function for the Hubbard model {is} given by
\eqsa{
	Z=\int\mathcal{D}\left[\hatd{c}{\sigma},\hatn{c}{\sigma}\right]e^{-\mathcal{S}\left[\hatd{c}{\sigma},\hatn{c}{\sigma}\right]}
}
where the action is  
\eqsa{
	\mathcal{S}\left[\hatd{c}{\sigma},\hatn{c}{\sigma}\right]
	&=&
	\sum_{ij\sigma}
	\int_{0}^{\beta}d\tau\ 
	\hatd{c}{i\sigma}(\tau)
	\left[\left(\partial_{\tau}-\mu\right)\delta_{ij}+t_{ij}\right]
	\hatn{c}{j\sigma}(\tau)
	\nn
	&+&U\sum_{i}
	\int_{0}^{\beta}d\tau\ \hatd{c}{i\uparrow}(\tau)\hatn{c}{i\uparrow}(\tau)\hatd{c}{i\downarrow}(\tau)\hatn{c}{i\downarrow}(\tau).
		\label{Hubbard_KR_action_0}
}
We use the same notation for both of the operators and the corresponding grassmann fields for simplicity.

On the other hand, the partition function in the slave-boson formalism is
\eqsa{
	Z=\int\mathcal{D}\left[\hatd{f}{\sigma},\hatn{f}{\sigma}\right]
	\mathcal{D}\left[\hat{\mbox{\boldmath$B$}}^{\dagger},\hat{\mbox{\boldmath$B$}}\right]
	\mathcal{D}\left[\mbox{\boldmath$\lambda$}\right]
	e^{-\mathcal{S}
	}.
	\label{KR_Z}
}
The action is rewritten as
\eqwd{
	\mathcal{S}
	&=&
	\int_{0}^{\beta}d\tau\ 
	\hatd{f}{i\sigma}(\tau)
	\left[\left(\partial_{\tau}-\mu+\lambda_{i\sigma}^{(2)}\right)\delta_{ij}+\hatn{\zeta}{ij\sigma}(\tau)t_{ij}\right]
	\hatn{f}{j\sigma}(\tau)
	+\int_{0}^{\beta}d\tau\ U\hatd{d}{i}(\tau)\hatn{d}{i}(\tau)
	\nn
	&+&
	\int_{0}^{\beta}d\tau\ 
	\left[
	\sum_{i}
	\left(
	\hatd{e}{i}(\tau)\partial_{\tau}\hatn{e}{i}(\tau)
	+
	\sum_{\sigma}\hatd{p}{i\sigma}(\tau)\partial_{\tau}\hatn{p}{i\sigma}(\tau)
	+\hatd{d}{i}(\tau)\partial_{\tau}\hatn{d}{i}(\tau)
	\right)
	\right.
	\nn
	&+&\sum_{i}\lambda_{i}^{(1)}\left(\hatd{e}{i}(\tau)\hatn{e}{i}(\tau)+\sum_{\sigma}\hatd{p}{i\sigma}(\tau)\hatn{p}{i\sigma}(\tau)+\hatd{d}{i}(\tau)\hatn{d}{i}(\tau)-1\right)
	\nn
	&-&
	\left.
	\sum_{i\sigma}\lambda_{i\sigma}^{(2)}\left(\hatd{p}{i\sigma}(\tau)\hatn{p}{i\sigma}(\tau)+\hatd{d}{i}(\tau)\hatn{d}{i}(\tau)\right)
	\right],\label{A_SFB}
}
where 
$\hat{\mbox{\boldmath$B$}}_{i}=(\hatd{e}{i},\hatd{p}{i\uparrow},\hatd{p}{i\downarrow},\hatd{d}{i})$,
$\mbox{\boldmath$\lambda$}_{i}=(\lambda_{i}^{(1)},\lambda_{i\uparrow}^{(2)},\lambda_{i\downarrow}^{(2)})$, and
$\hatn{\zeta}{ij\sigma}(\tau)=\hatn{z}{i\sigma}(\tau)\hatd{z}{j\sigma}(\tau)$.
The Lagrange multipliers $\lambda_{i}^{(1)}$, $\lambda_{i\uparrow}^{(2)}$, and $\lambda_{i\downarrow}^{(2)}$ are introduced to reinforce
the set of constraints Eqs. (\ref{cnst1}) and (\ref{cnst2}).
\subsection{Mean-field theory for KR formalism}
The mean-field approximation for Eq.(\ref{KR_Z})
corresponds to
replacing bosonic fields
$\hat{\mbox{\boldmath$B$}}_{i}$ and $\mbox{\boldmath$\lambda$}_{i}$
with the homogeneous saddle point
values
for them as
\eqsa{
&\hatd{e}{i},\ \hatn{e}{i}\rightarrow \cond{e},\quad 
\hatd{p}{i\sigma},\ \hatn{p}{i\sigma}\rightarrow \conp{p}{\sigma},\quad 
\hatd{d}{i},\ \hatn{d}{i}\rightarrow \cond{d},\nn
&\lambda_{i}^{(1)}\rightarrow \lambda^{(1)},\quad 
\lambda_{i\sigma}^{(2)}\rightarrow \lambda_{\sigma}^{(2)}.
\nonumber
}
These saddle point values are determined self-consistently
through minimizing the free energy $f$ given in
Eq.(\ref{Free_E0}) below.
Then the action {for the fermionic degrees of freedom}
contains {only} quadratic terms of fermionic fields $\hatd{f}{i\sigma}$
after the slave bosons are replaced with $c$-numbers $\cond{e}$, $\conp{p}{\sigma}$, and $\cond{d}$
as
\eqsa{
	S_{0}=
	\int_{0}^{\beta}d\tau
	\sum_{ij\sigma}
	\hatd{f}{i\sigma}(\tau)
	\left[\left(\partial_{\tau}-\mu\right)\delta_{ij}+\zeta_{0\sigma}t_{ij}\right]
	\hatn{f}{j\sigma}(\tau).\label{S_0}
}
We can 
easily integrate out
the remaining fermionic degrees of freedom
and obtain the mean-field free energy for homogeneous phases as
\eqsa{
	f&=&
	-\frac{T}{N_{s}}\sum_{k,\sigma}
	\ln
	\left[
	1+e^{-\beta\left(\zeta_{0\sigma}\epsilon_{k}-\mu+\lambda_{\sigma}^{(2)}\right)}
	\right]
	+U\cond{d}^{2}\nn
	&+&\lambda^{(1)}\left(\cond{e}^{2}+\sum_{\sigma}\conp{p}{\sigma}^{2}+\cond{d}^{2}-1\right)
	\nn
	&-&\sum_{\sigma}\lambda_{\sigma}^{(2)}\left(\conp{p}{\sigma}^{2}+\cond{d}^{2}\right)
	,\label{Free_E0}
}
where
$N_{{\rm s}}$ is the number of sites,
$\epsilon_{k}$ is the Fourier transformation of $t_{ij}$
and
the mean-field quasiparticle renormalization $\zeta_{0\sigma}$ is given as
\eqsa{
	\zeta_{0\sigma}&=&
	\left[
	g_{1\sigma}
	\left(
	\conp{p}{\sigma}\cond{e}+\cond{d}\conp{p}{\overline{\sigma}}
	\right)
	g_{2\sigma}
	\right]^{2},\label{zeta_0_MFKR}\\
g_{1\sigma}&=&\left(1-\conp{p}{\overline{\sigma}}^{2}-\cond{e}^{2}\right)^{p_{1}},
\\
g_{2\sigma}&=&\left(1-\conp{p}{\sigma}^{2}-\cond{d}^{2}\right)^{p_{2}}.
}
It is known that
with this drastic mean-field approximation,
the non-interacting limit is correctly reproduced
when we set $p_{1}$ and $p_{2}$ defined in Eqs.(\ref{p_1}) and (\ref{p_2}) as
$p_{1}=p_{2}={-}1/2$.
For $U=0$, the mean-field values of the density of bosons are given as,
$\cond{d}^{2}=(n/2)^{2}$, $\cond{e}^{2}=1-n+\cond{d}^{2}$,
and
$\conp{p}{\uparrow}^{2}=\conp{p}{\downarrow}^{2}=n/2-\cond{d}^{2}$.
Then the mean-field renormalization factor turns out to be 1, correctly.
Hereafter, the values of $p_{1}$ and $p_{2}$ are fixed as $p_{1}=p_{2}={-}1/2$. 

The saddle point values for bosonic fields,
$\cond{e}$, $\conp{p}{\sigma}$, and $\cond{d}$,
can {easily be} examined in some well-defined
limits.
In the strong coupling limit $U/ |t|\gg 1$, 
the density of doublon
is suppressed,  $\cond{d}^{2}\sim 0$,
at any doping level.
Then the spin density is roughly proportional to the spin-dependent electron density,
$\conp{p}{\sigma}^{2}\sim n_{\sigma}$
(please {recall} the mean-field constraint, $n_{\sigma}=\conp{p}{\sigma}^{2}+\cond{d}^{2}$).
In the hole-doping case, the density of empty site $\cond{e}^{2}$ is given by the doping concentration,
$\cond{e}^{2}\sim x$.

In this limit,
the doping dependences of $\zeta_{0\sigma}$ and L\bu{U}SW
are given as follows: 
By using the fact $n_{\sigma}=n/2=(1-x)/2$ in the paramagnetic phase, the mean-field renormalization factor $\zeta_{0}$
is given as
\eqsa{
	\zeta_{0\sigma}&=&\zeta_{0}=\frac{\cond{p}^{2}\left(\cond{e}+\cond{d}\right)^{2}}{\frac{n}{2}\left(1-\frac{n}{2}\right)}\nn
	&\simeq&\frac{\frac{1-x}{2}\cdot x}{\frac{1-x}{2}\cdot\frac{1+x}{2}}=\frac{2x}{1+x}.
}
Multiplying the number density of unoccupied states,
$1+x$ with $\zeta_{0}$,
we obtain the L\bu{U}SW as $2x$.
This doping-dependence of the L\bu{U}SW
is the same as that of {the} exact solution in the
strong coupling limit.

\subsection{Previous studies on charge fluctuations}
\subsubsection{Formation of upper and lower Hubbard bands}
The mean-field theory only accounts for the coherent quasiparticle excitations in the correlated electron systems.
Of course, momentum-dependent renormalizations
do not appear.
For overall description of the energy spectrum including incoherent Hubbard bands,
dynamics of the charge bosons
is known to be
essential
when one wishes to improve the slave boson mean-field theory.
For example, Castellani {\it et al}. have claimed that
the Gaussian fluctuations of charge bosons around the saddle point solution
can reproduce the structure of incoherent Hubbard bands\cite{Castellani92}.
To take into account the fluctuations of bosonic fields around the saddle point solutions,
the Bogoliubov prescription\cite{Bogoliubov47} is used, in which
the boson operators are divided into condensate components
and fluctuating components out of condensation as 
\eqsa{
	&\hatd{e}{i}=\cond{e}+\wdtd{e}{i},\ \hatn{e}{i}=\cond{e}+\wdtn{e}{i},
	\\
	&\hatd{p}{i\sigma}=\conp{p}{\sigma}+\wdtd{p}{i},\ 
	\hatn{p}{i\sigma}=\conp{p}{\sigma}+\wdtn{p}{i},
	\\
	&\hatd{d}{i}=\cond{d}+\wdtd{d}{i},\ 
	\hatn{d}{i}=\cond{d}+\wdtn{d}{i}.
}
\subsubsection{Bosonic propagators}
Propagators of the Gaussian fluctuations of the slave bosons
are given by the action $S_{{\rm B}}$ with quadratic terms including {only}
the bosonic fields $\widetilde{\mbox{\boldmath$B$}}^{\dagger}$,
$\widetilde{\mbox{\boldmath$B$}}$ as
\eqsa{
	S_{{\rm B}}=S_{{\rm B}}^{(0)}+S_{{\rm B}}^{(1{\rm c})}+S_{{\rm B}}^{(1{\rm s})}+S_{{\rm B}}^{(1{\rm cs})},
	\label{SB_qd}
}
{\eqsa{
	S_{{\rm B}}^{(0)}
	&=&
	\int_{0}^{\beta}d\tau\ 
	\sum_{i}
	\Large{\{}
	\wdtd{e}{i}(\tau)[\partial_{\tau}+\lambda_{i}^{(1)}]\wdtn{e}{i}(\tau)
	\nn
	&+&
	\sum_{\sigma}\wdtd{p}{i\sigma}(\tau)[\partial_{\tau}+\lambda_{i}^{(1)}-\lambda_{i\sigma}^{(2)}]
	\wdtn{p}{i\sigma}(\tau)
	\nn
	&+&
	\wdtd{d}{i}(\tau)[\partial_{\tau}+U+\lambda_{i}^{(1)}-\sum_{\sigma}\lambda_{i\sigma}^{(2)}
	]\wdtn{d}{i}(\tau)
	\},\label{B0}
	\\
	S_{{\rm B}}^{(1{\rm c})}
	&=&
	\int_{0}^{\beta}d\tau\sum_{ij}
	\bvec{\beta}_{i}(\tau)
	\bvec{L}^{{\rm c}}_{ij}
	\bvec{\beta}_{j}^{\dagger}(\tau),\label{B1c}
	\\
	S_{{\rm B}}^{(1{\rm s})}
	&=&
	\int_{0}^{\beta}d\tau\sum_{ij}
	{\bvec{\phi}_{i\uparrow}^{\dagger}}^{T}(\tau)
	\bvec{L}^{{\rm s}}_{ij}
	{\bvec{\phi}_{j\uparrow}}(\tau)^{T}\nn
	&=&
	\int_{0}^{\beta}d\tau\sum_{ij}
	{\bvec{\phi}_{i\downarrow}^{\dagger}}^{T}(\tau)
	\bvec{L}^{{\rm s}}_{ij}
	{\bvec{\phi}_{j\downarrow}}(\tau)^{T},\label{B1s}
	\\
	S_{{\rm B}}^{(1{\rm cs})}
	&=&
	\int_{0}^{\beta}d\tau
	\sum_{ij\sigma}
	\left[
	\bvec{\beta}_{i}(\tau)
	\bvec{L}^{{\rm cs}}_{ij\sigma}
	{\bvec{\phi}_{j\uparrow}}^{T}(\tau)
	\right.
	\nn
	&+&
	{\bvec{\phi}^{\dagger}_{i\sigma}}^{T}(\tau)
	{\bvec{L}^{{\rm cs}}_{ij\sigma}}^{T}
	{\bvec{\beta}_{j}^{\dagger}}(\tau)
	+
	\delta_{ij}
	K^{{\rm cs}}_{i\sigma}
	\nn
	&\times&
	\left.
	\left\{
	{\bvec{\beta}_{i}}(\tau)\cdot{\bvec{\phi}^{\dagger}_{i\sigma}}(\tau)
	+
	{\bvec{\phi}_{i\sigma}}(\tau)\cdot{\bvec{\beta}^{\dagger}_{i}}(\tau)
	\right\}
	\right],\label{B1cs}
}
where we use vector notations,
$\bvec{\beta}_{i}=
(\wdtn{e}{i}, \wdtd{d}{i})$,
$\bvec{\phi}_{i}=
(\wdtn{p}{i\sigma}, \wdtd{p}{i\overline{\sigma}})$,
and coefficients defined as
}
{\eqsa{
	\bvec{L}^{{\rm c}}_{ij}
	&=&
	g_{ij}
	\left(
	\begin{array}{cc}
	\conp{p}{\sigma}^2+\conp{p}{\overline{\sigma}}^2
	&2\conp{p}{\sigma}\conp{p}{\overline{\sigma}}\\
	2\conp{p}{\sigma}\conp{p}{\overline{\sigma}}
	&\conp{p}{\sigma}^2+\conp{p}{\overline{\sigma}}^2
	\end{array}
	\right),\\
	\bvec{L}^{{\rm s}}_{ij}
	&=&
	g_{ij}
	\left(
	\begin{array}{cc}
	\cond{e}^{2}+\cond{d}^{2}&2\cond{e}\cond{d}\\
	2\cond{e}\cond{d}&\cond{e}^{2}+\cond{d}^{2}
	\end{array}
	\right),\\
	\bvec{L}^{{\rm cs}}_{ij\sigma}
	&=&
	g_{ij}
	\left(
	\begin{array}{cc}
	\cond{e}\conp{p}{\sigma}&\cond{e}\conp{p}{\overline{\sigma}}\\
	\cond{d}\conp{p}{\sigma}&\cond{d}\conp{p}{\overline{\sigma}}
	\end{array}
	\right),\\
	K^{{\rm cs}}_{i\sigma}
	&=&
	\sum_{j}g_{ij}(\cond{e}\conp{p}{\sigma}+\cond{d}\conp{p}{\overline{\sigma}}).
}
Here the hopping parameter for bosons is given as
$g_{ij}=t_{ij}g_{1\sigma}^{2}g_{2\sigma}^{2}\avrg{\hatd{f}{i\sigma}\hatn{f}{j\sigma}}$, where static correlation functions for quasiparticles $\avrg{\hatd{f}{i\sigma}\hatn{f}{j\sigma}}$
are introduced.}
The average $\avrg{\cdots}$ is defined as
\eqsa{
	\avrg{X}
	=
	\frac
	{
	\int\mathcal{D}\left[\hatd{f}{\sigma},\hatn{f}{\sigma}\right]
	X e^{-S_{0}-S_{{\rm B}} }
	}
	{
	\int\mathcal{D}\left[\hatd{f}{\sigma},\hatn{f}{\sigma}\right]
	e^{-S_{0}-S_{{\rm B}} }
	},\label{avrg_0B_ff}
}
where $S_{0}+S_{{\rm B}}$ is the approximate action used in this subsection.
The term $\avrg{\hatd{f}{i\sigma}\hatn{f}{j\sigma}}$
represents the mean field that stands for
kinetic motions of quasiparticles surrounding bosons.
This mean field seems to be self-consistently determined through Eq.(\ref{avrg_0B_ff}).
However, by using the approximate action, $S_{0}+S_{{\rm B}}$,
we obtain $\avrg{ \hatd{f}{i\sigma}\hatn{f}{j\sigma} }$ as
\eqsa{
	\avrg{ \hatd{f}{i\sigma}\hatn{f}{j\sigma} }
	=
	\frac
	{
	\int\mathcal{D}\left[\hatd{f}{\sigma},\hatn{f}{\sigma}\right]
	\hatd{f}{i\sigma}\hatn{f}{j\sigma} e^{-S_{0}}
	}
	{
	\int\mathcal{D}\left[\hatd{f}{\sigma},\hatn{f}{\sigma}\right]
	e^{-S_{0} }
	},\label{static_correlation}
} 
because $S_{{\rm B}}$ does not contain $\hatn{f}{i\sigma}$ and/or $\hatd{f}{i\sigma}$. 
This mean field $\avrg{\hatd{f}{i\sigma}\hatn{f}{j\sigma}}$
is obtained through 
decoupling the ``interaction" term
$\hatn{z}{i\sigma}\hatd{f}{i\sigma}\hatn{f}{j\sigma}\hatd{z}{j\sigma}$
as
\eqsa{
\hatn{z}{i\sigma}\hatd{f}{i\sigma}\hatn{f}{j\sigma}\hatd{z}{j\sigma}
&\simeq
\hatn{z}{i\sigma}\hatd{z}{j\sigma}
\avrg{\hatd{f}{i\sigma}\hatn{f}{j\sigma}}
+
\avrg{\hatn{z}{i\sigma}\hatd{z}{j\sigma}}
\hatd{f}{i\sigma}\hatn{f}{j\sigma}
\nn
&-
\avrg{\hatn{z}{i\sigma}\hatd{z}{j\sigma}}
\avrg{\hatd{f}{i\sigma}\hatn{f}{j\sigma}}
\label{decouple}
}
by neglecting
fluctuations
\eqsa{
\left(
\hatn{z}{i\sigma}\hatd{z}{j\sigma}
-
\avrg{\hatn{z}{i\sigma}\hatd{z}{j\sigma}}
\right)
\left(
\hatd{f}{i\sigma}\hatn{f}{j\sigma}
-
\avrg{\hatd{f}{i\sigma}\hatn{f}{j\sigma}}
\right).
\nonumber
}
Averages such as $\avrg{\hatd{f}{i\sigma}\hatn{f}{j\sigma}}$ and
$\avrg{\hatn{z}{i\sigma}\hatd{z}{j\sigma}}$ are taken by using the resultant
action.
However,
we should note that,
in the previous studies\cite{Castellani92,Raimondi},
the average
$\avrg{\hatn{z}{i\sigma}\hatd{z}{j\sigma}}$
were treated
as $\zeta_{0}$
and the contribution from
static
correlation functions of bosons
such as
$\avrg{\wdtd{d}{i}\wdtd{e}{j}}$
{was} dropped.
We also note that, in the mean-field level, {the} dispersion of spin bosons
for {the} paramagnetic phase
{vanishes}
as $x\rightarrow 0$,
because of simultaneously vanishing condensation of charge bosons: $\cond{e},\ \cond{d}\rightarrow 0$
{in Eqs.(\ref{B1s}) and (\ref{B1cs})}.

In the close proximity to the Mott insulating states, where $\cond{e},\ \cond{d}\ll 1$,
propagators for the Gaussian fluctuations of charge bosons
{$\bvec{\beta}_{i}=({\beta}^{1}_{i},{\beta}^{2}_{i})=
(\wdtn{e}{i}, \wdtd{d}{i})$}
are approximately determined by the action
$S_{{\rm B}}^{(0)}
+
S_{{\rm B}}^{(1{\rm c})}
$
as
{\eqsa{
	-\avrg{{\beta^{a}}_{Q}{\beta^{b}}^{\dagger}_{Q}}
	=
	\frac{Z_{+}^{ab}(Q)}{i\omega_{m}-|\Lambda_{Q}|}
	-
	\frac{Z_{-}^{ab}(Q)}{i\omega_{m}+|\lambda_{Q}|},
}
where the coefficients $Z_{\pm}^{ab}(Q)$ are given by
\eqsa{
	&
	\left(
	\begin{array}{cc}
	Z_{\pm}^{11}(Q)&Z_{\pm}^{12}(Q)\\
	Z_{\pm}^{21}(Q)&Z_{\pm}^{22}(Q)
	\end{array}	
	\right)
	=
	\displaystyle
	\frac{\delta\lambda+\delta U/2}{2\sigma_{Q}}
	\left(
	\begin{array}{cc}
	1
	&0\\
	0
	&1
	\end{array}	
	\right)
	\nn
	&\quad\quad\quad\quad\quad\quad\pm
	\frac{1}{2}
	\left(
	\begin{array}{cc}
	1
	&0\\
	0
	&-1
	\end{array}	
	\right)
	-
	\frac{\cond{p}^{2}|\epsilon|\epsilon_{Q}}{2\sigma_{Q}}
	\left(
	\begin{array}{cc}
	1&1\\
	1&1
	\end{array}	
	\right).
}
}
Parameters used in the above equations are given as
\eqsa{
	\delta\lambda&=&
	\lambda^{(1)}-\lambda^{(2)},\\
	\sigma_{Q}&=&\sqrt{\left(\lambda^{(1)}+\frac{\delta U}{2}-\cond{p}^{2}|\epsilon|\epsilon_{Q}
	\right)^{2}-\cond{p}^{4}|\epsilon|^{2}\epsilon_{Q}^{2}},\\
	\Lambda_{Q}&=&\frac{\delta U}{2}+\sigma_{Q},\\
	\lambda_{Q}&=&-\frac{\delta U}{2}+\sigma_{Q},
}
where
$\delta U=U-2\lambda^{(2)}$ gives the amplitude of the Hubbard gap at half-filling $n=1$,
and
\eqsa{
	\displaystyle|\epsilon|=
	\left|
	\frac{1}{N_{s}}
	\sum_{k}
	\epsilon_{k}
	/\left[
	1+e^{+\beta \left(q\epsilon_{k}-\mu+\lambda^{(2)}\right)}
	\right]
	\right|.\nonumber
}

Raimondi and Castellani\cite{Raimondi} introduced the following approximate form of the single particle
{G}reen's function as
\eqsa{
	\mathcal{G}_{ij\sigma}(\tau)
	&=&
	-\avrg{T\hatn{c}{i\sigma}(\tau)\hatd{c}{j\sigma}(0)}
	\nn
	&=&
	-\avrg{T\hatd{z}{i\sigma}(\tau)\hatn{z}{j\sigma}(0)\hatn{f}{i\sigma}(\tau)\hatd{f}{j\sigma}(0)}
	\nn
	&\simeq&
	-\avrg{T\hatd{z}{i\sigma}(\tau)\hatn{z}{j\sigma}(0)}\avrg{T\hatn{f}{i\sigma}(\tau)\hatd{f}{j\sigma}(0)},
}
where vertex corrections are dropped.
A further approximation was introduced:
By focusing only on the charge dynamics,
contributions from the fluctuations of spin bosons were neglected as
\eqsa{
	\avrg{T\hatd{z}{i\sigma}(\tau)\hatn{z}{j\sigma}(0)}
	&\simeq&
	\zeta_{0\sigma}+\cond{p}^{2}g_{1\sigma}^{2}g_{2\sigma}^{2}
	\nn
	&\times&
	\avrg{T
	\wdtn{b}{i\sigma}(\tau)
	\wdtd{b}{j\sigma}(0)},
}
where $\wdtn{b}{i\sigma}(\tau)=\wdtd{e}{i\sigma}(\tau)+\wdtn{d}{i\sigma}(\tau)$.
As a result, the approximate Green's function is given as
\eqsa{
	\mathcal{G}_{ij\sigma}(\tau)
	&\simeq&
	-
	\zeta_{0\sigma}\avrg{T\hatn{f}{i\sigma}(\tau)\hatd{f}{j\sigma}(0)}
	\nn
	&-&
	\cond{p}^{2}g_{1\sigma}^{2}g_{2\sigma}^{2}\avrg{T
	\wdtn{b}{i\sigma}(\tau)
	\wdtd{b}{j\sigma}(0)}
	\nn
	&\times&
	\avrg{T\hatn{f}{i\sigma}(\tau)\hatd{f}{j\sigma}(0)}.\label{GFA}
}
The first term of the right hand side of Eq.(\ref{GFA})
gives the coherent band
and the second term gives the incoherent Hubbard bands.
%
\section{Cofermion theory}\label{Sec.III}
\subsection{Perspective}
In the above sections, we reviewed the mean-field KR theory and 
how fluctuations of charge bosons induce the incoherent bands.
These previous theories have drawbacks, in spite of an advantage
in the simplicity.
For example,
the momentum-independent
quasiparticle renormalization in the previous theories
cannot account for the ``Fermi arc" observed in ARPES measurements
of the cuprate superconductors\cite{Damascelli_RMP}.
\bu{Since} significant momentum-dependent
quasiparticle renormalizations have been captured 
in numerical works through the cluster type extension of the dynamical mean-field theory\cite{Imada_Onoda,Onoda_Imada,Senechal04,Stanescu06,Zhang07,Sakai09}, 
it is desirable to construct a theory that is physically transparent and can
examine
the observed singular self-energy, which results in, for examples,
the pseudogap and the Fermi arc.
We assume that the experimentally observed arc-like Fermi surface, in
the hole-underdoped
cuprates,
is a consequence of Fermi-surface reconstruction
caused by the divergence of the quasiparticle self-energy
or emergence of zeros of the Green's function.
The divergence means the breakdown of the
perturbation theory and hence the breakdown of
the Fermi liquid theory as well. 
Here we
extend
the previous theories
to understand this unconventional feature.
Our goal is to acquire a simple framework and an intuitive understanding.

The Fermi-surface reconstruction itself
is not an unconventional phenomenon as we have discussed in Sec.~\ref{Sec.I}
in the example of 
the antiferromagnetic order in the ordinary Slater's mean-field description.
In an antiferromagnetic metal on square lattices with the ordering vector $Q_{0}=(\pi,\pi)$, 
the $\uparrow$-spin electrons with momentum $k$
hybridize with the $\downarrow$-spin electrons with momentum $k+Q_{0}$,
as the term $\Delta_{\rm AFM}[\hatd{c}{k\uparrow}\hatn{c}{k+Q_{0}\downarrow}+{\rm H.c.}]$,
in the presence of the mean field $\Delta_{\rm AFM}$.
As a result,
the self-energy for the $\uparrow$-spin electron diverges
at the momentum $k$\bu{,} where the $\downarrow$-spin electron with momentum $k+Q_{0}$ has a pole.

\begin{figure}[h]
\begin{center}
\includegraphics[width=8.5cm]{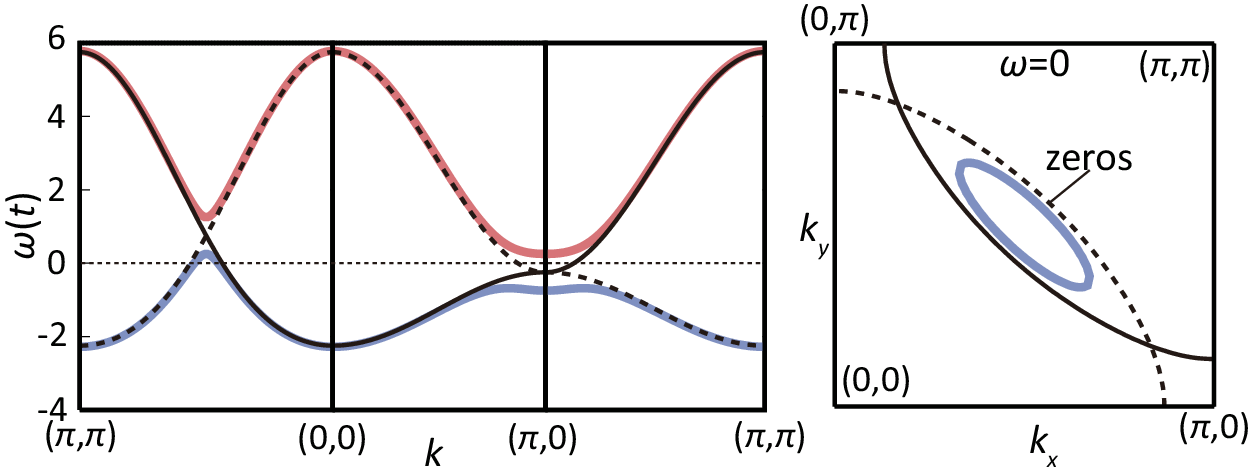}
\end{center}
\caption{(color online) Left panel shows band dispersions and zero surface
along lines running from $(\pi,\pi)$ to $(0,0)$,
from $(0,0)$ to $(\pi,0)$, and from $(\pi,0)$ to $(0,0)$.
Right panel shows bare and reconstructed Fermi surface, and zero surface at $\omega=0$.
Thin (black) solid curve stands for bare band dispersion $\omega=\xi_{k}$,
and thin dashed (black) curve stands for zero surface $\omega=\xi_{k+Q_{0}}$.
\bu{Bold}
(red and blue) solid curves stand for reconstructed bands.
\label{AFM_band}}
\end{figure}

When
the bare band dispersion of $\hatd{c}{k\sigma}$
before the hybridization is switched on is given by
\eqsa{
	\xi_{k}&=\epsilon_{1k}+\epsilon_{2k}-\mu,\nn
	\epsilon_{1k}&=-2t(\cos k_{x}+\cos k_{y}),\nn
	\epsilon_{2k}&=4t'\cos k_{x}\cos k_{y},\nonumber
}
the Green's function of 
the electron $\hatd{c}{k\sigma}$
that
hybridizes with the electron $\hatd{c}{k+Q_{0}\overline{\sigma}}$
is obtained
as
\eqsa{
	G(k,\omega)
	&=&
	\frac
	{1}
	{\displaystyle\omega-\xi_{k}-\frac{\Delta_{{\rm AFM}}^{2}}{\omega-\xi_{k+Q_{0}}}}
	\label{AFM_zero}
	\\
	&=&
	\frac{
	\displaystyle
	\frac{1}{2}
	-
	\frac
	{ \epsilon_{1k} }
	{ 2\sqrt{ \epsilon_{1k}^{2}+\Delta_{{\rm AFM}}^{2} } }
	}{\omega-\epsilon_{2k}+\mu-\sqrt{ \epsilon_{1k}^{2}+\Delta_{{\rm AFM}}^{2} }}
	\nn
	&+&
	\frac{
	\displaystyle
	\frac{1}{2}
	+
	\frac
	{ \epsilon_{1k} }
	{ 2\sqrt{ \epsilon_{1k}^{2}+\Delta_{{\rm AFM}}^{2} } }
	}{\omega-\epsilon_{2k}+\mu+\sqrt{ \epsilon_{1k}^{2}+\Delta_{{\rm AFM}}^{2} }}
	\label{AFM_2bands}.
}
The Green's function written in Eq.(\ref{AFM_zero})
shows that $G(k,\omega)$ vanishes 
at $\omega-\xi_{k+Q_{0}}=0$. 
In other words,
$G(k,\omega)$ has zero at $\omega-\xi_{k+Q_{0}}=0$.
This emergence of zeros is nothing but the
divergence of the self-energy of the electron $\hatd{c}{k\sigma}$,
$\Sigma(k,\omega)=\Delta_{{\rm AFM}}^{2}/(\omega-\xi_{k+Q_{0}})$
at $\omega-\xi_{k+Q_{0}}=0$,
as is seen in Eq.(\ref{AFM_zero}). 
As a result, the zero surface defined by $\omega=\xi_{k+Q_{0}}$
splits the pole surface defined by $\omega=\xi_{k}$ into 
the two pole surfaces $\omega=\epsilon_{2k}-\mu\pm
\sqrt{ \epsilon_{1k}^{2}+\Delta_{{\rm AFM}}^{2} }$, as is seen in
Eq.(\ref{AFM_2bands}).
Then the Fermi surface is disconnected into pockets at the gap edge.
As an example, a zero surface,
reconstructed band dispersion, and Fermi surface are depicted in Fig.\ref{AFM_band},
for $t'=0.25t$, $\mu=-0.75t$, and $\Delta_{{\rm AFM}}$.
This has to some extent a qualitative similarity to what is observed in
the cuprates. 
Another example is the case of {the} BCS superconductivity where the quasiparticle $\hatd{c}{k\uparrow}$
has a ``particle-particle hybridization" proportional to $\hatd{c}{k\uparrow}\hatd{c}{-k\downarrow}$
with $\hatd{c}{-k\downarrow}$\cite{AGD}.
 
However, as is already remarked in Sec.~\ref{Sec.I}, symmetry breakings such as 
the antiferromagnetic orders have not been universally observed 
in the hole-underdoped
cuprates where the ``Fermi arc" or the truncated Fermi surface
has been experimentally suggested.
Therefore, we need a mechanism for the emergence of zeros without any symmetry breakings.
In this context, the truncated Fermi surface scenario reviewed in Ref.\onlinecite{Rice_PTP}
gives an interesting example inspired by a result of the renormalization group methods,
which attributes the emergence of zeros to
Umklapp scatterings without the symmetry breakings\cite{Furukawa}. The pseudogap in this framework
is the precursor of the Hubbard gap.
We employ an alternative framework, where a zero surface with a gap formation naturally emerges by a hybridization with some additional fermionic excitations,
as is discussed in the above example of the antiferromagnetic ordered phase,
although our additional fermions are different from the quasiparticle at $k+Q_{0}$ in the antiferromagnetic case. Since our pseudogap (the hybridization gap) will turn out to be different from the remnant of the Hubbard gap, our framework will yield results qualitatively different from 
the scenario by the Umklapp scattering as we see in this paper.

Is there such fermionic excitations in proximity to Mott insulators
in the absence of symmetry breakings?
Or how do such fermionic excitations emerge?
A hint for \bu{the} existence of such additional fermionic excitations
comes from examination\bu{s} of the LUSW illustrated in Fig.~\ref{SW_HB} in the doped
\bu{Mott insulators}.
Hereafter we restrict our discussion to the hole-doped
\bu{Mott insulator}
as in Fig.~\ref{SW_HB},
with the hole-underdoped cuprates in mind.
The LUSW is defined as \bu{the} spectral weight above $\mu$ within the coherent band
near the top of the LHB.

We first recall the origin of the LUSW discussed
in the literature\cite{Meinders93,Phillips09}.
First we begin with the atomic limit, where $t=0$ and $U\neq 0$.
Then the LUSW of the hole-doped Mott insulator consists only of states
created by adding an $\sigma$-spin electron to an empty site, namely, a holon site,
to avoid creating a doubly occupied site, doublon, and \bu{to} avoid
the cost of the on-site Coulomb repulsion $U$.
An electron added to an empty site is confined tightly, in this limit. 
If the electron escapes from the holon sites, it inevitably
creates a doublon and costs $U$.
In other words, an electron created at
a holon site can contribute to the LUSW,
although this electron does not propagate coherently.

In the KR formalism, creation (annihilation) operators for the
electron at the holon site are
given by composite {\it fermion} operators
$\hatn{e}{i}\hatd{f}{i\sigma}$ ($\hatn{f}{i\sigma}\hatd{e}{i}$).
This is just a $\sigma$-spin electron
in the original Hubbard model
and is definitely fermionic.

When $t/U$ becomes nonzero, we have tightly bound doublons and holons even in the Mott insulating phase.
Therefore, holons in the hole-doped systems consist of both of the doped holons and the preexisting holons
already present in the insulators.
In the hole doped systems, these two types of holons should not be distinguished and should constitute the
same object.
Then an electron added to these holons should constitute LUSW
and may constitute a novel composite
particle distinguished from the original electron and offers a 
hint for the additional fermionic excitations
which brings about zeros to the quasiparticle Green's functions,
if this composite particle hybridizes with an original quasiparticle.

Originally the binding energy of a doublon and a holon
in the Mott insulator is the order of $U$.
However, for finite $t$ and finite doping $x$, indeed, quantum fluctuations 
induced by coherent carriers
dramatically weaken bindings between doublon-holon pairs.
{Then an electron added to this holon bound to a doublon requires only small energy and
merges into the excitation of an added electron to a doped holon site.}
Actually, remnants of doublon-holon pairs, namely, weakly-bound doublon-holon pairs
are known to play an important role in correlated electron systems, especially
in the context of variational wave function theories\cite{Kaplan,Yokoyama}.

A part of the tightly-bound doublon-holon pair, $\cond{e}\hatd{f}{i\sigma}$,
has {already been} taken into account
in the mean-field level {in the} previous theories
(please note $\hatn{e}{i}\hatd{f}{i\sigma}=\cond{e}\hatd{f}{i\sigma}+\wdtn{e}{i}\hatd{f}{i\sigma}$).
However,
it represents nothing but the renormalized quasiparticle,
which propagates in homogeneous mean fields and has nothing to do with
the above composite particle.
To substantiate our picture,
we need to take into account
the composite operator including bosonic fluctuations, $\wdtn{e}{i}\hatd{f}{i\sigma}$.
If we treat such \bu{a}
composite operator $\wdtn{e}{i}\hatd{f}{i\sigma}$ as a fermion, 
overlaps between
tightly-bound doublon-holon pairs
and quasiparticle states cause \bu{a} hybridization of fermions between two types,
where one is the quasiparticle $\hatd{f}{i\sigma}$ and the other is
the ``composite fermion"
or ``cofermion" 
$\wdtn{e}{i}\hatd{f}{i\sigma}$.
From such a hybridization between the cofermion and quasiparticle,
weakly-bound pairs discussed above
will naturally be born as a result.

If such a hybridization between the quasiparticle and the composite fermion
really exists,
this hybridization {would} contribute to the self-energy of quasiparticles.
Depending on the dynamics of the composite fermions,  
such self-energy {would} have {a} strong momentum-dependence
and possibly have poles near the zero energy.
In such cases, the hybridization
between the quasiparticles and the composite
particles cause\bu{s} {a} hybridization gap
near the Fermi level.
This gap is expected to account for the
pseudogap phenomena.
As is discussed below, we 
construct \bu{an} action containing both of
the quasiparticles and the composite fermions.

In the following sections,
we \bu{concretely} give our theoretical treatment {by}
introducing the composite fermions discussed above.
First,
we list up shortcomings of
the previous theories
and
present
our guiding principle
to overcome \bu{the}
previous KR's mean-field
theory.

As is mentioned in the above section,
the KR formalism gives exact results
if we thoroughly
take into account the fluctuations of slave bosons and
Lagrange multipliers introduced to keep the
constraint Eqs.(\ref{cnst1}) and (\ref{cnst2})
(
Hereafter we call this constraint
the local conservation
).
However,
the fluctuations of slave bosons
are neglected in the original
KR's mean-field theory.
In addition,
the constraint is also
treated just on average.
\bu{The o}riginal
constraint
is for the operators or the dynamical bosonic fields.
In the mean-field theory,
the constraint is replaced with averaged density of boson
condensations.
When we include the fluctuations of slave bosons,
we should take care of the constraint
or the local conservation of the densities of the slave bosons.
In the previous theories,
by including the fluctuations of charge bosons,
the incoherent Hubbard bands
are reproduced.
However, there exists a difficulty in conserving
spectral weight, because the fluctuations
violate the local conservation of \bu{the} boson density\cite{Raimondi}.

\begin{figure}[h]
\begin{center}
\includegraphics[width=6cm]{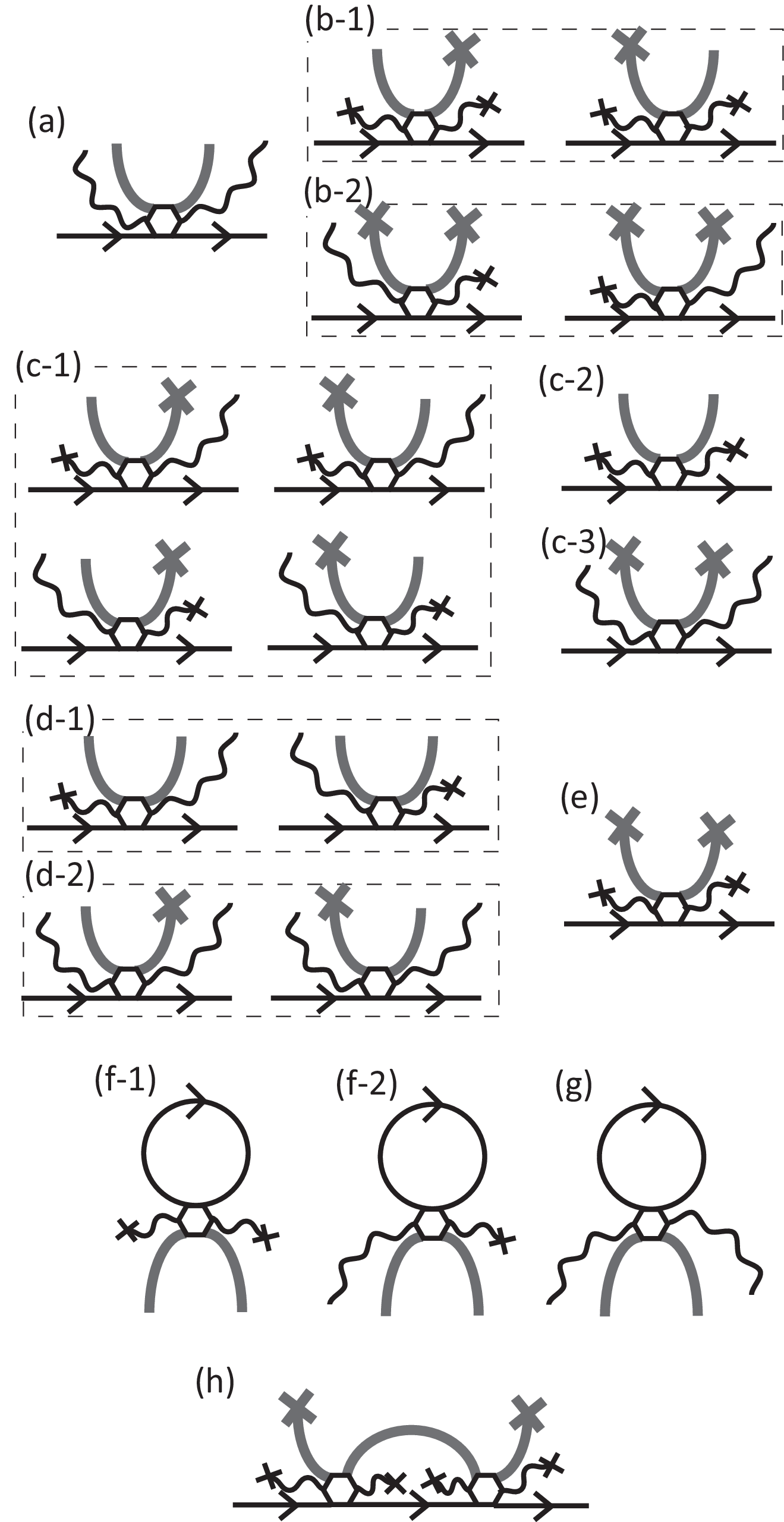}
\end{center}
\caption{(a)-(e)\bu{:}
Diagrams representing terms in $\hatd{f}{i\sigma}\hat{\zeta}_{ij}\hatn{f}{j\sigma}$.
(f)-(h)\bu{:} Examples of \bu{various} diagrams
\bu{on the one-loop level generated from the coupling (polygons)
and derived}
from
the
term\bu{s}
(a)-(e).
Solid lines with arrows represent propagators of the
quasiparticles.
Wavy lines stand for the charge bosons,
and \bu{bold} solid lines
are the spin bosons.
Condensations of bosons
are represented by lines terminated at crosses.
Coupling constant $g_{1\sigma}^{2}g_{2\sigma}^{2}t_{ij}$
is represented by open polygons.
Here,
we do not distinguish
holons and doublons.
Spins are also not distinguished in the diagram,
for simplicity.
\label{diagrams_gs}}
\end{figure}

When we impose
the local constraints
more strictly
for fluctuating bosons beyond the mean-field level, 
it turns out that the term
\eqsa{
	\hat{\zeta}_{ij\sigma}^{(1)}=
	g_{1\sigma}^{2}
	g_{2\sigma}^{2}
	(\wdtd{p}{i\sigma}\wdtn{e}{i}+\wdtd{d}{i}\wdtn{p}{i\overline{\sigma}})
	(\wdtd{e}{j}\wdtn{p}{j\sigma}+\wdtd{p}{j\overline{\sigma}}\wdtn{d}{j})
}
represented by the diagram in Fig.\ref{diagrams_gs}\bu{(a)}
is dominating among all the possible diagrams
for the kinetic term
in the action Eq. (\ref{A_SFB}), \bu{namely}
\eqsa{
	\int_{0}^{\beta}d\tau\ 
	\sum_{ij}
	\hatd{f}{i\sigma}(\tau)t_{ij}\hat{\zeta}_{ij\sigma}(\tau)\hatn{f}{j\sigma}(\tau)\bu{.}\label{VTX}
}
{Here we employ $g_{1\sigma}^{2}=(1-\overline{p}_{0\overline{\sigma}}^{2}
-\overline{e}^{2}_0)^{-1}$ and $g_{1\sigma}^{2}=(1-\overline{p}_{0\sigma}^{2}
-\overline{d}^{2}_0)^{-1}$ by following Ref.\onlinecite{KR}.}

To elucidate why we retain $\hat{\zeta}_{ij\sigma}^{(1)}$,
we
classify
the diagrams illustrated in Fig.\ref{diagrams_gs}
into four types,
categorized by
\bu{the}
time dependence (or frequency dependence)
of quasiparticles and the local conservation
of the boson densities:
\begin{description}
\item[\textsf{(T-1)}]
Diagrams
containing external propagators of quasiparticles,
in addition to
bosonic propagators
violating the local conservation
of boson densities
(Figures \ref{diagrams_gs}b,\ref{diagrams_gs}c,\ref{diagrams_gs}d, and \ref{diagrams_gs}h).
Here the violation means
that before and after
the interactions (represented by hexagons),
the number of bosons expressed
by external boson propagators
is not the same.
\item[\textsf{(T-2)}]
Diagrams that contain
time dependence
of quasiparticles,
but that do not violate the local conservation
(Figures \ref{diagrams_gs}a and \ref{diagrams_gs}e).
\item[\textsf{(T-3)}]
Diagrams that do not contain
time dependence
of quasiparticles
but do
violate the local conservation
(Fig.\ref{diagrams_gs}f).
\item[\textsf{(T-4)}]
Diagrams that
neither
include
time dependence
of quasiparticles
nor
violate the local conservation
(Fig.\ref{diagrams_gs}g).
\end{description}

{Here} we present our guiding principle to
take account of
boson fluctuations:
We exclude ({\bf \textsf{T-1}}) because it violates
the local conservation when the quasiparticles
dynamically fluctuate.
On the other hand we retain diagrams
belonging to the categories
({\bf \textsf{T-2}}), ({\bf \textsf{T-3}}), and ({\bf \textsf{T-4}}).
The reason to retain these diagram is as follows.
The diagrams in the category
({\bf \textsf{T-2}}) do not violate the local conservation,
when bosons fluctuate.
Therefore, we take the diagrams in this category
into account.
On the other hand,
the diagrams in the category
({\bf \textsf{T-3}}) do violate the local conservation.
However,
in these diagrams, quasiparticles
enter as time averaged Green's functions.
Therefore,
quasiparticles feel the time averaged bosonic motions.
The real violation of the local conservation
occurs only when a dynamical quasiparticle process
is induced by
fluctuating boson hoppings.
On the contrary,
the real
violation
does not occur when the quasiparticles emerge as
the time averaged quantities
as in the case of ({\bf \textsf{T-3}}).
This is the reason to retain the diagrams in the category ({\bf \textsf{T-3}}).
Since ({\bf \textsf{T-4}}) does not violate local conservation, we retain it.

For the slave-particle formalism of correlated fermion systems,
it is wellknown that
fluctuations of
gauge fields play an important role on
reinforcing the local constraint imposed on slave particles\cite{Lee_RMP}.
It was pointed out by Jolicoeur and Le Guillou that
the Kotliar-Ruckenstein formalism has
the $U$(1)$\times$$U$(1)$\times$$U$(1)
gauge symmetry\cite{gauge_KR}.
It comes from the phase symmetry of the slave bosonic particles, namely, $\hatn{e}{i}$, $\hatn{p}{i\sigma}$, and
$\hatn{d}{i}$.

In our theory, we will treat fluctuations of such phases together with fluctuations of the amplitude of the
condensation fraction of these slave particles, by using the Bogoliubov prescription.
Therefore, the phase fluctuations are taken into account, although 
the $U$(1)$\times$$U$(1)$\times$$U$(1)
gauge structure is not strictly conserved.
	\subsection{{Stratonovich-Hubbard trasformation}}\label{A1c}
	We introduce Grassmannian valuables (or fermionic fields)
$\hat{\mbox{\boldmath$\Upsilon$}}_{i\sigma}^{\ }
	=
	(\hatn{\psi}{i\sigma}, \hatn{\chi}{i\sigma})^{T}$
that stands for the cofermions as are discussed in
\bu{conceptually in Sec.~\ref{Sec.III}A.}
by using
\bu{the} following identity
\eqsa{
	\int
	\prod_{i\sigma}
	d\hat{\mbox{\boldmath$\Upsilon$}}_{i\sigma}^{\dagger}
	d\hat{\mbox{\boldmath$\Upsilon$}}_{i\sigma}^{\ }
	e^{
	\mathcal{A}
	}
	=
	\det
	\left[
	\widetilde{\mbox{\boldmath$T$}}_{\uparrow}\widetilde{\mbox{\boldmath$T$}}_{\downarrow}
	\right],\label{Id}
}
where matrices $\widetilde{\mbox{\boldmath$T$}}_{\sigma}$ are defined as
\eqsa{
(\widetilde{\mbox{\boldmath$T$}}_{\sigma})_{ij}={g_{1\sigma}^{2}g_{2\sigma}^{2}}{t_{ij}}
\left[
\begin{array}{cc}
\wdtd{p}{i\sigma}\wdtn{p}{j\sigma}&\wdtd{p}{i\sigma}\wdtd{p}{j\overline{\sigma}}\\
\wdtn{p}{i\overline{\sigma}}\wdtn{p}{j\sigma}&\wdtn{p}{i\overline{\sigma}}\wdtd{p}{j\overline{\sigma}}
\end{array}
\right],
}
and
\eqsa{
	\mathcal{A}
	&=&
	\int_{0}^{\beta}d\tau
	\sum_{ij\sigma}
	\left[
	\left(\hat{\mbox{\boldmath$\Upsilon$}}_{i\sigma}^{\dagger}(\tau)
	-
	\hat{\mbox{\boldmath$C$}}_{i\sigma}^{\dagger}(\tau)
	\right)
	\right.
	\nn
	&\times&
	\left.
	(\widetilde{\mbox{\boldmath$T$}}_{\sigma})_{ij}
	\left(\hat{\mbox{\boldmath$\Upsilon$}}_{j\sigma}^{\ }(\tau)
	-
	\hat{\mbox{\boldmath$C$}}_{j\sigma}^{\ }(\tau)
	\right)
	\right].
}
Here we use vector notations as
\eqsa{
	\hat{\mbox{\boldmath$C$}}_{i\sigma}^{\dagger}
	&=
	\left(\wdtn{e}{i},\wdtd{d}{i}\right)\hatd{f}{i\sigma},\ 
	\hat{\mbox{\boldmath$C$}}_{i\sigma}^{\ }
	=
	\hatn{f}{i\sigma}\left(\wdtd{e}{i},\wdtn{d}{i}\right)^{T}.
}

\begin{figure}
\begin{center}
\includegraphics[width=7cm]{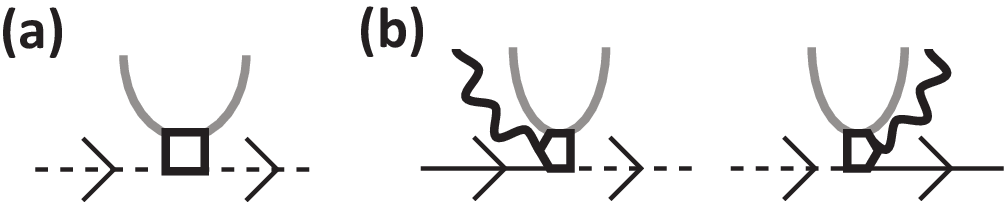}
\end{center}
\caption{
Diagrams for
transformed Lagrangians
\bu{defined in Eqs.(\ref{TL1})-(\ref{S_2})}.
Solid and dashed lines with arrows represent propagators of the
quasiparticles
and cofermions, respectively.
Wavy lines stand for the charge bosons,
and \bu{bold} solid lines
are the spin bosons.
(a)
The diagram represents the Lagrangian $\mathcal{L'}_{a1}$ (Eq.(\ref{S_1})).
(b)
The diagrams stand for terms in the Lagrangian $\mathcal{L'}_{a2}$ (Eq.(\ref{S_2})).
\label{diagrams_2}}
\end{figure}

The identity Eq.(\ref{Id}) gives the transformation
for a coupling term of the quasiparticles and fluctuating bosons depicted
in Fig.\ref{diagrams_gs}a,
\eqsa{
	\mathcal{L}_{{\rm a}}
	=
	\sum_{ij\sigma}
	\hat{\mbox{\boldmath$C$}}_{i\sigma}^{\dagger}(\tau)
	(\widetilde{\mbox{\boldmath$T$}}_{\sigma})_{ij}
	\hat{\mbox{\boldmath$C$}}_{j\sigma}^{\ }(\tau),
}
as
\eqsa{
	\exp\left[
	-\int_{0}^{\beta}d\tau
	\mathcal{L}_{{\rm a}}
	\right]
	=
	\frac{
	\displaystyle
	\int
	\prod_{i\sigma}
	d\hat{\mbox{\boldmath$\Upsilon$}}_{i\sigma}^{\dagger}
	d\hat{\mbox{\boldmath$\Upsilon$}}_{i\sigma}^{\ }
	e^{
	-\int_{0}^{\beta}d\tau
	\mathcal{L'}_{a}
	}}{\det [\wdtn{\mbox{\boldmath$T$}}{\uparrow}\wdtn{\mbox{\boldmath$T$}}{\downarrow}]}
,
}
where
\eqsa{
	\mathcal{L'}_{a}
	=
	\mathcal{L'}_{a1}
	+
	\mathcal{L'}_{a2}
	\label{TL1},
}
\eqsa{
	\mathcal{L'}_{a1}
	=
	\sum_{ij\sigma}
	\hat{\mbox{\boldmath$\Upsilon$}}_{i\sigma}^{\dagger}(\tau)
	(\widetilde{\mbox{\boldmath$T$}}_{\sigma})_{ij}
	\hat{\mbox{\boldmath$\Upsilon$}}_{j\sigma}^{\ }(\tau)
	\label{S_1},
}
\eqsa{
	\mathcal{L'}_{a2}
	&=&
	-
	\sum_{ij\sigma}
	\left\{
	\hat{\mbox{\boldmath$C$}}_{i\sigma}^{\dagger}(\tau)
	(\widetilde{\mbox{\boldmath$T$}}_{\sigma})_{ij}
	\hat{\mbox{\boldmath$\Upsilon$}}_{j\sigma}^{\ }(\tau)
	\right.
	\nn
	&+&
	\left.
	\hat{\mbox{\boldmath$\Upsilon$}}_{i\sigma}^{\dagger}(\tau)
	\widetilde{\mbox{\boldmath$T$}}_{ij}
	\hat{\mbox{\boldmath$C$}}_{j\sigma}^{\ }(\tau)
	\right\}	
	.\label{S_2}
}
These transformed Lagrangian $\mathcal{L'}_{a1}$ (Fig.\ref{diagrams_2}a)
and $\mathcal{L'}_{a2}$ (Fig.\ref{diagrams_2}b)
lead to the cofermions' self-energy and the hybridization between the quasiparticles
and cofermions, respectively after integrating out the fluctuating bosonic degrees of freedom.
It will be
discussed below, by using a set of the Dyson equations.
	\subsection{{Prescription for self-consistent procedure and Green's functions}}\label{A1d}
\begin{figure}[h]
\begin{center}
\includegraphics[width=6cm]{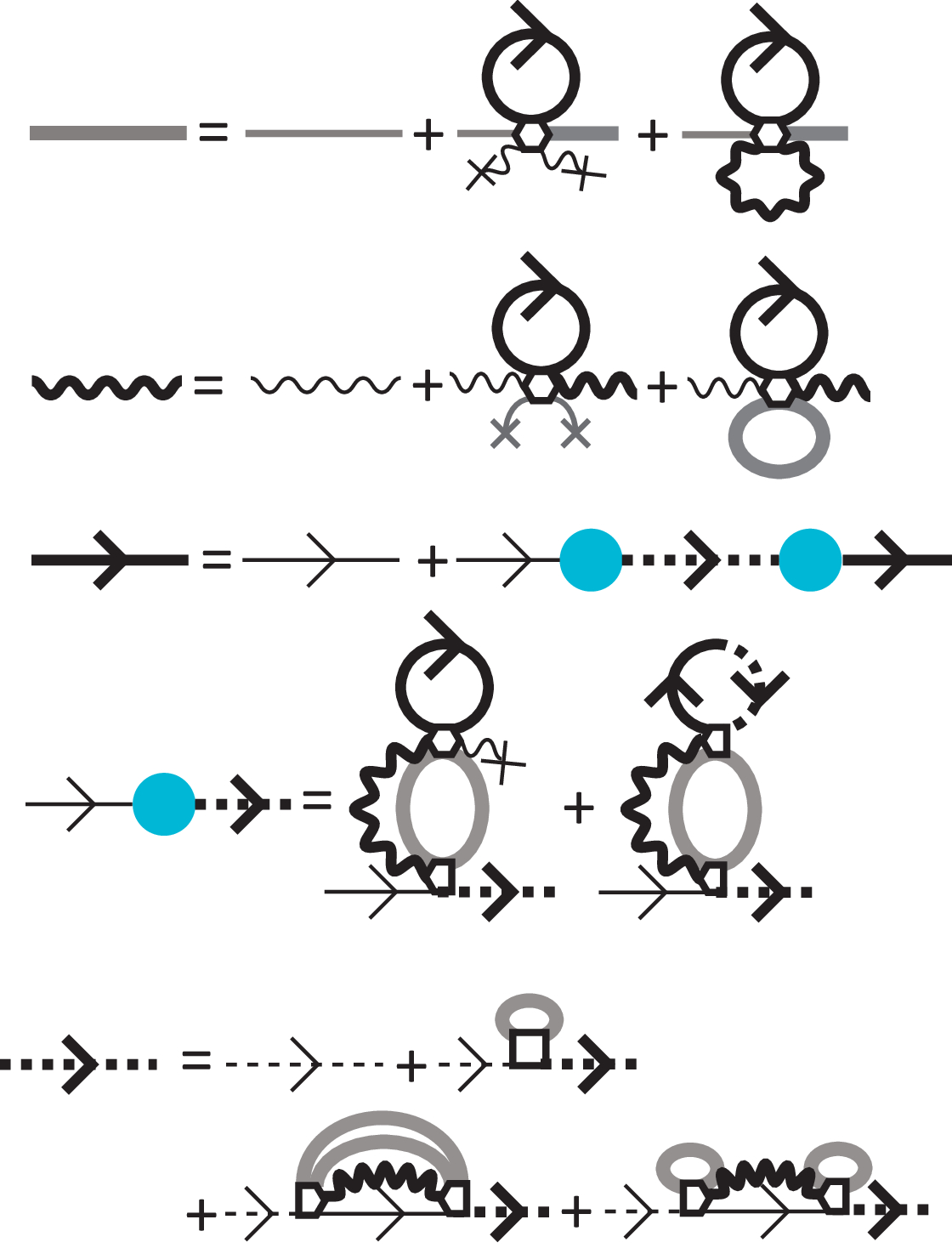}
\end{center}
\caption{
Diagrams for
Dyson equations.
Solid and dashed lines with arrows represent propagators of the
quasiparticles
and cofermions, respectively.
\bu{Other notations are the same as Fig.\ref{diagrams_gs}.
Bold (thin) lines are for renormalized (bare) propagators.
Filled (blue) circles are amplitudes of the hybridization
between quasiparticles and cofermions.}
\label{Dyson_2}}
\end{figure}

Here we construct approximated Green's functions for the Gaussian fluctuations of the bosons,
quasiparticles, and cofermions by using a set of
Dyson equations as is depicted in Fig.\ref{Dyson_2}:
\bu{Wavy} lines and \bu{bold}
\bu{grey} lines
stand for the Green's functions of the charge bosons
$
\mathcal{A}^{ab}(r,\tau)
=-\avrg{T\beta_{i}^{a}(\tau){\beta_{j}^{b}}^{\dagger}(0)},
$
and the spin bosons
$
\mathcal{C}^{ab}(r,\tau)
=-\avrg{T\phi_{i}^{a}(\tau){\phi_{j}^{b}}^{\dagger}(0)},
$
respectively, where $a,b=1,2$, $r=i-j$, $(\beta_{i}^{1},\beta_{i}^{2})=(\wdtn{e}{i},\wdtn{d}{i})$,
and $(\phi_{i}^{1},\phi_{i}^{2})=(\wdtn{p}{i\sigma},\wdtd{p}{i\overline{\sigma}})$.
\bu{Bold}
lines with arrows represent
the quasiparticles \bu{propagator} $\mathcal{G}_{\sigma}^{(f)}(r,\tau)$.
On the other hand, Thin \bu{wavy} lines and thin lines represent
bare propagators of the \bu{charge} bosons $\mathcal{A}_{0}^{ab}(r,\tau)$,
the \bu{spin} bosons $\mathcal{C}_{0}^{ab}(r,\tau)$, respectively,
determined by $\hat{\mathcal{L}}_{{\rm B}}^{(0)}$,
in which self-energy effects are not taken into account.
Thin lines with arrows stand for bare propagators of the quasiparticles
$\mathcal{G}_{0\sigma}^{(f)}(r,\tau)$ determined by
\eqsa{
	\hat{\mathcal{L}}_{0}=\sum_{ij}
	\hatd{f}{i\sigma}(\tau)
	[
	(\partial_{\tau}-\mu)\delta_{ij}
	+
	\zeta_{0\sigma}t_{ij}
	]
	\hatn{f}{j\sigma}(\tau),
}
where $\zeta_{0\sigma}=g_{1\sigma}^{2}g_{2\sigma}^{2}
(\conp{p}{\sigma}\cond{e}+\cond{d}\conp{p}{\overline{\sigma}})^{2}$.
The Lagrangian $\hat{\mathcal{L}}_{0}$ is obtained by decoupling the fluctuating bosons
from the KR's action
(see Eq.(\ref{A_SFB})).
\bu{Bold}
and thin dashed lines stand for the cofermions' propagators $\mathcal{F}^{ab}$ and bare propagators {$\mathcal{F}^{ab}_{0}=\delta_{a,b}/\epsilon$ $(\epsilon\rightarrow 0)$}, respectively.

In the set of Dyson equations (Fig.\ref{Dyson_2}),
we neglect the coupling between charge and spin bosons described by
propagators
such as
$\avrg{\wdtd{p}{i\sigma}\wdtn{e}{i}}$,
at the Gaussian level, since these coupling terms
are higher order contributions.
Below we explain that the coupling gives higher order contribution\bu{s}
with respect to \bu{the} hole-doping rate $x$, in proximity to Mott insulating states:
Since operators including both charge and spin such as $\wdtd{p}{i\sigma}\wdtn{e}{i}$
do not conserve the electric charge, propagators such as
$\avrg{\wdtd{p}{i\sigma}\wdtn{e}{i}}$ should vanish in the Mott insulating phase, where the charge
can not fluctuate.
Therefore,
the charge and spin excitations
are well separated in the Mott insulating phase.

When hole carriers are doped, $\wdtd{p}{i\sigma}\wdtn{e}{i}$ can have a
nonzero expectation value, at most, scaled by \bu{the}
condensate fraction of holons $\overline{e}_{0}$
which gives a rough estimate of the amplitude of charge fluctuations.
From a relation $\overline{e}_{0}^2\propto x$ held in the KR theory for the hole-doped case,
we obtain $\avrg{\wdtd{p}{i\sigma}\wdtn{e}{i}}\propto \sqrt{x}$.
Futhermore,
there is an additional constraint for the coupling terms such as $\avrg{\wdtd{p}{i\sigma}\wdtn{e}{i}}$:
they do not appear alone in calculations of physical quantities.
To conserve charge and spin on average,  $\avrg{\wdtd{p}{i\sigma}\wdtn{e}{i}}$ appears with $\avrg{\wdtd{e}{i}\wdtn{p}{i\sigma}}$ in pair, for example.
Therefore, the contribution of the coupling between charge and spin bosons to
physical quantities is scaled by $(\sqrt{x})^2$.
It concludes that the coupling between the charge and spin bosons gives contributions
as a higher order in terms of $x$ in physical quantities.

By solving the set of
Dyson equations,
we obtain the propagators for the quasiparticles and cofermions.
Here the bosonic degrees of freedom are taken into account in \bu{a} self-consistent fashion, 
through the cofermion self-energy
$\mbox{\boldmath$\Sigma$}^{({\rm cf})}_{\sigma}(r,\tau)$, and the amplitude
$\mbox{\boldmath$\Delta$}_{ij}$ of hybridization
between the quasiparticles and cofermions,
each of which
we detail below.

The Lagrangian for the cofermions is given by
{\eqsa{
	\hat{\mathcal{L}}_{{\rm cf}}
	=
	-
	\sum_{ij\sigma}
	\hat{\mbox{\boldmath$\Upsilon$}}_{i\sigma}^{\dagger}(\tau)
	\left[
	\mbox{\boldmath$\Sigma$}^{({\rm cf})}_{\sigma}(r,\tau)
	\right]
	\hat{\mbox{\boldmath$\Upsilon$}}_{j\sigma}(\tau),
}
where $\hat{\mbox{\boldmath$\Upsilon$}}_{i\sigma}^{\dagger}=(\hatd{\psi}{i\sigma}, \hatd{\chi}{i\sigma})$
is a vector notation for the cofermions, as is defined in the main article, and $r=i-j$.
The cofermion self-energy $\mbox{\boldmath$\Sigma$}^{({\rm cf})}_{\sigma}(r,\tau)$ is a 2$\times$2 symmetric matrix,
\eqsa{
	\mbox{\boldmath$\Sigma$}^{({\rm cf})}_{\sigma}
	=
	\left[
	\begin{array}{cc}
	\Sigma_{\sigma}^{11}&\Sigma'_{\sigma}\\
	\Sigma'_{\sigma}&\Sigma_{\sigma}^{22}\\
	\end{array}
	\right].\label{self-energy_matrix}
}
The details for the self-energy matrix are given in Appendix.\ref{app_self_ene_sec}.
On the other hand,
the hybridization between the quasiparticles and cofermions is described by
\eqsa{
	\hat{\mathcal{L}}_{{\rm hyb}}
	=
	\sum_{i,j,\sigma}
	[
	\hat{\mbox{\boldmath$\Upsilon$}}_{i\sigma}^{\dagger}(\tau)
	\mbox{\boldmath$\Delta$}_{ij}
	\hatn{f}{j\sigma}(\tau)
	+
	\hatd{f}{i\sigma}(\tau)
	\mbox{\boldmath$\Delta$}_{ij}^{T}
	\hat{\mbox{\boldmath$\Upsilon$}}_{j\sigma}(\tau)
	],
}
where $\mbox{\boldmath$\Delta$}_{ij}^{T}=(\Delta_{ij}^{(\psi)}, \Delta_{ij}^{(\chi)})$.
For details for $\mbox{\boldmath$\Delta$}_{ij}$,
see Appendix.\ref{Appendix_hyb}. 

As a result,
the effective Lagrangian for the quasiparticles and cofermions, $\hat{\mathcal{L}}_{{\rm eff}}$, is 
given as
\eqsa{
\hat{\mathcal{L}}_{{\rm eff}}=\hat{\mathcal{L}}_{0}+\hat{\mathcal{L}}_{{\rm cf}}+\hat{\mathcal{L}}_{{\rm hyb}}.
}

When the charge gap is relatively small, $\Sigma_{\sigma}^{11}\simeq\Sigma_{\sigma}^{22}$
and $\Delta_{ij}^{(\psi)}\simeq\Delta_{ij}^{(\chi)}$
hold approximately. When the charge gap collapses, $\Sigma_{\sigma}^{11}=\Sigma_{\sigma}^{22}$
and $\Delta_{ij}^{(\psi)}=\Delta_{ij}^{(\chi)}$ hold exactly.}
In our results, we employ
approximate relations
$\Sigma=\Sigma_{\sigma}^{11}\simeq\Sigma_{\sigma}^{22}$ and
$\Delta_{ij}=\Delta_{ij}^{(\psi)}\simeq\Delta_{ij}^{(\chi)}$.
Then a cofermion mode $(\hatn{\psi}{k\sigma}+\hatn{\chi}{k\sigma})/\sqrt{2}$
hybridize\bu{s} with quasiparticles through the amplitude $\Delta (k)$, which is
depicted in Fig.\ref{Dyson_2} as closed (blue) circles, where $k$ is a momentum.
The inverse of cofermion propagator (namely, the cofermion self-energy)
for $(\hatn{\psi}{k\sigma}+\hatn{\chi}{k\sigma})/\sqrt{2}$
is given as
\eqsa{
\frac{-1}{2}[\Sigma_{\sigma}(k,i\varepsilon_{n})+\Sigma'_{\sigma}(k,i\varepsilon_{n})]
=
\gamma_{k}i\varepsilon_{n}-\alpha_{k}+O(\varepsilon_{n}^{2}),
}
where $\varepsilon_{n}$ is a fermionic Matsubara frequency.

Then,
the Green's function for the quasiparticles is
given as
\eqsa{
	&\mathcal{G}_{\sigma}^{(f)}(k,i\varepsilon_{n}\rightarrow \omega +i\delta)
	=
	G_{\sigma}^{(f)}(k,\omega+i\delta)
	\nn
	&\simeq
	\displaystyle
	\left[\omega+i\delta
	-\zeta_{0\sigma}\epsilon_{k}
	+\mu-
	\frac{
	\Delta(k)^{2}
	}{
	\gamma_{k}
	(\omega
	+i\delta)
	-\alpha_{k}
	}
	\right]^{-1},
}
where $\epsilon_{k}$ is the Fourier transformation of $t_{ij}$, and $\mu$
is the chemical potential.
Here we note that
the weights of the two quasiparticle bands split by the zero surface defined by $\omega=\alpha_{k}/\gamma_{k}$
are not the same in our theory.

In our calculations, we define the doping rate $x$ by using the quasiparticle Green's function as
\eqsa{
	1-x=\lim_{T\rightarrow 0+}\frac{T}{N_{s}}\sum_{k,i\varepsilon_{n},\sigma}\mathcal{G}_{\sigma}^{(f)}(k,i\varepsilon_{n}),
}
where $T$ stands for temperature and $N_{s}$ is the number of sites.

The Green's function for the electrons, instead of the quasiparticles,
is given as
\eqsa{
	\mathcal{G}_{ij\sigma}(\tau)
	&=&-\avrg{T \hatn{c}{i\sigma}(\tau)\hatd{c}{j\sigma}(0)}
	\nn
	&\simeq&-\avrg{T \hatd{z}{i\sigma}(\tau)\hatn{z}{j\sigma}(0)}
	\nn
	&\times&\avrg{T \hatn{f}{i\sigma}(\tau)\hatd{f}{j\sigma}(0)}
	\nn
	&=&\avrg{T \hatd{z}{i\sigma}(\tau)\hatn{z}{j\sigma}(0)}
	\mathcal{G}^{(f)}_{ij\sigma}(\tau),\label{GREEN_16}
}
where the bosonic and fermionic degrees of freedom are
decoupled, because the resultant action in our theory
does not contain the hybridization between bosons and fermions.
The quasiparticle Green's function is defined, as in
the previous sections, as
\eqsa{
	\mathcal{G}^{(f)}_{ij\sigma}(\tau)
	=-\avrg{T \hatn{f}{i\sigma}(\tau)\hatd{f}{j\sigma}(0)}.
}
The bosonic part in Eq.(\ref{GREEN_16}) is given by
\eqsa{
	\avrg{T \hatd{z}{i\sigma}(\tau)\hatn{z}{j\sigma}(0)}
	&\simeq& 
	g_{1\sigma}^{2}g_{2\sigma}^{2}
	\left\langle
	T [\hatd{\bvec{b}}{i}(\tau)\cdot \hatn{\bvec{p}}{i\sigma}(\tau)]
	\right.
	\nn
	&\times&
	\left.
	[\hatd{\bvec{p}}{j\sigma}(\tau)\cdot \hatn{\bvec{b}}{j}(\tau)]
	\right\rangle,
}
where we use vector notation as
$\hatd{\bvec{b}}{i}=(\hatd{e}{i},\hatn{d}{i})$, 
$\hatd{\bvec{p}}{i\sigma}=(\hatd{p}{i\sigma},\hatn{p}{i\overline{\sigma}})$.
Because
we adopt the boson dynamics in which
charge and spin bosons are decoupled,
this bosonic part of the Green's function is
rewritten as
\eqsa{
	&&\avrg{T \hatd{z}{i\sigma}(\tau)\hatn{z}{j\sigma}(0)}
	\nn
	&&\simeq 
	g_{1\sigma}^{2}g_{2\sigma}^{2}
	\left\langle
	T [\overline{\bvec{b}}_{0}\cdot \overline{\bvec{p}}_{0\sigma}^{T}]
	[\overline{\bvec{p}}_{0\sigma}\cdot \overline{\bvec{b}}_{0}^{T}]
	\right\rangle
	\nn
	&&+
	g_{1\sigma}^{2}g_{2\sigma}^{2}
	\left\langle
	T [\wdtd{\bvec{b}}{i}(\tau)\cdot \overline{\bvec{p}}_{0\sigma}^{T}]
	[\overline{\bvec{p}}_{0\sigma}\cdot \wdtn{\bvec{b}}{j}(\tau)]
	\right\rangle
	\nn
	&&+ 
	g_{1\sigma}^{2}g_{2\sigma}^{2}
	\left\langle
	T [\overline{\bvec{b}}_{0}\cdot \wdtn{\bvec{p}}{i\sigma}(\tau)]
	[\wdtd{\bvec{p}}{j\sigma}(\tau)\cdot \overline{\bvec{b}}_{0}^{T}]
	\right\rangle
	\nn
	&&+ 
	g_{1\sigma}^{2}g_{2\sigma}^{2}
	\left\langle
	T [\wdtd{\bvec{b}}{i}(\tau)\cdot \wdtn{\bvec{p}}{i\sigma}(\tau)]
	[\wdtd{\bvec{p}}{j\sigma}(\tau)\cdot \wdtn{\bvec{b}}{j}(\tau)]
	\right\rangle,\label{B_PART}
}
where
$\overline{\bvec{b}}_{0}=(\cond{e},\cond{d})$ and
$\overline{\bvec{p}}_{0\sigma}=(\conp{p}{\sigma},\conp{p}{\overline{\sigma}})$.
If we retain only the first and second lines of the right hand side of Eq.(\ref{B_PART}),
the electron Green's function is reduced to that already obtained in Ref.\onlinecite{Raimondi}. 
The contribution of the fourth line
\bu{in}
the right hand side of Eq.(\ref{B_PART}) is
small compared with these from other lines
in Eq.(\ref{B_PART})
\bu{because it is the fourth order in the terms
of the fluctuation denoted by tildes},
and we ignore the fourth term.
\subsection{{Superconductivities mediated by \bu{cofermions} and charge bosons}
\label{subsec_SC}}
\begin{figure}[t]
\begin{center}
\includegraphics[width=8cm]{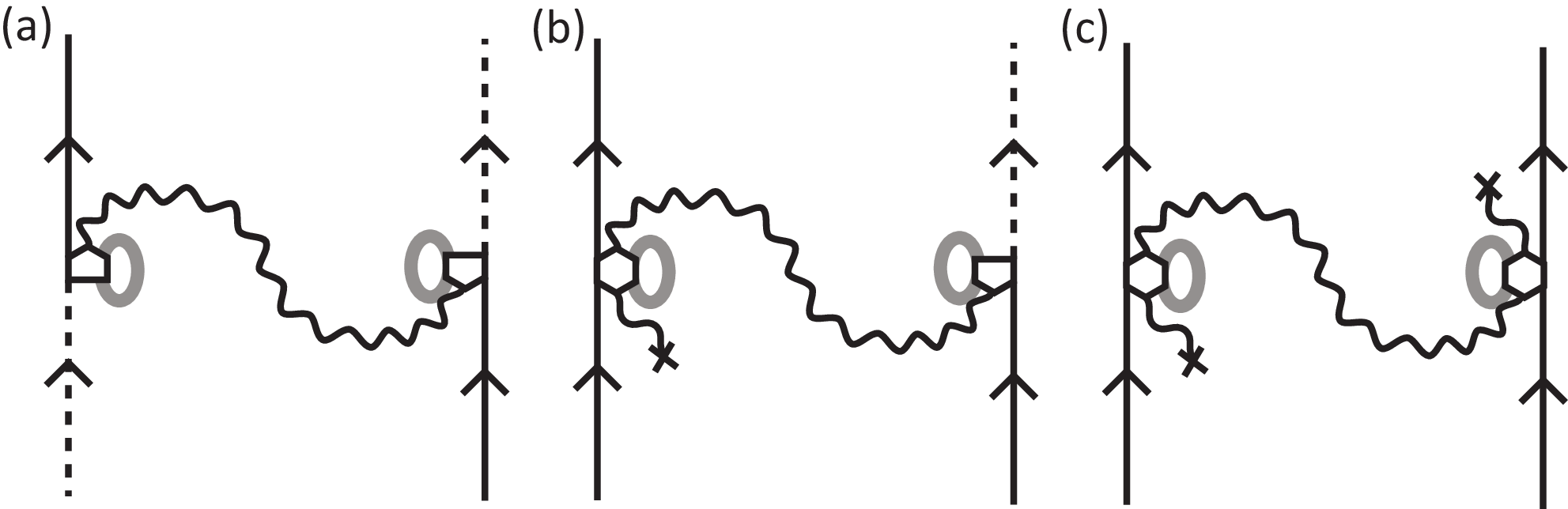}
\end{center}
\caption{Interactions
\bu{between quasiparticles and cofermions mediated}
by exchanging one charge boson.
\label{Fig_EX}}
\end{figure}

In this subsection,
we examine
how the present insight from the novel cofermion excitations offers
the mechanism for
the emergence of superconductivity.
Here we examine the singlet superconductivity induced by
exchanging one charge boson (see Fig.\ref{Fig_EX}).

We start from
the mean-field action
for the singlet superconductivity.
Here the effective interaction of quasiparticles and cofermions is
induced by
exchanging one charge boson.
The mean-field action $S_{{\rm SC}}$ is defined as
\eqsa{
	S_{{\rm SC}}
	=
	\sum_{k,i\varepsilon_{n}}
	\mbox{\boldmath$\Phi$}^{\dagger}_{k}(i\varepsilon_{n})
	\left[
	\mbox{\boldmath$\mathcal{G}$}(k,i\varepsilon_{n})
	\right]^{-1}
	\mbox{\boldmath$\Phi$}^{\ }_{k}(i\varepsilon_{n}),
	\label{S_SC_ff}
}
where $\mbox{\boldmath$\Phi$}^{\dagger}_{k}(i\varepsilon_{n})$ and 
$\mbox{\boldmath$\Phi$}^{\ }_{k}(i\varepsilon_{n})$ are defined as
\begin{widetext}
\eqsa{
	\mbox{\boldmath$\Phi$}^{\dagger}_{k}(i\varepsilon_{n})
	&=&
	\left[
	\hatd{f}{k\uparrow}(i\varepsilon_{n}), \hatd{\psi}{k\uparrow}(i\varepsilon_{n})
	+\hatd{\chi}{k\uparrow}(i\varepsilon_{n}), \hatn{f}{-k\downarrow}(-i\varepsilon_{n}),
	\hatn{\psi}{-k\downarrow}(-i\varepsilon_{n})
	+\hatn{\chi}{-k\downarrow}(-i\varepsilon_{n})
	\right],
	\\
	\mbox{\boldmath$\Phi$}^{\ }_{k}(i\varepsilon_{n})
	&=&
	\left[
	\hatn{f}{k\uparrow}(i\varepsilon_{n}), \hatn{\psi}{k\uparrow}(i\varepsilon_{n})
	+\hatn{\chi}{k\uparrow}(i\varepsilon_{n}),
	\hatd{f}{-k\downarrow}(-i\varepsilon_{n}),
	\hatd{\psi}{-k\downarrow}(-i\varepsilon_{n})
	+\hatd{\chi}{-k\downarrow}(-i\varepsilon_{n})
	\right]^{T}.
}
The inverse of the matrix form of the Green's function
is given by
\eqsa{
	&&\left[
	\mbox{\boldmath$\mathcal{G}$}(k,i\varepsilon_{n})
	\right]^{-1}
	=
	\left[
	\begin{array}{cccc}
	i\varepsilon_{n}-\xi_{k}&\Delta_{k}&M_{k}&\Lambda_{k}\\
	\Delta_{k}& \gamma_{k}i\varepsilon_{n}-\alpha_{k}& \Lambda_{k}& 0\\
	M_{k}& \Lambda_{k}& i\varepsilon_{n}+\xi_{k}&-\Delta_{k}\\
	\Lambda_{k}& 0 & -\Delta_{k}& \gamma_{k}i\varepsilon_{n}+\alpha_{k}\\
	\end{array}
	\right]\nn
	&&=
	\left[
	\begin{array}{cccc}
	1&0&0&0\\
	0&\sqrt{\gamma_{k}}&0&0\\
	0&0&1&0\\
	0&0&0&\sqrt{\gamma_{k}}\\
	\end{array}
	\right]
	\left[
	\begin{array}{cccc}
	i\varepsilon_{n}-\xi_{k}&\wdtn{\Delta}{k}&M_{k}&\wdtn{\Lambda}{k}\\
	\wdtn{\Delta}{k}& i\varepsilon_{n}-\wdtn{\alpha}{k}& \wdtn{\Lambda}{k}& 0\\
	M_{k}& \wdtn{\Lambda}{k}& i\varepsilon_{n}+\xi_{k}&-\wdtn{\Delta}{k}\\
	\wdtn{\Lambda}{k}& 0 & -\wdtn{\Delta}{k}& i\varepsilon_{n}+\wdtn{\alpha}{k}\\
	\end{array}
	\right]
	\left[
	\begin{array}{cccc}
	1&0&0&0\\
	0&\sqrt{\gamma_{k}}&0&0\\
	0&0&1&0\\
	0&0&0&\sqrt{\gamma_{k}}\\
	\end{array}
	\right],
}
where $\xi_{k}=\zeta_{0}\epsilon_{k}-\mu$, and $M_{k}$ and $\Lambda_{k}$ are superconducting order parameters
{as is explicitly given in Eqs.(\ref{SCE_M}) and (\ref{SCE_L})}.
Here the superconducting order parameters
are given as
\eqsa{
	M_{k}(i\varepsilon_{n})
	&=&\frac{T}{N_{s}}\sum_{i\omega_{m},Q}
	\widetilde{t}_{k}\widetilde{t}_{k+Q}
	\avrg{\wdtd{b}{Q}(i\omega_{m})\wdtn{b}{Q}(i\omega_{m})}
	\nn
	&\times&
	\left[
	\left(\frac{\cond{e}+\cond{d}}{\sqrt{2}}\right)^{2}
	\left[\mbox{\boldmath$\mathcal{G}$}(k+Q,i\varepsilon_{n}+i\omega_{m})\right]_{13}
	+
	\frac{\cond{e}+\cond{d}}{\sqrt{2}}
	\left[\mbox{\boldmath$\mathcal{G}$}(k+Q,i\varepsilon_{n}+i\omega_{m})\right]_{14}
	\right]
	\label{gap_01}
}
and
\eqsa{
	\Lambda_{k}(i\varepsilon_{n})
	&=&\frac{T}{N_{s}}\sum_{i\omega_{m},Q}
	\widetilde{t}_{k}\widetilde{t}_{k+Q}
	\avrg{\wdtd{b}{Q}(i\omega_{m})\wdtn{b}{Q}(i\omega_{m})}
	\nn
	&\times&
	\left[
	\left[\mbox{\boldmath$\mathcal{G}$}(k+Q,i\varepsilon_{n}+i\omega_{m})\right]_{14}
	+
	\frac{\cond{e}+\cond{d}}{\sqrt{2}}
	\left[\mbox{\boldmath$\mathcal{G}$}(k+Q,i\varepsilon_{n}+i\omega_{m})\right]_{13}
	\right],
	\label{gap_02}
}
where $
\left[\mbox{\boldmath$\mathcal{G}$}(k+Q,i\varepsilon_{n}\right]_{ij}$
is the ($i,j$)-th element of the matrix $\mbox{\boldmath$\mathcal{G}$}$.  
We should note that there are quasiparticle-quasiparticle and cofermion-quasiparticle
singlet pairings, which are described by following anomalous Green's functions:
\eqsa{
	\left[\mbox{\boldmath$\mathcal{G}$}(k,i\varepsilon_{n})\right]_{13}
	&=&
	\frac{
	2\wdtn{\alpha}{k}\wdtn{\Delta}{k}\wdtn{\Lambda}{k}
	-
	M_{k}\left[\left(i\varepsilon_{n}\right)^{2}-\widetilde{\alpha}_{k}^{2}\right]
	}
	{\left[\left(i\varepsilon_{n}\right)^{2}-\lambda_{k+}^{2}\right]
	\left[\left(i\varepsilon_{n}\right)^{2}-\lambda_{k-}^{2}\right]}
	,
	\\
	\left[\mbox{\boldmath$\mathcal{G}$}(k,i\varepsilon_{n})\right]_{14}
	&=&
	\frac{1}{\sqrt{\gamma_{k}}}
	\frac{-M_{k}\wdtn{\Delta}{k}\left(i\varepsilon_{n}-\wdtn{\alpha}{k}\right)
	+\wdtn{\Lambda}{k}
	\left[
	\widetilde{\Lambda}_{k}^{2}+\widetilde{\Delta}_{k}^{2}-
	\left(i\varepsilon_{n}-\wdtn{\alpha}{k}\right)
	\left(i\varepsilon_{n}+\xi_{k}\right)
	\right]
	}
	{\left[\left(i\varepsilon_{n}\right)^{2}-\lambda_{k+}^{2}\right]
	\left[\left(i\varepsilon_{n}\right)^{2}-\lambda_{k-}^{2}\right]},
}
where
\eqsa{
	\lambda_{k\pm}
	&=&
	\frac{
	\sqrt{X_{k}\pm\sqrt{X_{k}^{2}-4Y_{k}}}
	}
	{\sqrt{2}}
	,
	\\
	X_{k}&=&\widetilde{\alpha}_{k}^{2}+\xi_{k}^{2}
	+2\widetilde{\Delta}_{k}^{2}+2\widetilde{\Lambda}_{k}^{2}+2M_{k}^{2}
	,
	\\
	Y_{k}&=&
	\left(
	\widetilde{\alpha}_{k}M_{k}+2\widetilde{\Delta}\widetilde{\Lambda}_{k}
	\right)^{2}
	+
	\left[
	\widetilde{\alpha}_{k}\xi_{k}+\left(\widetilde{\Lambda}_{k}^{2}-\widetilde{\Delta}_{k}^{2}\right)
	\right]^{2}.
}
For simplicity,
we introduce the cut off frequency
\eqsa{
	\omega_{c}={\rm max}\left\{\zeta_{0}t, \sqrt{\widetilde{\Lambda}_{k}^{2}+M_{k}^{2}}\right\},
}
to the bosonic propagators for charge fluctuations $\wdtd{b}{i}=\wdtd{e}{i}+\wdtn{d}{i}$ and
apply a quasi-static approximation to solve the gap equations as 
\eqsa{
	M_{k}(i\varepsilon_{n})
	&\times&
	\left[
	\left(\frac{\cond{e}+\cond{d}}{\sqrt{2}}\right)^{2}
	\left[\mbox{\boldmath$\mathcal{G}$}(k+Q,i\varepsilon_{n}+i\omega_{m})\right]_{13}
	+
	\frac{\cond{e}+\cond{d}}{\sqrt{2}}
	\left[\mbox{\boldmath$\mathcal{G}$}(k+Q,i\varepsilon_{n}+i\omega_{m})\right]_{14}
	\right]
	\\
	&\simeq&
	\frac{1}{N_{s}}\sum_{Q}
	\widetilde{t}_{k}\widetilde{t}_{k+Q}
	\avrg{\wdtd{b}{Q}(\omega_{c})\wdtn{b}{Q}(\omega_{c})}
	T\sum_{i\omega_{m}}
	\left[
	\left(\frac{\cond{e}+\cond{d}}{\sqrt{2}}\right)^{2}
	\left[\mbox{\boldmath$\mathcal{G}$}(k+Q,i\varepsilon_{n}+i\omega_{m})\right]_{13}
	\right.
	\nn
	&
	+&
	\left.
	\frac{\cond{e}+\cond{d}}{\sqrt{2}}
	\left[\mbox{\boldmath$\mathcal{G}$}(k+Q,i\varepsilon_{n}+i\omega_{m})\right]_{14}
	\right]
	\label{gap_1}
}
and
\eqsa{
	\widetilde{\Lambda}_{k}(i\varepsilon_{n})
	&\times&
	\left[
	\left[\mbox{\boldmath$\mathcal{G}$}(k+Q,i\varepsilon_{n}+i\omega_{m})\right]_{14}
	+
	\frac{\cond{e}+\cond{d}}{\sqrt{2}}
	\left[\mbox{\boldmath$\mathcal{G}$}(k+Q,i\varepsilon_{n}+i\omega_{m})\right]_{13}
	\right]\\
	&\simeq&
	\frac{1}{\sqrt{\gamma_{k}}}\cdot
	\frac{1}{N_{s}}\sum_{Q}
	\widetilde{t}_{k}\widetilde{t}_{k+Q}
	\avrg{\wdtd{b}{Q}(\omega_{c})\wdtn{b}{Q}(\omega_{c})}
	\nn
	&\times&
	T\sum_{i\omega_{m}}
	\left[
	\left[\mbox{\boldmath$\mathcal{G}$}(k+Q,i\varepsilon_{n}+i\omega_{m})\right]_{14}
	+
	\frac{\cond{e}+\cond{d}}{\sqrt{2}}
	\left[\mbox{\boldmath$\mathcal{G}$}(k+Q,i\varepsilon_{n}+i\omega_{m})\right]_{13}
	\right].
	\label{gap_2}
}
{Then} we obtain the superconducting states induced with
electron-electron and cofermion-electron pairs formed
by exchanging one charge boson.
\end{widetext}

We solve the gap equations Eqs.(\ref{gap_1}) and (\ref{gap_2}) with an assumption
for the symmetry of the superconducting order parameters.
Here we assume a simple $d_{x^{2}-y^{2}}$-symmetry,
which has experimentally been suggested for the hole-doped cuprates,
for the singlet
electron-electron and cofermion-electron pairs as
\eqsa{
	M_{k}(i\varepsilon_{n})
	&=&
	\frac{M_{d_{x^{2}-y^{2}}}(i\varepsilon_{n})}{2}
	\left(\cos k_{x} - \cos k_{y}\right),
	\label{SCE_M}
	\\
	\widetilde{\Lambda}_{k}(i\varepsilon_{n})
	&=&
	\frac{
	\widetilde{\Lambda}_{d_{x^{2}-y^{2}}}(i\varepsilon_{n})
	}{2}
	\left(\cos k_{x} - \cos k_{y}\right).
	\label{SCE_L}
}
As a consequence of the quasi-static approximation,
the gap amplitudes
$M_{d_{x^{2}-y^{2}}}(i\varepsilon_{n})$ and $\widetilde{\Lambda}_{d_{x^{2}-y^{2}}}(i\varepsilon_{n})$
become constant independently of the Matsubara frequency.
Of course, we expect that the amplitudes
of $M_{d_{x^{2}-y^{2}}}(i\varepsilon_{n})$ and $\widetilde{\Lambda}_{d_{x^{2}-y^{2}}}(i\varepsilon_{n})$ will damp for $\omega_{{\rm c}}\ll |i\varepsilon_{n}|$.

Here we note how the cofermion-quasiparticle pairs,
namely, $\left[\mbox{\boldmath$\mathcal{G}$}\right]_{14}$,
enhance the quasiparticle-quasiparticle pairing\bu{.}
First, it is clearly seen in Eq.(\ref{gap_01}) or (\ref{gap_1}),
that  the cofermion-quasiparticle pairs contribute
to the superconducting order parameters of
the quasiparticle-quasiparticle channel, $M_{k}(i\varepsilon_{n})$.
This contribution becomes constructive one,
only when the phases of the cofermion-quasiparticle  
and the quasiparticle-quasiparticle pairs
are \bu{the} same, namely, $M_{k}\cdot \Lambda_{k}>0$.
In this case, the interaction 
illustrated in Fig.\ref{Fig_EX}(b)
behaves as an attractive interaction between
the cofermion-quasiparticle  
and the quasiparticle-quasiparticle pairs.
Then
the formation of the cofermion-quasiparticle pairs
enhances the quasiparticle-quasiparticle pairing
through this attraction,
and, indeed, enhances the pairing in our self-consistent solution of Eqs.(\ref{gap_1}) and (\ref{gap_2}).
%
%

%
\section{Results}\label{Sec.IV}
In this section,
we show how our theory
predicts physical properties of our interest.
We estimate
the quasiparticle {and electron} 
Green's function obtained
in the above section, Sec.\ref{A1d}.
We show the results for the Hubbard model
defined in Eq.(\ref{Hubbard})
on a square lattice at $U=12t$ and $t'=0.25t$ 
to get insight into the cuprate superconductors\bu{.}
\bu{We} restrict 
the mean-field solutions to the homogeneous and paramagnetic ones.
All the calculations are done at zero temperature.
At this parameter and within the present calculation,
the amplitude of the Hubbard gap $\delta U$ at half filling $n=1$ is estimated to be $3.6t$.

First, we give the spectral functions calculated from
the electron Green's function given in Eq.(\ref{GREEN_16}), and show
the global structure of the spectra obtained from
the electron Green's function.
We define the retarded Green's function at the frequency $\omega$ 
as
\eqsa{
	G_{\sigma}(k,\omega)=\mathcal{G}_{\sigma}(k,i\varepsilon_{n}
	\rightarrow \omega+\delta),
}
where $\delta \rightarrow +0$.
Then, the spectral function is given by
\eqsa{
	A(k,\omega)=-\frac{1}{\pi}{\rm Im}\left[G_{\sigma}(k,\omega)\right].
}

\begin{figure}[h]
\begin{center}
\includegraphics[width=7cm]{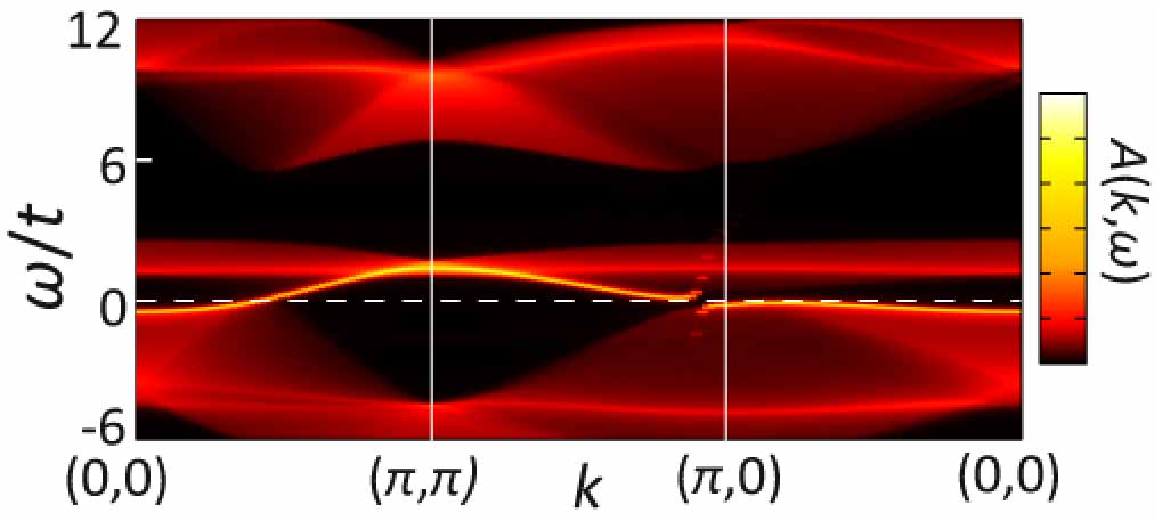}
\end{center}
\caption{(color online)
Spectral function $A(k,\omega)$ along lines running from $(0,0)$ to $(\pi,\pi)$,
from $(\pi,\pi)$ to $(\pi,0)$, and from $(\pi,0)$ to $(0,0)$.
Here we use a finite broadening factor $\delta=0.05t$.
\label{Akw_GL}}
\end{figure}
In Fig.\ref{Akw_GL}, we show the result for $x=0.05$.
There are two main features.
The first is the coherent band seen around the Fermi level, {\it i.e.}, $\omega=0$.
The second is the incoherent spectrum induced by
the charge and spin boson dynamics.
As is already discussed in Ref.\onlinecite{Raimondi},
the remnant of the upper Hubbard band is seen above $\omega\simeq 6t$,
while the lower Hubbard band is seen between the coherent band and $\omega\simeq -8t$.
The gap seen in the spectral function 
at $2t\lesssim \omega \lesssim 6t$ is nothing but
the remnant of the Hubbard gap.
Here an additional incoherent band is seen just above the Fermi level up to
$\omega\simeq 2t$.
This additional incoherent band originates from the dynamics of the spin bosons.
The spectral function shown in Fig.\ref{Akw_GL}
is \bu{qualitatively} consistent with the results of the quantum Monte Carlo
simulations done by Preuss {\it et al.}\cite{Preuss},
although their simulations were done for
$t'/t=0$, $U/t=8$ and $T\neq 0$.
We note that the dispersion of the coherent band
is \bu{given by poles of}
the quasiparticle
Green's function.
\bu{Below,}
we
\bu{concentrate on}
the low-energy
\bu{excitations}
within
the coherent band\bu{.} 
\bu{Therefore,}
we focus on the quasiparticle
Green's function\bu{s}
\bu{instead of electrons Green's functions}.

\begin{figure}[h]
\begin{center}
\includegraphics[width=6cm]{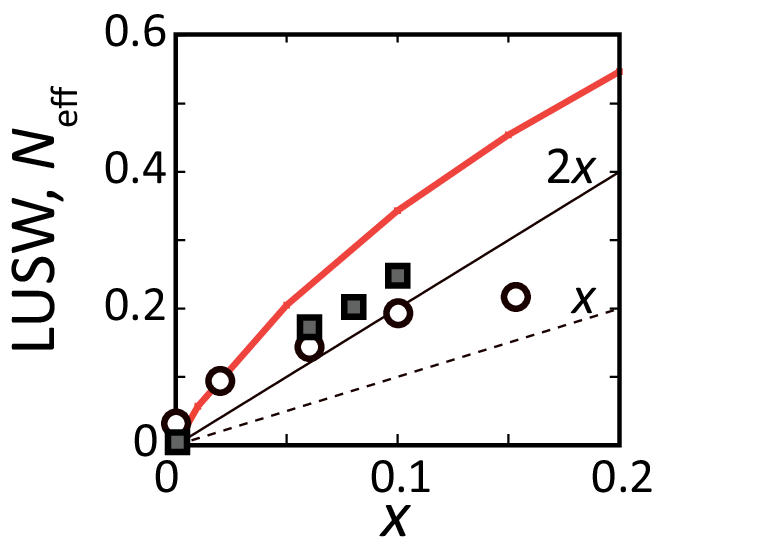}
\end{center}
\caption{(color online)
\bu{Bold} (red) solid curve
shows
doping dependence of 
low-energy spectral weight.
Thin (black) solid line represents $2x$ and
thin (black) dashed line stand for $x$.
Open circles {and closed squares} illustrate $N_{{\rm eff}}$
given in Ref.\onlinecite{Uchida91} {for La$_{2-x}$Sr$_x$CuO$_4$,
and Ref.\onlinecite{Waku04} for Ca$_{2-x}$Na$_x$CuO$_2$Cl$_2$,respectively}. 
\label{LESW_x}}
\end{figure}
In Fig.\ref{LESW_x},
we show the LUSW defined by
$
	(1+x)\zeta_{0}
$
as a function of $x$.
The quick increase of the LUSW larger than $2x$
is clearly seen in Fig.\ref{LESW_x}.
We refer to the data on the effective electron number $N_{{\rm eff}}$
estimated from the optical conductivity measurement\cite{Uchida91,Waku04},
which shows nice agreement with the present LUSW at the small doping as is expected.
For larger doping the LUSW and $N_{{\rm eff}}$ become trivially different because $N_{{\rm eff}}$ is also proportional to the population of the occupied states and decreases
when the number of occupied states becomes small, while the LUSW expresses only the population of  
unoccupied states and should always be larger than $2x$.
\bu{LUSW and $N_{{\rm eff}}$}
become identical in the small doping asymptotically. The quick increase of LUSW indicates that the Mott gap collapses and interrupted by the quick emergence of the low-energy unoccupied states.  This LUSW comes from the quasiparticle band hybridized with the cofermions.
\bu{The} quick increase of the LUSW is \bu{also} related with a positive feedback of the doping, where the hole doping introduces additional screening of carriers leading to a further increase of unbound holon and doublon.  Although the Mott transition is continuous in the low-energy limit at the Fermi level as the ground state properties, this positive feedback gives a character close to the first-order transition in the energy scale of the LUSW, namely in the low-energy excitation spectra. It may be related to the tendency
for the phase separation universally suggested in
the doped Mott insulators ( including the cuprate superconductors )
which has been discussed from several different viewpoints\cite{Emery93}.

\subsection{\bu{Fermi-surface topology}}
\label{FST_A}
\begin{figure}[h]
\begin{center}
\includegraphics[width=8cm]{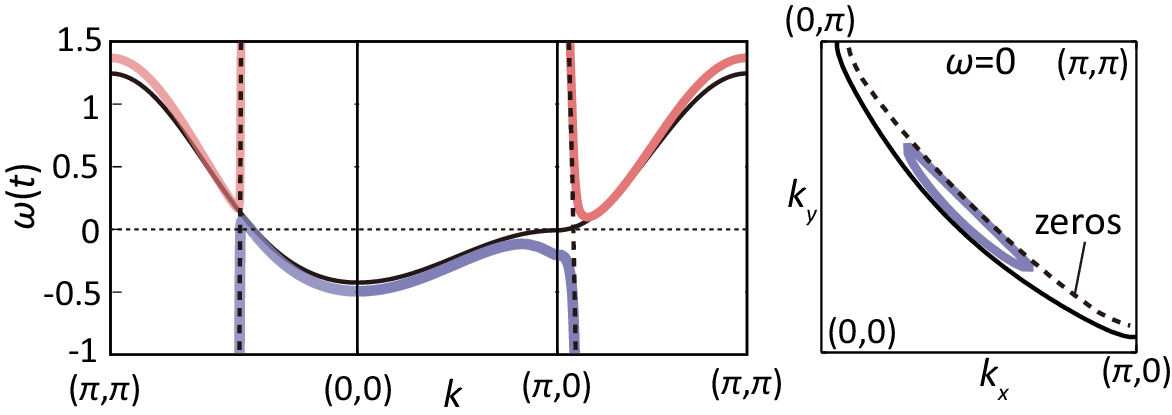}
\end{center}
\caption{(color online)
Left panel shows
band dispersions and zero surface calculated for $x=0.05$
along lines running from $(\pi,\pi)$ to $(0,0)$,
from $(0,0)$ to $(\pi,0)$, and from $(\pi,0)$ to $(0,0)$.
Right panel shows bare and reconstructed Fermi surface, and zero surface at $\omega=0$.
Thin solid (black) curve gives the bare band dispersion $\omega=\zeta_{0\sigma}\epsilon_{k}-\mu$,
and thin dashed (black) curve
represents the zero surface $\gamma_{k}\omega=\alpha_{k}$.
Thick (blue and red) solid curves stand for reconstructed bands.
\label{D_cofermion}}
\end{figure}
\begin{widetext}
Now we show how the reconstruction of the Fermi surface
occurs in our theory.
The quasiparticle Green's function
\eqsa{
G_{\sigma}^{(f)}(k,\omega)=\mathcal{G}_{\sigma}^{(f)}(k,
{i}\varepsilon_{n}
\rightarrow \omega +i\delta),
\nonumber
}
is given as
\eqsa{
	G_{\sigma}^{(f)}(k,\omega)
	&=&	
	\left[\omega+i\delta-\zeta_{0\sigma}\epsilon_{k}
	+\mu-\frac{\Delta_{k}^{2}}{\gamma_{k}
	{\left(\omega+i\delta\right)}-\alpha_{k}}\right]^{-1}
	\label{QDF_F1}
	\\
	&=&
	\left[\frac{1}{2}+\frac{d^{-}_{k}}{\sqrt{(d^{-}_{k})^{2}+\Delta_{k}^{2}/\gamma_{k}}}\right]
	\frac{1}{\omega+i\delta-d^{+}_{k}-\sqrt{(d^{-}_{k})^{2}+\Delta_{k}^{2}/\gamma_{k}}}
	\nn
	&+&
	\left[\frac{1}{2}-\frac{d^{-}_{k}}{\sqrt{(d^{-}_{k})^{2}+\Delta_{k}^{2}/\gamma_{k}}}\right]
	\frac{1}{\omega+i\delta-d^{+}_{k}+\sqrt{(d^{-}_{k})^{2}+\Delta_{k}^{2}/\gamma_{k}}},
	\label{QDF_F2}
}
where $d^{\pm}_{k}=\frac{1}{2}\left(\zeta_{0}\epsilon_{k}-\mu\pm\alpha_{k}/\gamma_{k}\right)$.
\end{widetext} 
Green's function given in Eq.(\ref{QDF_F1})
shows the divergence of
the quasiparticle self-energy
given by $\Delta_{k}^{2}/(\gamma_{k}\omega-\alpha_{k})$,
at $\gamma_{k}\omega-\alpha_{k}=0$.
In other words, the zero surface
defined by $\gamma_{k}\omega=\alpha_{k}$
emerges.
Then, the zero surface splits the band dispersion 
defined by
$\omega=\zeta_{0\sigma}\epsilon_{k}-\mu$
into two bands as $\omega=d^{+}_{k}\pm\sqrt{(d^{-}_{k})^{2}+\Delta_{k}^{2}/\gamma_{k}}$, as is depicted in Fig.\ref{D_cofermion}.
For small doping such as $x=0.05$,
our theory predicts that the reconstructed Fermi surface becomes
a small pocket, as is seen in the right panel of Fig.\ref{D_cofermion}.

\begin{figure}[h]
\begin{center}
\includegraphics[width=7cm]{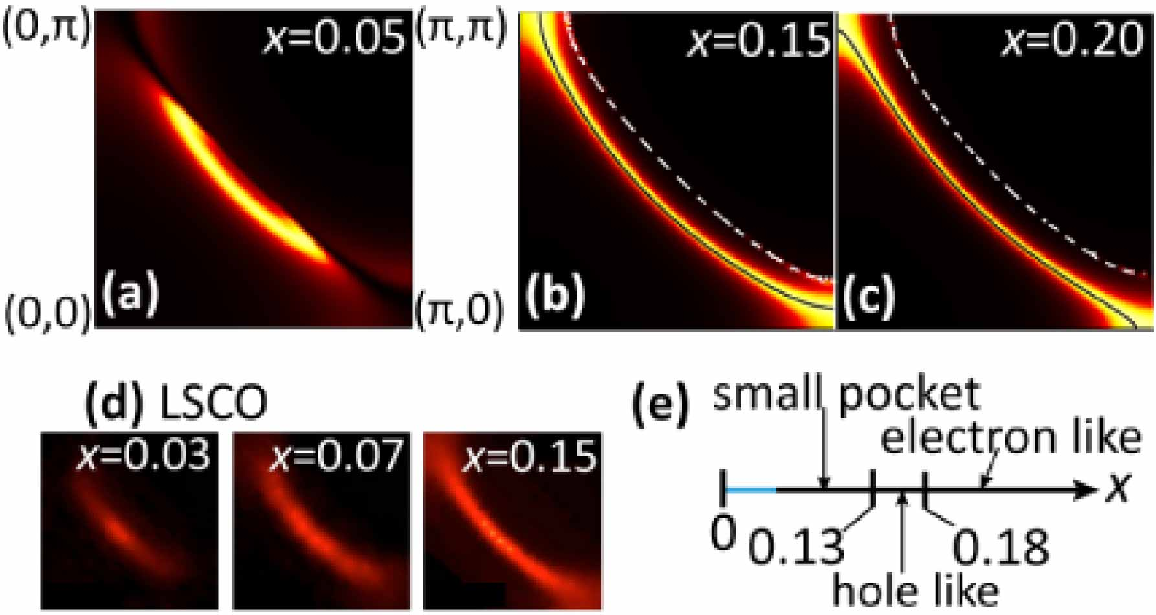}
\end{center}
\caption{(color online)
(a),(b),(c) Single particle spectral function at $\omega=0$.
Solid and dashed lines illustrate the poles of quasiparticles.
(d) Fermi surface observed by ARPES\cite{Yoshida06}.
(e) \bu{Phase diagram of}
Fermi-surface topology in our theory.
\bu{Topological phase boundaries exist at the doping
$x=0.13$ and 0.18.}
\label{Akw}}
\end{figure}
To show changes in the Fermi-surface topology with increasing doping,
the single particle spectral function at $\omega=0$
is given for the hole-doping rate, $x=0.05,\ 0.15,\ 0.20$ in Fig.\ref{Akw}(a)-(c),
with $\delta=0.05t$.
The topological transitions occur at $x\simeq 0.13$ and $x\simeq 0.18$,
as is depicted in Fig.\ref{Akw}(e).
\bu{In the region $0.13\lesssim x$, only small Fermi
pockets exist, where the zero surface smears out the outer
part of the pocket yielding an arc structure as we see in Fig.\ref{Akw}(a)
(note that the damping is large near the zero surface because of an enhanced self-energy),}
consistently with the experimental signature in Fig.~\ref{Akw}(d) as we discuss below.
For \bu{$0.13\gtrsim x$, a completely different topology with}
large Fermi surfaces appear, instead of Fermi pockets.
For $0.13\lesssim x \lesssim 0.18$,
a hole-like surface
centered at $(\pi,\pi)$ (depicted by the solid curve
in Fig.\ref{Akw}(b)) and, 
an electron-like one 
centered at $(\pi,\pi)$ (depicted by the dashed curve
in Fig.\ref{Akw}(b)) \bu{coexist}.
These two surfaces together figure out an enclosed hole (electron unoccupied) strip between the solid and dashed curves.  
However, the electron-like surface (the dashesd curve) is hardly seen
because the zero surface exists near this electron-like surface. 
On the other hand,
for \bu{$0.18\gtrsim x$},
there exist
an electron-like surface centered at $(0,0)$ (depicted by the solid curve
in Fig.\ref{Akw}(c)), and 
an electron-like one centered at $(\pi,\pi)$ (depicted by the dashed curve
in Fig.\ref{Akw}(c)).
The electron-like surface centered at $(\pi,\pi)$
becomes less-visible for the used broadening factor $\delta=0.05t$
than the electron-like surface centered at $(\pi,\pi)$ for $x\lesssim 0.18$.

The quantum transition at $x\simeq 0.18$ is a trivial one expected from
the single particle picture.
The topology of the Fermi surface at this quantum transition point
changes from a hole-like surface centered at $(\pi,\pi)$ depicted 
for $x=0.15$
in Fig.\ref{Akw} by {the} solid curve, 
to an electron-like surface centered at $(0,0)$
depicted 
for $x=0.20$
in Fig.\ref{Akw} by {the} solid curve.
This topology change is essentially understood from the non-interacting picture,
where the band dispersion 
$\epsilon_{k}
=
-2t(\cos k_{x}+\cos k_{y})+4t'\cos k_{x}\cos k_{y}
$
with $t'/t=0.25$ shows the transition from the
electron-like to the hole-like
Fermi surfaces with the increasing electron concentrations.
When the Fermi level is shifted by doping and touches 
the 
saddle points at $(\pm\pi,0)$ and $(0,\pm\pi)$,
such topological changes occur.

However,
our solution shows another nontrivial topological transition at $x\simeq 0.13$.
The topology in the phase $x < 0.13$ is highly nontrivial,
\bu{and is not adiabatically connected with the conventional Fermi liquid.}
Such a topological change never occurs without the zeros of the
Green's function.

For comparison with experiments, we refer to the ARPES spectrum of
La$_{2-x}$Sr$_{x}$CuO$_4$\cite{Yoshida06} in Fig.\ref{Akw}(d).
Arc-like Fermi surfaces observed for $x=0.03$ and $x=0.07$
show overall consistency with our result for $x=0.05$.
On the other hand, so-called nodal metallic behaviors, {\it i.e.},
\bu{large amplitude of} the spectral weight
\bu{allowed}
only
around the nodal direction (line running from $(0,0)$ to $(\pi, \pi)$)
are not seen in our results.
Instead\bu{,} in our theory, the weight is rather larger
near the endpoint of the arc than the nodal point
as we see in Fig.\ref{Akw}(a).
This subtle discrepancy
\bu{is likely to}
originate from
additional self-energy effects, which are not included in the present theory.
For example, fluctuations of $d$-wave superconductivity and/or
short range antiferromagnetic fluctuations are possible candidates.
However, we note that these fluctuations with finite correlation length
by themselves
never induce changes in \bu{the} Fermi-surface topology.

\begin{figure}[h]
\begin{center}
\includegraphics[width=7cm]{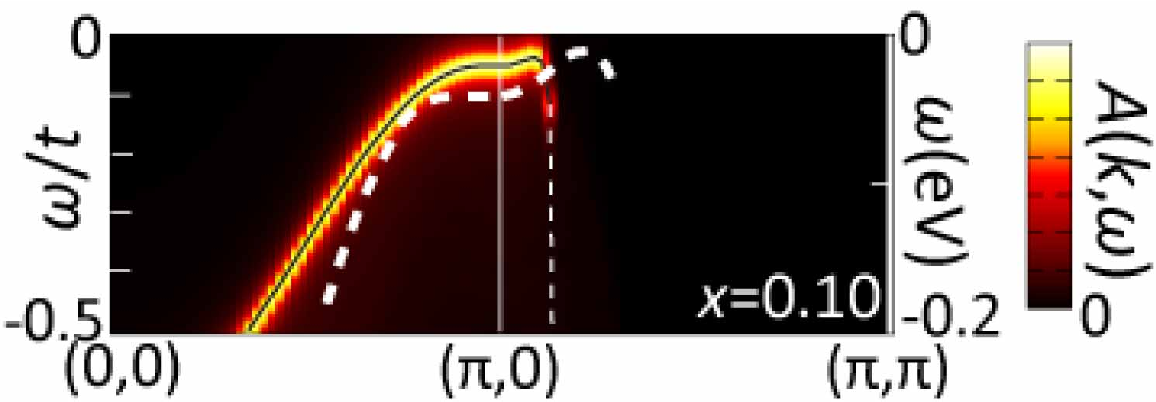}
\end{center}
\caption{(color online):
Single particle spectral function near antinodal point for $x=0.10$.
\bu{T}hin solid and dashed lines illustrate poles of quasiparticles.
\bu{White bold} dashed line stands for
ARPES spectrum observed in La$_{2-x}$Sr$_{x}$CuO$_4$ for $x=0.10$\cite{Ino}.
\label{GXM}}
\end{figure}
When the large Fermi surface appears {for $x\gtrsim 0.13$},
the Fermi surface always extends across the
Brillouin zone boundary or the lines which connect the $\Gamma$ point, $(0,0)$, and the
antinodal points $(\pm\pi,0)$ or $(0,\pm\pi)$.
However,
the single-particle excitation from the Fermi level now has
a gap due to the hybridization gap $\Delta_{k}$ in (\ref{QDF_F1})
in the quantum phase characterized by the nontrivial Fermi surface topology
for $x\lesssim 0.13$.
In Fig.\ref{GXM}, the single particle spectral function is depicted along
the symmetry lines in the Brillouin zone.
A gap measured from the Fermi level
emerges, which
corresponds to
the pseudogap in the ARPES measurements.
Hereafter, we define the amplitude of the pseudogap in the single particle spectrum,
$\Delta$, as the gap amplitude between
the Fermi level $\mu$ and the
maximum of the single particle dispersion below $\mu$ along
the line running from $(0,0)$ to $(\pi, 0)$ and
the line running from $(\pi,0)$ to $(\pi, \pi)$.
We also reproduce the ARPES dispersion observed in La$_{2-x}$Sr$_{x}$CuO$_4$ for $x=0.10$
\cite{Ino} for comparison
in Fig.\ref{GXM} (indicated by bold dashed curve).
Here we {employ a widely accepted parameter} $t=$400 meV.
Indeed the dispersion has an excellent similarity.
The quasiparticle dispersion does not touch the Fermi level
along the Brillouin zone boundary, here the line connecting $(\pi,0)$
and $(\pi,\pi)$, in both of our result and the ARPES measurement.
In other words, the pseudogap amplitude $\Delta$ is finite for both cases.
However,
the quasiparticle state at $(\pi,0)$ in the ARPES data
has lower energy than our result and
the discrepancy is up to 20 meV.
We expect some additional factors to push the dispersion down around the
antinodal points $(\pm\pi,0)$ and $(0,\pm\pi)$.
As is discussed above,
fluctuations of $d$-wave superconductivity and/or
short range antiferromagnetic fluctuations are possible candidates
of the origin of this discrepancy. 

\subsection{Pseudogap}
The pseudogap in single particle excitations defined in the above section
characterizes a quantum phase with the nontrivial Fermi surface,
because a nonzero pseudogap induces the breakdown of the
trivial large Fermi surface.
In our theory, the pseudogap is induced by the hybridization gap between
the cofermions and quasiparticles.
Therefore, the amplitude of the pseudogap is roughly determined by
the hybridization gap
{$\Delta_{k}/\sqrt{\gamma_{k}}$}.
The explicit evaluation
\bu{of}
the amplitude of $\Delta_{k}$ is
done by using Eq.(\ref{app_hyb_delta_jan}) in Appendix.\ref{Appendix_hyb},
\bu{in the lowest order of $t_{ij}$}.
The amplitude of $\Delta_{k}$ is roughly
estimated as $t^{2}/W$
multiplied by a numerical factor,
where $W$ is the band width of original electrons.
Here we note that the denominators in Eq.(\ref{app_hyb_delta_jan})
include the dispersion of the charge and spin bosons.
The band width of the charge and spin bosons are 
roughly equal to the band width of original electrons, $W\simeq 8t$,
and the band width of the spin wave, $J\sim 4t^{2}/U$, respectively.
For $U=12t$, the band width of the charge bosons dominates
that of the spin bonosons, namely, $W>J$.
Therefore, the denominators appearing in Eq.(\ref{app_hyb_delta_jan})
are roughly scaled by $W$.
On the other hand, the amplitude of $\gamma_{k}$
is roughly
estimated as $t^{2}/W^{2}$
multiplied by a numerical factor,
through
Eq.(\ref{app_self_ene_der_cf_jan}).
Therefore, the amplitude of the hybridization gap $\Delta_{k}/\sqrt{\gamma_{k}}$
is roughly
estimated as $(t^{2}/W)/\sqrt{t^{2}/W^{2}}=t$
multiplied by a numerical factor.
The present result for the energy scale of the
pseudogap is consistent
with previous studies
\bu{based on}
the cluster perturbation theory\cite{Senechal04}
and, later, cellular DMFT\cite{Kyung06}.

\begin{figure}[h]
\begin{center}
\includegraphics[width=7cm]{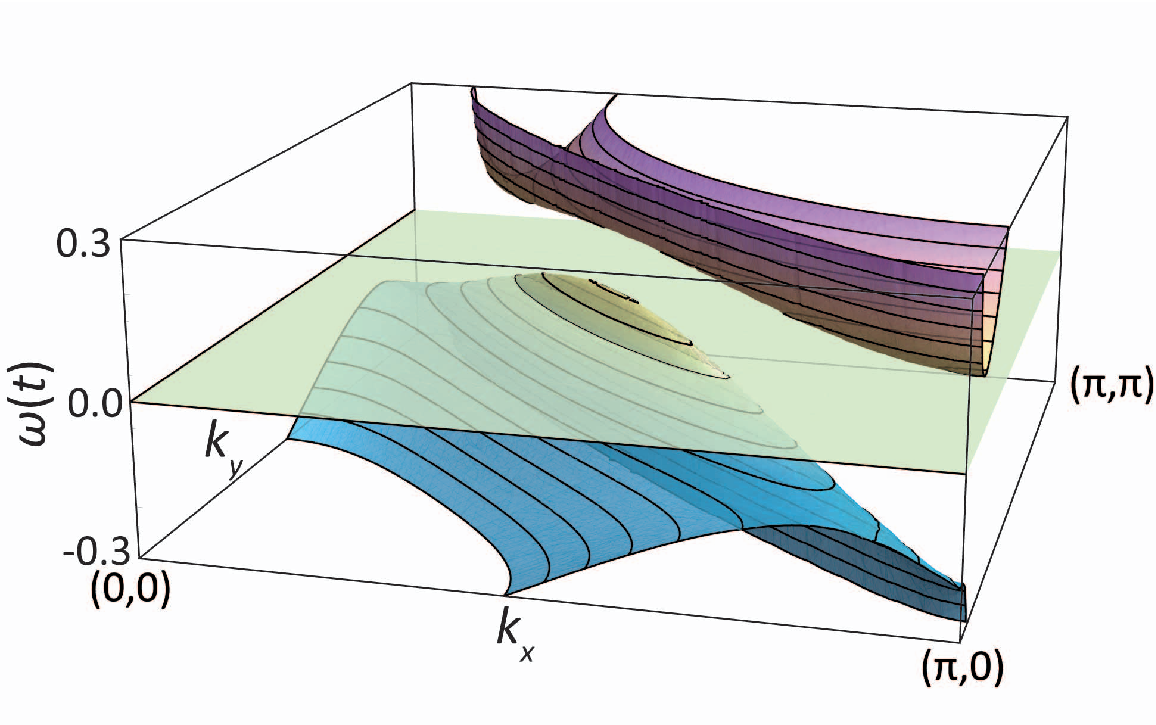}
\end{center}
\caption{
Band dispersion \bu{of quasiparticles} for $x=0.05$.
\bu{A full gap structure is clear.}
\label{3D_disp}}
\end{figure}

We discussed the pseudogap $\Delta$ along the definition often used
in ARPES measurements,
\bu{in \ref{FST_A}.}
Here we discuss the behaviors of the
hybridization gap between two bands given in Eq.(\ref{QDF_F2}) and
illustrate in Fig.\ref{D_cofermion}.
First, the
hybridization gap $\Delta_{k}/\gamma_{k}$ in our theory does not have nodes, and
is $s$-wave-like in the momentum space,
although,
as is illustrated in Fig.\ref{3D_disp},
the gap seems to be smaller in the nodal direction (the line running from $(0,0)$ to $(\pi,\pi)$)
than around the antinodal points $(\pm\pi,0)$ and $(0,\pm\pi)$. 

\begin{figure}[ht]
\begin{center}
\includegraphics[width=7cm]{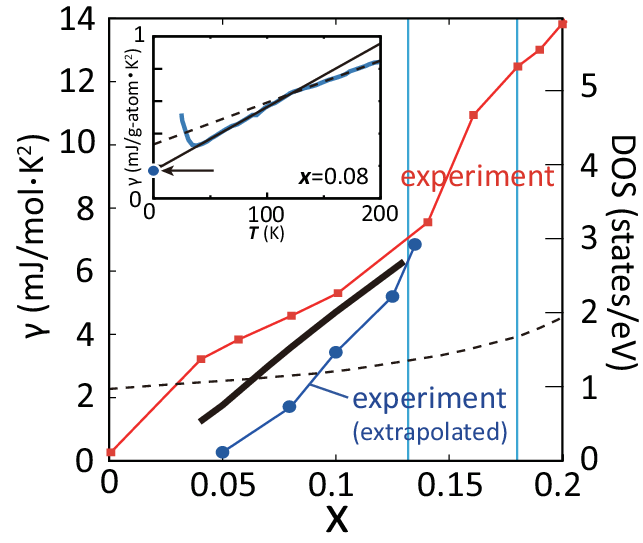}
\end{center}
\caption{
Density of states (DOS) of electrons
given as a function of hole doping rate $x$.
The thin dashed curve stands for the DOS of the noninteracting case.
Closed (blue) circles show $\gamma$ extrapolated
from experimental data of La$_{2-x}$Sr$_{x}$CuO$_4$
in Ref.\onlinecite{Loram01}.
Closed (red) squares stand for 
experimental data of La$_{2-x}$Sr$_{x}$CuO$_4$
in Ref.\onlinecite{Momono}.
Inset shows an example for extrapolation used here{,
see the text below Eq.(\ref{DOS_f}) for details.}
\label{PG_LSCO}}
\end{figure}
Below the transition point $x\simeq 0.13$,
the density of states (DOS) of the original electrons $\hatd{c}{k\sigma}$ at the Fermi level,
$\rho_{{\rm F}}$,
{defined as}
\eqsa{
\rho_{{\rm F}}=\zeta_{0}\frac{1}{N_{s}}
\sum_{k\sigma}\left[-\frac{1}{\pi}{\rm Im}G_{f\sigma}(k,\omega=0)\right],
\label{DOS_f}
}
is clearly
suppressed, as is
{illustrated} in Fig.\ref{PG_LSCO}.
We compare $\rho_{{\rm F}}$ with {the} specific heat measure{d} for
La$_{2-x}$Sr$_{x}$CuO$_4$\cite{Momono,Loram01}.
The specific heat coefficient $\gamma$ is expected
to be proportional to the DOS.
In the Sommerfeld's free electron model,
the specific heat coefficient $\gamma$ at $T=0$ is given by
$
	\gamma=\pi^{2}\rho_{F}/3.
$
Here we use this relation and $t=$400 meV to calculate
the expected values for $\gamma$ with our theory.

We note that $\gamma$ in Fig.\ref{PG_LSCO} indicated
by closed (blue) circles is obtained through a linear extrapolation 
from low-temperature data in Ref.\onlinecite{Loram01},
and shows the possible lower limit for $\gamma$.
From the data in Ref.\onlinecite{Loram01},
$\gamma$ in the underdoped region shows three characteristic
temperature ranges.
For example, the data for $x=0.08$ (see the inset of Fig.\ref{PG_LSCO}) 
show (1) for 100 K$\lesssim T\lesssim 200$ K, \bu{a} linear-temperature
dependence,
(2) for 50 K $\lesssim T\lesssim 100$ K, \bu{another} linear-temperature
dependence but with different slope from (1), and (3)
for $T\lesssim 50$ K,
the BCS-like jump.
We linearly extrapolate $\gamma$ for 50 K $\lesssim T\lesssim 100$ K
and obtain $\gamma$ at $T=0$.
On the other hand,
data for $\gamma$ indicated
by closed (red) squares\cite{Momono} in Fig.\ref{PG_LSCO}
\bu{may} set \bu{an}
upper limit for $\gamma$,
because,
to exclude
the influences from superconductivities,
the authors of Ref.\onlinecite{Momono} used
extrapolation from Zn-doped samples,
which tend to show larger $\gamma$
than samples without Zn-doping.

Our result for $\gamma$ is consistent with \bu{the} experiments,
although we do not choose parameters specific to La$_{2-x}$Sr$_{x}$CuO$_4$.
On the other hand,
the extrapolated $\gamma$ (closed (blue) circles in Fig.\ref{PG_LSCO})
indicates further reduction of the density of states at the Fermi level.
It \bu{suggests}
possible roles of
\bu{fluctuations from the}
$d$-wave superconductivity and/or
antiferromagneti\bu{sm}
\bu{in}
lower energy scale\bu{, which is}
left for future studies beyond the scope of this paper.

\subsection{Superconductivities}
In studies on \bu{high-$T_{\rm c}$ superconducting cuprates,}
the {mechanism of the}
superconductivity
{is of course the central issue}
\cite{Norman_05}.
However,
\bu{the} 
consensus on the mechanism
has not been reached after
\bu{more than}
twenty years of the discovery.
\bu{Here we show how
the novel cofermion mechanism for superconductivities
developed in Sec.~\ref{Sec.III}D
works and reproduces experimental observations
on the high-$T_{\rm c}$ superconducting cuprates.}

\begin{figure}[h]
\begin{center}
\includegraphics[width=8cm]{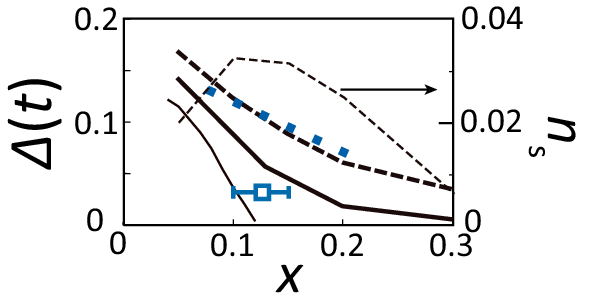}
\end{center}
\caption{Doping dependence of excitation gaps.
The thin (black) solid line stands the pseudogap amplitude in the normal state.
This amplitude shows the contribution that has nothing to do with
the superconductivity but rather to do with the precursor of the Mott gap.
The bold (black) dashed line stands the minimum of the single particle excitation gap in the superconducting state
along the Brillouin zone boundary or the \bu{diagonal}
line connecting the $\Gamma$ point and the antinodal points.
The bold (black) solid line
\bu{shows}
\bu{a refined estimate of the}
single particle excitation gap in the superconducting state
\bu{obtained from}
the pair potential
\bu{mediated}
by exchanging one charge boson\bu{, where the potential is
multiplied by a scale factor}
$\left(1-\zeta_{0}\right)$.
The $x$-dependence of excitation gaps observed by ARPES for Bi$_2$Sr$_2$CaCu$_2$O$_{8+\delta}$\cite{Ding01}
are indicated by the thick (blue) dotted line.
The open square stands for
the superconducting gap
of optimally doped La$_{2-x}$Sr$_{x}$CuO$_4$
estimated by ARPES measurements\cite{Ino99}.
The thin (black) dashed curve stands for
the density of
superconducting electrons
$n_{{\rm s}}$.
\label{Fig_SC}}
\end{figure}
\bu{We examine t}he doping dependence of
the single particle excitation gap in the superconducting state
based on the formalism developed in Sec.~\ref{Sec.III}D.
\bu{The result is shown
in Fig.\ref{Fig_SC} (bold dashed line).}
We adopt, as a definition of the amplitude of the gap
\bu{in the superconducting state},
the minimum of the single particle excitation gap
along the Brillouin zone boundary
\bu{and}
the line connecting the $\Gamma$ point (0,0) and the antinodal points
$(\pm\pi,0)$ and $(0,\pm\pi)$,
which is the same definition for the pseudogap in the normal state.
This definition of the amplitude of the gap
is consistent with the data-analysis of the ARPES
measurements.
As is shown in Fig.\ref{Fig_SC} (bold dashed line), the single particle excitation gap
is
still finite around $x=0.3$
for the $d$-wave superconducting state in the
present treatment,
which is quantitatively inconsistent with
the experimental fact that
the superconductivity disappears
for $x\gtrsim0.26$
in the high-$T_{{\rm c}}$ cuprates.
To highlight a possible origin of this inconsistency,
we also show a result for the single particle excitation gap in Fig.\ref{Fig_SC} (thick solid line)
calculated by rescaling
the quasi-static boson propagator as 
\eqsa{
	\avrg{\wdtd{b}{Q}(\omega_{c})\wdtn{b}{Q}(\omega_{c})}
	\rightarrow
	\left(1-\zeta_{0}\right)\avrg{\wdtd{b}{Q}(\omega_{c})\wdtn{b}{Q}(\omega_{c})}.\nonumber
}
By the rescaling, the single particle excitation gap
is clearly suppressed around $x=0.3$.
The reason why we consider such \bu{a} rescaling is the following:
In our treatment, the charge fluctuations, especially holon fluctuations remain finite even
in the dilute limit $x\rightarrow 1$, where
holons
fill all the sites and should be
localized as hard core bosons.
Holons are completely localized and form the ``Mott insulating" state for $x=1$. 
The fluidity of the holons for $x\lesssim1$ originates from
finite electron density, $1-x\neq 0$,
because
the finite
electron density
introduces ``vacancies" in the holon's Mott insulating
state,
and
causes the fluidity of the holons.
On the other hand, doublons vanish at the limit $x\rightarrow 1$. Near this limit, the density of the doublons are scaled by
$(1-x)^{2}$.
Therefore, for large doping $x$,
the weight of fluctuating charge bosons
should vanish as a function of
$1-x$.
We note that, for $x\simeq 1$, $1-\zeta_{0}\simeq 1-x$.
On the other hand, for $x\simeq 0$,
the charge fluctuating bosons
have
the weight $1-\zeta_{0}$\bu{,}
because the incoherent band induced by fluctuating bosons share the total spectral weight
with the coherent band, which has the weight $\zeta_{0}$.
Therefore for all the doping range,
the rescaling of the boson propagators is naturally interpolated
by $1-\zeta_{0}$.

Then we compare our result with experimentally observed \bu{gap} $\Delta$.
The result of
$\Delta$ without rescaling of the propagators
given by the thick
dashed
curve in Fig.\ref{Fig_SC}
shows a rough qualitative agreement with the ARPES result obtained
from the
single particle excitation gap at low temperature\bu{s} ($T$=14 K) for
Bi$_2$Sr$_2$CaCu$_2$O$_{8+\delta}$\cite{Ding01}
(dotted (blue) line in Fig.\ref{Fig_SC}).
However, our result without the rescaling overestimates
$\Delta$ around $x=0.3$, where the superconducting phase
is not observed in
\bu{any of the}
cuprates.
On the other hand,
the single particle excitation gap calculated by using 
the rescaled boson propagator
shows
\bu{a}
suppression of $\Delta$ around $x=0.3$.
The single particle excitation gap $\Delta$ for $x\sim 0.1$
is consistent with 
the superconducting gap
of optimally doped La$_{2-x}$Sr$_{x}$CuO$_4$
estimated by ARPES measurements\cite{Ino99}
(open square in Fig.\ref{Fig_SC}).
On the other hand,
the rescaled boson propagator appears
to be still overestimated for the overdoped region, $x\sim 0.3$.
The origin of such overestimated
fluctuating charge bosons
in our theory
may be
ascribed to \bu{the} approximation on
the bosonic propagators given in 
the Dyson equations shown in Fig.\ref{Dyson_2}:
Lifetime of the coherent propagation of the
fluctuating charge bosons
is not
taken into account in our theory.
There exist two possible origins of the finite lifetime.
The first is damping caused by single particle excitations,
namely,
the Landau damping. 
When the doping increases, the motion of the coherent carriers 
may disturb the propagation of the
fluctuating charge bosons
seriously.
The second is
the separation of charge and spin fluctuations, which becomes
a good approximation for $x\simeq 0$.
Such separation of charge and spin fluctuations
may become worse for large $x$,
because the motion of the coherent carriers
mixes the charge and spin modes.
The mixing of charge and spin fluctuations also
disturb the propagation of the
fluctuating charge bosons
and give a finite lifetime.
Therefore, these factors neglected in our theory will
suppress the charge fluctuations and consequently, the superconductivities.

The single particle gap $\Delta$
decreases
monotonically, as the doping $x$ increases.
It is consistent with ARPES measurements.
On the other hand, the critical temperature of the
superconductivity, $T_{{\rm c}}$, shows
a peak around the $x\sim 0.15$ in La$_{2-x}$Sr$_{x}$CuO$_4$,
and exhibits a dome-like structure as a function of $x$.
To estimate the tendency of $T_{{\rm c}}$ from our results
for $T=0$, we show the $x$-dependences of
the density of
superconducting electrons
$n_{{\rm s}}$ in Fig.\ref{Fig_SC}(thin dashed curve) defined as
\eqsa{
n_{{\rm s}}=
\left|\zeta_{0}\frac{T}{N_{{\rm s}}}\sum_{k,i\varepsilon_{n}}
\left(\cos k_{x}-\cos k_{y}\right)
\left[\mbox{\boldmath$\mathcal{G}$}(k,i\varepsilon_{n})\right]_{13}
\right|.
}

The pairing mechanism proposed here includes
the contribution of the cofermion-quasiparticle pairs
in addition to the quasiparticle-quasiparticle pairs.
This contribution of the cofermion-quasiparticle pairs
itself is a novel perspective introduced in this \bu{paper.}
Furthermore, it offers an insight
\bu{into}
one of the most interesting issues
in physics of the cuprate superconductors,
namely, relationship between the pseudogap formation
and the high-$T_{{\rm c}}$ superconductivity.
The cofermions induce the hybridization gap around the Fermi level, and
as a result, the pseudogap.
The pseudogap formation itself reduces the DOS around the Fermi level
and
\bu{is}
harmful for the high-$T_{{\rm c}}$ superconductivity.
However, the cofermions support the high-$T_{{\rm c}}$ superconductivity
through the cofermion-quasiparticle pairing, simultaneously.
Therefore the pseudogap structure does not necessarily destroy the
superconducting pairing.
The pseudogap formation is the other side of the
coin of the cofermion pairing contributing
to the superconductivity.
This dual character offers an in sight into the recent controversy
\bu{called dichotomy}
observed in ARPES measurements
\cite{Tanaka06,Yang08}:
There exist two major pictures for the relationship
between the pseudogap and high-$T_{{\rm c}}$ superconductivity.
One tells us that the pseudogap is a precursor
of the Mott gap\cite{Furukawa},
and coexists with the superconducting gap.
The other claims that
preformed pairs of strong coupling superconductivities
induce
the pseudogap\cite{Yanase03}.
\bu{The present theory offers a new alternate route of understanding
that reconciles the dichotomy.}

By exchanging one charge boson,
attractive interactions are induced for forward scattering channels.
Even though there is the strong on-site repulsion $U$,
such attractive interactions may favor anisotropic superconductivities
not only the $d_{x^2-y^2}$-wave superconductivity but \bu{also},
for example, $d_{xy}$-wave and extended $s$-wave superconductivities.
In the present theory, however,
a simple extended $s$-wave superconducting parameter proportional to $\cos k_x + \cos k_y$
does not develop.
\subsection{{Metal-insulator transitions}}
We propose a scenario for the filling control metal-Mott insulator transitions
based on our theory.
Before going into our scenario,
we note that there is
a difficulty in our theory
in the small doping limit $x\rightarrow +0$.
{In this limit, we have Fermi pockets with small
but finite volume depending on parameters such as
$t'/t$ and $U/t$.
It is in contrast to the zero-doping limit
of the ordered insulators.}
In our theory, the weights of the two bands split by the zero surface illustrated in Fig.\ref{D_cofermion}
are not the same.
From Eq.(\ref{QDF_F2}), these weights are
given as
\eqsa{
	\frac{1}{N_{{\rm s}}}
	\sum_{k}
	\left[\frac{1}{2}\pm\frac{d^{-}_{k}}{\sqrt{(d^{-}_{k})^{2}+\Delta_{k}^{2}/\gamma_{k}}}\right].
}
In the case of the antiferromagnetic ordered state,
the counterparts of these weights are given by
\eqsa{
	\frac{1}{N_{{\rm s}}}
	\sum_{k}
	\left[
	\frac{1}{2}
	\pm
	\frac
	{ \epsilon_{1k} }
	{ 2\sqrt{ \epsilon_{1k}^{2}+\Delta_{{\rm AFM}}^{2} } }
	\right],
}
from Eq.(\ref{AFM_2bands}),
and are the same, because of
\eqsa{
\frac{1}{N_{{\rm s}}}
	\sum_{k}\frac
	{ \epsilon_{1k} }
	{ 2\sqrt{ \epsilon_{1k}^{2}+\Delta_{{\rm AFM}}^{2} } }=0.
}
The weight compensation between these two bands split by the zero surface
is needed to reproduce ordered insulating phases.
On the other hand, our theory does not offer this weight compensation.
Therefore, our theory does not necessarily predict
that the small hole pockets seen in Fig.\ref{Akw}(a)
shrink to points
at the small doping limit $x\rightarrow +0$.

\bu{In addition, in the limit $x\rightarrow +0$,
the imaginary part of the quasiparticle and cofermion self-energy,
which is not taken into account in the present theory,
may affect the weight compensation.
Higher order corrections to the cofermion dynamics
will also affect the structure of the zero surface in this limit.}

Despite these limitations,
\bu{o}ur result \bu{still} suggest a possible scenario
for the filling control metal-Mott insulator transitions.
If the small Fermi pockets seen in Fig.\ref{Akw}(a)
shrink to the Fermi points,
the filling control metal-insulator transition
occurs as a topological transition.
On the other hand,
if
the small Fermi pockets
do not
shrink to the Fermi points,
the quasiparticle weight of the Fermi pockets
becomes zero, in spite of the finite volume of pockets.
Then the insulating state appears,
because of the fading out of the coherent band.
In both of the two cases,
a small but finite DOS is expected in
the limit $x\rightarrow +0$. 
It should be noted that,
anyway, a nontrivial topological transition
is expected
to occur at a finite doping,
before the system becomes insulating.

Here we note that the order of the three quantum phase transitions, namely, the Mott transition at $x=0$, and the two topological transitions at $x\sim 0.13$ and $x\sim 0.18$ are all continuous within the present calculated results, although in general, the first order transition is not a priori excluded.  If a topological quantum phase transition takes place as a continuous one at $T=0$, it becomes just a crossover at nonzero temperature, because the Fermi surface is
in the strict sense not well defined at $T>0$ and the toplogical character of the Fermi surface loses well-defined meaning at $T>0$ any more.  It has been stressed before that the sharp change between the overdoped and underdoped regions is associated with some kind of quantum criticality triggered by a quantum phase transition of the symmetry breaking order.  In this conventional quantum criticality, the phase transition usually extends to nonzero temperature as the border of symmetry broken phase.  However, the present quantum criticality is completely different from this category, where not the symmetry breaking but the topology change drives the transition\cite{Imada05,Misawa06,Yamaji06,Misawa07}.
\section{Summary and Discussions}\label{Sec.VI}
In this paper, 
we have proposed a microscopic
mechanism for the Fermi-surface reconstruction and non-Fermi-liquid behaviors
emerging in proximity to the Mott insulators,
motivated by puzzles experimentally observed
in the underdoped cuprates.
We construct our theory
starting from
one of the simplest theory for
correlated metals, namely, the slave-boson mean-field theory for the Hubbard model by Kotliar and Ruckenstein (KR).
Then a special emphasis is placed on the role of extra charge dynamics on
low-energy spectra.
We find that hidden cofermionic particle\bu{s} constructed from 
composite fermions play a crucial role in addition to the quasiparticles described by
the previous KR theory.
The additional cofermions called
{\it holo-electrons} and {\it doublo-holes}
represent substantial low-energy part of \bu{the} charge dynamics.

Thus introduced new cofermions hybridize with the
quasiparticles
and cause a hybridization gap represented by zeros of the quasiparticle Green functions.
As a result, although the explicit symmetry breaking is absent,
the gap identified as the pseudogap of the cuprates naturally comes out.
Although the origin of the gap has an apparent \bu{similarity} to the case of the
symmetry-broken phases such as commensurate antiferromagnetic metals, in the sense that
both of them can be ascribed to a hybridization gap, 
the mechanism of the hybridization and the partner of the hybridization are very different from the simple symmetry breaking.  
The origin of the gap is also different from the Mott gap itself.
The large Mott gap in the insulating phase is quickly replaced with the present much smaller hybridization gap upon doping,
because the \bu{LUSW} above the hybridization gap emerges.
In the result of the cellular dynamical mean-field theory\cite{Sakai09}, this
hybridization gap coexists with the clear remnant of
Mott gap structure in a much larger energy scale.
This indicates that the two gaps are separated phenomena.
Thus the pseudogap or the hybridization gap in the underdoped region is not really the precursor of the Mott gap.

When the cofermion strongly hybridizes with the quasiparticles and the resultant hybridization gap splits the quasiparticle band around the Fermi level, Fermi-surface reconstruction occurs in a natural way.
The hybridization gap indeed leads to a break-up of 
the Fermi surface into small hole pockets around
the nodal direction, and a phase topologically different from
the prediction of the weak coupling picture emerges.
This phase realized in the underdoped region is separated
by a topological phase transition
from the normal Fermi liquid phase
in the overdoped region.
Our result clearly shows that an unconventional non-Fermi liquid
phase emerges in the underdoped region.
Such topological changes and an emergence of a new phase 
are consistent with experimentally observed
``Fermi-arc" formation in hole-underdoped cuprates.

The topological quantum phase transition between the conventional Fermi liquid and the non Fermi liquid is a continuous one at $T=0$ within the present approximation. This implies that it transforms to a crossover at nonzero temperature, because the continuous topological transition is well defined only at zero temperature.  In principle, it does not exclude a possibility of the first order transition of the topological change, which extends to nonzero temperatures with the critical end point at a nonzero temperature\cite{Yamaji06}.
The present results do not support the existence of such a first order transition in agreement with the absence of the clear indication of the first order transition in the experiments.

Our formulation offers a new concept
for the ``pseudogap" phenomena observed in hole-underdoped cuprates.
It proposes that the ``pseudogap" in the single particle spectrum
results from the hybridization gap between the cofermions and the quasiparticles.
The part below the Fermi level has \bu{the}
maximum in the antinodal region and has a structure similar to the $d_{x^2-y^2}$ symmetry in agreement with the experimental observation.  The main part of the gap lies, however, above the Fermi level. The total hybridization gap has the structure of the $s$-wave gap. 

Consequently, the characteristic energy scale of the pseudogap is identified as 
that of the hybridization gap amplitude and thus basically scaled by the bandwidth (kineic energy) of the bare electrons, $t$ multiplied by a numerical factor.
Therefore, the energy scale of \bu{the} ``pseudogap" has something to do neither with
the Hubbard gap or the on-site Coulomb repulsion $U$ nor with the superexchange interaction 
$J\sim t^2/U$.

We have also numerically studied the relevance of our theory in the realistic condition.
For the on-site Coulomb repulsion $U$ large enough to create the Hubbard
gap $U\gtrsim 10t$ and small amount of hole doping $x\lesssim0.1$, regarded as a relevant parameter
for the high-$T_c$ cuprates, 
the structure of the momentum dependence and the amplitude of the pseudogap below the Fermi level have \bu{a} quantitative consistency between the ARPES measurements
for the hole-doped cuprates and the present results. 
Doping dependence of the pseudogap
amplitude is also consistent with \bu{the} experimental
observation.

Reduction of the electronic density of states induced by
the pseudogap formation gives another consistency with
\bu{the} experiments as
observed as specific heat coefficients.
The pseudogap formation in our theory also
induces asymmetry of the DOS around the Fermi level,  
which naturally explains the asymmetric
STM spectra\cite{Renner98,Hanaguri04}.
The overall consistency supports that the cofermions, holo-electrons and dublo-holes,
are indeed a relevant object to be considered in physics of the cuprate superconductors.

The present theory has further consequences and predictions for experiments.
The hybridization gap basically has a $s$-wave-like symmetry and a major part of the gap
lies above the Fermi level for the model of the hole doped cuprates.  
Although the present resolution limit of the inverse photoemission does not allow determination of the detailed 
structure of the unoccupied electronic states, we propose that our prediction of this $s$-wave-like gap structure accompanied by the main part of the \bu{LUSW} lying above the hybridization gap can be tested  experimentally if some high-resolution measurements of the unoccupied spectra are provided.
The optical conductivity $\sigma (\omega)$, especcially mid-infrared peak and a long tail of 
$\sigma (\omega)$ observed in the underdoped cuprates indeed supports the existence of such \bu{LUSW}\cite{Uchida91,Waku04}.

So far, direct measurement of the cofermions appears to be difficult,
because the dispersion of the cofermions corresponds to the zeros of the 
quasiparticles, while the zeros are in general hard to detect. 
In addition, the cofermion does not have a charge and does not allow an 
electromagnetic detection.  
However, it contributes to the entropy and thermal transport.
Since the electric conduction is contributed only from the quasiparticle, while
the thermal transport can arise from the cofermion as well, 
we expect a serious breakdown of the Wiedeman-Franz (WF) law.
{The WF law
predicts that the ratio of \bu{the} thermal conductivity $\kappa$
to \bu{the} electric conductivity $\sigma$ is equal to $L_{0}T$,
namely, $L_{0}=\kappa/T\sigma$,
where $L_{0}$ is a universal constant given by $L_{0}=(\pi^{2}/3)\cdot\left(k_{{\rm B}}/e\right)^{2}$, namely,
the Lorenz number.
On the other hand, in our theory, the ratio $L=\kappa/T\sigma$
is predicted to be larger than $L_{0}$
in proximity to Mott insulators,
because the cofermions carry additional energy,
and contribute to $\kappa$.
Indeed the breakdown of the Wiedeman-Franz law reported recently
in hole-underdoped cuprates\cite{Proust05} supports 
the present prediction.}

We propose a scenario for
the filling control metal-Mott insulator transitions
in two dimensions
based on the newly introduced cofermions.
We predict a pseudogap formation
and consequently a non-trivial change in \bu{the} Fermi-surface topology
in the underdoped region.
Then the criticality of the small Fermi pockets determines 
the nature of the Mott transition, if the transition is continuous.
Two scenarios remain possible: 
On the verge of the Mott transition, it may either shrink to points before vanishing by keeping the 
quasiparticle weight nonzero, or the quasiparticle weight decreases to vanish by keeping 
a finite volume of the Fermi surface.
In the former case,
the filling control metal-Mott insulator transitions occur as a topological one. 
The density of states remain nonzero and finite on the verge of the transition
because of the two deminsionality.
In the latter case,
the effective mass diverges.  These two possibilities are essentially identical to the 
two types discussed in the literature\cite{Imada93}.
Although the present numerical accuracy is not sufficient for determining the ultimate
criticality, the shrinkage of the pocket and arc overall supports the former scenario,
while the reduction of the DOS suggests only a part of the pocket contributes to low-energy
excitations as the arc. This agrees with the experimental results. 
More detailed and accurate determination of the criticality is left for future studies.    

We also propose a novel mechanism for
high-temperature superconductivity driven by the cofermion-quasiparticle pairing.
It offers a new insight into the relationship between the pseudogap formation
and the high-$T_{{\rm c}}$ superconductivity.
The cofermions newly introduced in the present paper induce
the hybridization gap around the Fermi level, and
as a result, the pseudogap.
This pseudogap formation itself reduces the DOS around the Fermi level
and destroys superconductivity.
On the other hand,
the cofermions enhance the high-$T_{{\rm c}}$ superconductivity
through the cofermion-quasiparticle pairing, simultaneously.
At the present level of approximation, 
in the overdoped region,
a process exchanging one charge boson
overestimates the charge fluctuations and
predicts much larger single particle gap
in superconducting phases, than experimentally observed gaps.
It is left for future studies to correct
such an overestimate of charge fluctuations.
\bu{In addition to one charge boson exchange processes,
cofermion polarization will help
the pairing between quasiparticles, which is not taken into
account as a higher order contribution in the present paper.} 
However, the superconducting gap obtained from the present approximation already reproduces
the dome-like structure as a function of the doping concentration with a right order of magnitude if we compare with the cuprates.

\bu{Although we focus on the hole-doped cuprates in this paper,
electron-underdoped
cuprates are also interesting from the viewpoint of the present cofermion theory.
In the electron-underdoped systems, our theory will predict the 
emergence of two electron pockets
centered at $(\pi, 0)$ and $(0, \pi)$ even in the absence of antiferromagnetic
long-range orders.
Such a Fermi-surface topology may reduce instabilities towards antiferromagnetic
orders, and will make $d_{x^{2}-y^{2}}$-wave superconductivities stable:
The electron pocket formation prevents the nesting of the Fermi surface,
and allows an emergence of full-gapped superconductors
but with the $d_{x^{2}-y^{2}}$-wave symmetry,
because the node lines run the momenta
where the original electron pockets are absent.}

Here, we make some remarks
\bu{in regard}
to the relationship with other theoretical approaches.
In contrast to our approach,
the high-energy
charge degrees of freedom to do with the UHB
have often been integrated out
to extract the low-energy physics from the Hubbard model.
The $t$-$J$ model is a typical effective model derived from such a treatment,
and has been studied as a canonical model describing the doped Mott insulator.
The gauge theory
based on the $t$-$J$ model
\bu{has been}
intensively studied to
explain the low-energy physics of cuprate
superconductors
as doped spin liquids\cite{Lee_RMP}.
Wen and Lee\cite{Wen96} proposed that
a hole-doped spin liquid state
may exhibit hole pockets.
According to this theory,
the optical or direct gap of this phase
is $d$-wave-like: There is no optical gap along
the nodal direction.
Although the gapless excitation is present in the
nodal direction by the electron-hole
excitation through the hole pocket even in our theory,
the gauge-theory scenario by Wen and Lee is in
contrast to our result predicting
that
a relatively small but nonzero amplitude of the optical gap
even along
the nodal direction is superimposed,
namely,
a $s$-wave-like optical gap should be
visible between the two bands separated by
the hybridization gap.
The $s$-wave-like pseudogap
is also supported in the previous numerical studies
by Stanescu and Kotliar\cite{Stanescu06},
and Sakai {\it et al.}\cite{Sakai09}, in support of the present theory.
\bu{An earlier exact diagonalization study without any bias
has indicated that even the doped $t$-$J$ model has a similar
$s$-wave-like pseudogap structure in agreement with our theory\cite{Tohyama04}.}
As we mentioned already, it is desired to test against the two contradicting predictions of our theory and the gauge theory by
inverse photoemission spectroscopies
or by more sophisticated and high-resolution
experimental tools to measure the unoccupied states
in the future.

Another crucial difference between the present theory and the gauge theory can be 
tested by the breakdown of the Wiedeman-Franz law. In the gauge theory, the breakdown is due to the contribution from the spinons\cite{Kim09}.
However, if the arc structure is observed, the $d$-wave like gap needs to be developed by the flux fluctuations, which leads to the confinement of a holon and a spinon generating a quasiparticle in the nordal direction.  It should recover the Wiedeman-Franz law in this region of the arc formation.  Therefore, at low temperatures in the underdoped region, the Wiedeman-Franz law should eventually be \bu{followed.}
On the contrary, in the present case of the cofermions, the contribution continues even at low temperatures and the breakdown of the Wiedeman-Franz law is robust.
Recent experimental results appear to support our prediction\cite{Hill01,Proust05}.

Integrating out the ``high-energy"
charge degrees of freedom represented by the existence of doublon
thoroughly
{leads to the ignorance and overlook of important aspects of}
the low-energy spectrum of the Hubbard model.
There
exists a variety of
experimental facts which are not accounted for by the $t$-$J$ physics:
An example is
\bu{the} estimate of the \bu{LUSW} by using $N_{{\rm eff}}(\omega)$
observed in \bu{the} optical conductivity measurement\cite{Uchida91}.
Choy {\it et al}. \bu{have made}
a careful
treatment
to extract the low-energy effective action, with
the failure of the $t$-$J$ model \bu{kept} in mind\cite{Choy08}.
They claim the weight transfer among the coherent band, the LHB, and the UHB described by
hidden $2e$-boson\bu{s}.
In spite of the illuminating proposal,
their effective action could not be solved exactly.
Approximate evaluations of their theory show a
soft gap behaviors, and predict a semiconducting
behaviors in the ``pseudogap phase,"
where resistivity $\rho$
is expected to diverge as $\rho\propto T^{-1}$
with lowering temperatures.
In the present theory,
the Fermi pocket
appears in the ``pseudogap phase"
in the underdoped region, say,
for $x\lesssim0.13$.
Therefore, we predict metallic conduction in
the ``pseudogap phase," in sharp contrast to 
the hidden $2e$-boson theory.

\bu{As the authors have already discussed in
Ref.\onlinecite{Yamaji11},
the formation of the cofermions shares profound similarity with
that of the excitions in semiconductors.
In the so-called $d$-$p$ model for the cuprates, which contains both
$d$-electrons on the copper sites and $p$-electrons on the oxygen sites,
it has been claimed that the excitionic effects due to $d$-$p$ interactions
strongly affects its excitation spectrum\cite{Wagner91}.
Relationship between exciton-like features of the cofermions and
excitionic effects studied in the $d$-$p$ model
are also desired to be clarified in the future.}

Our theory has been constructed to account for the charge dynamics
more seriously than the literature
and proposed
the topological changes of \bu{the}
Fermi surface in proximity to the Mott insulators.
The reconstructed Fermi surface also offers an
unexplored avenue at smaller energy scale, 
if it is combined with other possible fluctuations 
such as spin and superconducting fluctuations.
Clarifying
possible emergence of antiferromagnetic orders
in
the small doping region $x\lesssim 0.02$
\bu{by} using a unified scheme is left for
future studies.
It is also left for future studies how the topological changes
affect various possible symmetry breakings such as
time reversal symmetry breakings, stripe formation,
and incommensurate charge orders including phase separations. 
Experimentally suggested nodal metallic behaviors
should also be examined in more detail. 
Our theory will give a new insight into 
emergence of 
these competing orders, the high-$T_{{\rm c}}$ superconductivity,
and the anomalous metallic state.

\begin{acknowledgements}
The authors thank Yukitoshi Motome and Shiro Sakai for useful discussions.
Y.Y. is supported by the Japan Society for the Promotion of Science.
\end{acknowledgements}

\appendix

\section{Self-energy for the cofermions\label{app_self_ene_sec}}
Here we calculate the self-energy for the cofermions in
Dyson equations (see Fig.\ref{Dyson_2}).
\begin{widetext}
Then the self-energy matrix in Eq.(\ref{self-energy_matrix}) is calculated by
\eqsa{
	\Sigma^{cd} (k)
	&=&
	\frac{T^{3}}{N_{{\rm s}}^{3}}
	\sum_{P,Q,R}
	\sum_{a,b=1,2}
	\left(g_{1\sigma}g_{2\sigma}\right)^{4}
	\avrg{\beta_{Q}^{a}{\beta^{b}}^{\dagger}_{Q}}
	\left[
	t_{\bvec{k}+\bvec{P}}
	t_{\bvec{k}+\bvec{R}}
	\avrg{{\phi^{a}}^{\dagger}_{P\sigma}{\phi^{c}}_{P\sigma}}
	\avrg{\phi_{R\sigma}^{b}{\phi^{d}}^{\dagger}_{R\sigma}}
	\avrg{\hatn{f}{k+Q\sigma}\hatd{f}{k+Q\sigma}}
	\right.
	\nn
	&+&
	\left.
	\left(t_{\bvec{k}+\bvec{R}}\right)^{2}
	\avrg{{\phi^{a}}^{\dagger}_{P\sigma}{\phi^{b}}_{P\sigma}}
	\avrg{\phi_{R\sigma}^{c}{\phi^{d}}^{\dagger}_{R\sigma}}
	\avrg{\hatn{f}{k+Q-P+R\sigma}\hatd{f}{k+Q-P+R\sigma}}
	\right].
}
When we use the mean-field propagators for
the quasiparticles, $\avrg{\hatn{f}{k\sigma}\hatd{f}{k\sigma}}$,
we obtain \bu{a} simple analytic expression for the self-energy matrix as
\eqsa{
	\Sigma^{cd} (k)
	&=&
	-
	\frac{1}{N_{{\rm s}}^{3}}
	\sum_{\bvec{P},\bvec{R}}
	\sum_{a,b=1,2}
	\left(g_{1\sigma}g_{2\sigma}\right)^{4}
	t_{\bvec{k}+\bvec{P}}
	t_{\bvec{k}+\bvec{R}}
	\avrg{{\phi^{a}}^{\dagger}_{\bvec{P}\sigma}{\phi^{c}}_{\bvec{P}\sigma}}
	\avrg{\phi_{\bvec{R}\sigma}^{b}{\phi^{d}}^{\dagger}_{\bvec{R}\sigma}}
	\nn
	&\times&
	\sum_{\bvec{Q}}
	\left[
	\frac{\theta(\epsilon_{\bvec{k}+\bvec{Q}}-\mu)
	Z^{ab}_{-}(Q)}
	{i\varepsilon_{n}-|\epsilon_{\bvec{k}+\bvec{Q}}-\mu|
	-|\lambda_{Q}|}
	+
	\frac{\theta(\mu-\epsilon_{\bvec{k}+\bvec{Q}})
	Z^{ab}_{+}(Q)}
	{i\varepsilon_{n}+|\epsilon_{\bvec{k}+\bvec{Q}}-\mu|
	+|\Lambda_{Q}|}
	\right]
	\nn
	&+&
	\frac{1}{N_{{\rm s}}^{3}}
	\sum_{\bvec{P},\bvec{Q},\bvec{R}}
	\sum_{a,b=1,2}
	\left(g_{1\sigma}g_{2\sigma}\right)^{4}
	\left(t_{\bvec{k}+\bvec{R}}\right)^{2}
	\left[
	\frac
	{\theta(\epsilon_{\bvec{k}+\bvec{Q}-\bvec{P}+\bvec{R}}-\mu)
	Z^{ab}_{-}(Q)W^{ab}_{+}(P)W^{cd}_{-}(R)}
	{i\varepsilon_{n}-|\epsilon_{\bvec{k}+\bvec{Q}-\bvec{P}+\bvec{R}}-\mu|
	-|\lambda_{Q}|-|\ell_{P}|-|\ell_{R}|}
	\right.
	\nn
	&+&
	\left.
	\frac
	{\theta(\mu-\epsilon_{\bvec{k}+\bvec{Q}-\bvec{P}+\bvec{R}})
	Z^{ab}_{+}(Q)W^{ab}_{-}(P)W^{cd}_{+}(R)}
	{i\varepsilon_{n}+|\epsilon_{\bvec{k}+\bvec{Q}-\bvec{P}+\bvec{R}}-\mu|
	+|\Lambda_{Q}|+|\ell_{P}|+|\ell_{R}|}
	\right]
	,\label{app_self_ene_cf_jan}
}
where the propagators for the spin bosons are
given by
\eqsa{
	-\avrg{{\phi^{a}}^{\dagger}_{Q}{\phi^{b}}_{Q}}
	&=&
	\frac{W_{+}^{ab}(Q)}{i\omega_{m}-|\ell_{Q}|}
	-
	\frac{W_{-}^{ab}(Q)}{i\omega_{m}+|\ell_{Q}|},
	\\
	-\avrg{{\beta^{a}}_{Q}{\beta^{b}}^{\dagger}_{Q}}
	&=&
	\frac{Z_{+}^{ab}(Q)}{i\omega_{m}-|\Lambda_{Q}|}
	-
	\frac{Z_{-}^{ab}(Q)}{i\omega_{m}+|\lambda_{Q}|},
}
\bu{where the coefficient matrices, $W_{\pm}^{ab}$ and $Z_{\pm}^{ab}$,
are given as}
\eqsa{
	\left(
	\begin{array}{cc}
	W_{\pm}^{11}(Q)&W_{\pm}^{12}(Q)\\
	W_{\pm}^{21}(Q)&W_{\pm}^{22}(Q)
	\end{array}	
	\right)
	&=&
	\displaystyle
	\pm
	\frac{1}{2}
	\left(
	\begin{array}{cc}
	1
	&0\\
	0
	&-1
	\end{array}	
	\right)
	-
	\frac{
	\delta\lambda-
	a_1
	\frac{|\epsilon|}{2}\epsilon_{Q}}{2\ell_{Q}}
	\left(
	\begin{array}{cc}
	1&0\\
	0&1
	\end{array}	
	\right)
	-
	\frac{
	b_1
	\frac{|\epsilon|}{2}\epsilon_{Q}}{2\ell_{Q}}
	\left(
	\begin{array}{cc}
	0&1\\
	1&0
	\end{array}	
	\right),
	\\
	\left(
	\begin{array}{cc}
	Z_{\pm}^{11}(Q)&Z_{\pm}^{12}(Q)\\
	Z_{\pm}^{21}(Q)&Z_{\pm}^{22}(Q)
	\end{array}	
	\right)
	&=&
	\displaystyle
	\frac{\delta\lambda+\delta U/2}{2\sigma_{Q}}
	\left(
	\begin{array}{cc}
	1
	&0\\
	0
	&1
	\end{array}	
	\right)
	\pm
	\frac{1}{2}
	\left(
	\begin{array}{cc}
	1
	&0\\
	0
	&-1
	\end{array}	
	\right)
	-
	\frac{
	c_1
	\frac{|\epsilon|}{2}\epsilon_{Q}}{2\sigma_{Q}}
	\left(
	\begin{array}{cc}
	1&0\\
	0&1
	\end{array}	
	\right)
	-
	\frac{
	d_1
	\frac{|\epsilon|}{2}\epsilon_{Q}}{2\sigma_{Q}}
	\left(
	\begin{array}{cc}
	0&1\\
	1&0
	\end{array}	
	\right),
}
\end{widetext}
\bu{The parameters used in the above equations are given as
$\delta\lambda=
	\lambda^{(1)}-\lambda^{(2)},$
$\delta U=U-2\lambda^{(2)}$,
$|\epsilon|=
	\left|
	\frac{T}{N_{s}}
	\sum_{k,i\varepsilon_{n}}
	\epsilon_{k}
	\mathcal{G}_{\sigma}^{(f)}(k,i\varepsilon_{n})
	\right|$,
\eqsa{
	\ell_{Q}&=&
	\sqrt{
	\left( \delta\lambda - a_{1}\frac{|\epsilon|}{2}\epsilon_{Q}\right)^2
	- b_{1}^{2} \frac{|\epsilon|^2}{4}\epsilon_{Q}^2
	},\\
	\sigma_{Q}&=&\sqrt{
	\left(
	\lambda^{(1)}+\frac{\delta U}{2}-c_{1}\frac{|\epsilon|}{2}\epsilon_{Q}
	\right)^{2}-d_{1}^{2}\frac{|\epsilon|^{2}}{4}\epsilon_{Q}^{2}},\\
	\Lambda_{Q}&=&\frac{\delta U}{2}+\sigma_{Q},\\
	\lambda_{Q}&=&-\frac{\delta U}{2}+\sigma_{Q}.
}
}
\bu{The coefficients, $a_1$, $b_1$, $c_1$, and $d_1$ are determined as,}
\eqsa{
	a_1 &=& \cond{e}^2 + \cond{d}^2 + 
	\avrg{\wdtn{e}{i}\wdtd{e}{j}} + \avrg{\wdtd{d}{i}\wdtn{d}{j}},\\
	b_1 &=& 2 \cond{e}\cond{d} + 
	\avrg{\wdtn{e}{i}\wdtn{d}{j}} + \avrg{\wdtd{d}{i}\wdtd{e}{j}},\\
	c_1 &=& 2 \cond{p}^2 + 
	\avrg{\wdtd{p}{i\sigma}\wdtn{p}{j\sigma}} 
	+ \avrg{\wdtn{p}{i\overline{\sigma}}\wdtd{p}{j\overline{\sigma}}},\\
	d_1 &=& 2 \cond{p}^2 + 
	\avrg{\wdtn{p}{i\sigma}\wdtn{p}{j\overline{\sigma}}} 
	+ \avrg{\wdtd{p}{i\overline{\sigma}}\wdtd{p}{j\sigma}},
}
\bu{where we only take into account nearest-neighbor pairs for $(i,j)$.}
\begin{widetext}
\bu{To derive Eq.(\ref{app_self_ene_cf_jan}),} we \bu{also} use following relations,
\eqsa{
	T\sum_{i\omega_{m}}
	\frac{1}{i\varepsilon_{n}+i\omega_{m}-\xi}
	\frac{1}{i\omega_{m}-\lambda_{1}}
	=
	\frac{\theta(\xi)\theta(-\lambda_{1})-\theta(-\xi)\theta(\lambda_{1})}
	{i\varepsilon_{n}-\xi+\lambda},
}
and
\eqsa{
	&&T^{3}\sum_{i\omega_{\ell}}
	\sum_{i\omega_{m}}
	\sum_{i\omega_{n}}
	\frac{1}{i\varepsilon_{n}+i\omega_{m}-i\omega_{\ell}
	+i\omega_{n}-\xi}
	\cdot
	\frac{1}{i\omega_{m}-\lambda_{1}}
	\cdot
	\frac{1}{i\omega_{\ell}-\lambda_{2}}
	\cdot
	\frac{1}{i\omega_{n}-\lambda_{3}}
	\nn
	&&=
	\frac{
	\theta(\xi)
	\theta(-\lambda_{1})\theta(\lambda_{2})\theta(-\lambda_{3})
	-
	\theta(-\xi)
	\theta(\lambda_{1})\theta(-\lambda_{2})\theta(\lambda_{3})
	}
	{i\varepsilon_{n}-\xi+\lambda_{1}-\lambda_{2}+\lambda_{3}}.
}
Derivatives of $\Sigma^{cd} (k)$ with respect to $i\varepsilon_{n}$
give $\gamma_{k}$ after taking a limit $i\varepsilon_{n}\rightarrow 0$, as is
mentioned in Sec.\ref{A1d}.
Here we give the derivatives as
\eqsa{
	\left.\frac{\partial \Sigma^{cd}}{\partial i\varepsilon_{n}}
	\right|_{i\varepsilon_{n}\rightarrow 0}
	&=&
	\frac{1}{N_{{\rm s}}^{3}}
	\sum_{\bvec{P},\bvec{R}}
	\sum_{a,b=1,2}
	\left(g_{1\sigma}g_{2\sigma}\right)^{4}
	t_{\bvec{k}+\bvec{P}}
	t_{\bvec{k}+\bvec{R}}
	\avrg{{\phi^{a}}^{\dagger}_{\bvec{P}\sigma}{\phi^{c}}_{\bvec{P}\sigma}}
	\avrg{\phi_{\bvec{R}\sigma}^{b}{\phi^{d}}^{\dagger}_{\bvec{R}\sigma}}
	\nn
	&\times&
	\sum_{\bvec{Q}}
	\left[
	\frac{\theta(\epsilon_{\bvec{k}+\bvec{Q}}-\mu)
	Z^{ab}_{-}(Q)}
	{\left(|\epsilon_{\bvec{k}+\bvec{Q}}-\mu|
	+|\lambda_{Q}|\right)^{2}}
	+
	\frac{\theta(\mu-\epsilon_{\bvec{k}+\bvec{Q}})
	Z^{ab}_{+}(Q)}
	{\left(|\epsilon_{\bvec{k}+\bvec{Q}}-\mu|
	+|\Lambda_{Q}|\right)^{2}}
	\right]
	\nn
	&-&
	\frac{1}{N_{{\rm s}}^{3}}
	\sum_{\bvec{P},\bvec{Q},\bvec{R}}
	\sum_{a,b=1,2}
	\left(g_{1\sigma}g_{2\sigma}\right)^{4}
	\left(t_{\bvec{k}+\bvec{R}}\right)^{2}
	\left[
	\frac
	{\theta(\epsilon_{\bvec{k}+\bvec{Q}-\bvec{P}+\bvec{R}}-\mu)
	Z^{ab}_{-}(Q)W^{ab}_{+}(P)W^{cd}_{-}(R)}
	{\left(|\epsilon_{\bvec{k}+\bvec{Q}-\bvec{P}+\bvec{R}}-\mu|
	+|\lambda_{Q}|+|\ell_{P}|+|\ell_{R}|\right)^{2}}
	\right.
	\nn
	&+&
	\left.
	\frac
	{\theta(\mu-\epsilon_{\bvec{k}+\bvec{Q}-\bvec{P}+\bvec{R}})
	Z^{ab}_{+}(Q)W^{ab}_{-}(P)W^{cd}_{+}(R)}
	{\left(|\epsilon_{\bvec{k}+\bvec{Q}-\bvec{P}+\bvec{R}}-\mu|
	+|\Lambda_{Q}|+|\ell_{P}|+|\ell_{R}|\right)^{2}}
	\right].\label{app_self_ene_der_cf_jan}
}
\section{Hybridization\label{Appendix_hyb}}
The amplitudes of hybridization between
quasiparticles and cofermions,
\eqsa{
	\mbox{\boldmath$\Delta$}_{ij}^{T}
	=
	\left(
	\Delta_{ij}^{(\psi)},
	\Delta_{ij}^{(\chi)}
	\right)^{T},
}
are given as
$
	(\Delta^{1}_{\bvec{k}},\Delta^{2}_{\bvec{k}})
	=
	(\Delta^{(\psi)}_{\bvec{k}},\Delta^{(\chi)}_{\bvec{k}}).
$
Its Fourier transformation is given as
\eqsa{
	\Delta^{d}_{\bvec{k}}
	&=&
	\frac{T^{2}}{N_{{\rm s}}^{2}}
	\sum_{P,R}
	\sum_{\bvec{q}}
	\sum_{a,b,c=1,2}
	t_{\bvec{k}+\bvec{R}}\left(g_{1\sigma}^{2}g_{2\sigma}^{2}\right)^{2}
	\avrg{\beta_{P-R}^{a}{\beta^{b}}^{\dagger}_{P-R}}
	\avrg{{\phi^{a}}^{\dagger}_{P\sigma}{\phi^{b}}_{P\sigma}}
	\avrg{\phi_{R\sigma}^{d}{\phi^{c}}^{\dagger}_{R\sigma}}b^{c}_{0}
	t_{\bvec{q}+\bvec{R}}
	\avrg{\hatd{f}{\bvec{q}\sigma}\hatn{f}{\bvec{q}\sigma}},
}
where we use vector notations,
$\bvec{b}_{0}=({b}^{1}_{0},{b}^{2}_{0})=(\cond{e},\cond{d})$,
$\bvec{\beta}_{i}=({\beta}^{1}_{i},{\beta}^{2}_{i})=
(\wdtn{e}{i}, \wdtd{d}{i})$,
and
$\bvec{\phi}_{i}=({\phi}^{1}_{i},{\phi}^{2}_{i})=
(\wdtn{p}{i\sigma}, \wdtd{p}{i\overline{\sigma}})$.
Here, if we use the mean-field propagators for
the quasiparticle,
we obtain
\eqsa{
\Delta^{d}_{\bvec{k}}
	&=&
	-
	\frac{1}{N_{{\rm s}}^{2}}
	\sum_{\bvec{P},\bvec{R},\bvec{q}}
	\sum_{a,b,c=1,2}
	t_{\bvec{k}+\bvec{R}}\left(g_{1\sigma}^{2}g_{2\sigma}^{2}\right)^{2}
	t_{\bvec{q}+\bvec{R}}
	n_{\bvec{q}}
	\nn
	&\times&
	\left[
	\frac{
	Z_{+}^{ab}(\bvec{P}-\bvec{R})
	W_{-}^{ab}(\bvec{P})
	W_{+}^{dc}(\bvec{R})}{|\Lambda_{\bvec{P}-\bvec{R}}|+|\ell_{\bvec{P}}|
	+|\ell_{\bvec{R}}|}
	+
	\frac{
	Z_{-}^{ab}(\bvec{P}-\bvec{R})
	W_{+}^{ab}(\bvec{P})
	W_{-}^{dc}(\bvec{R})}{|\lambda_{\bvec{P}-\bvec{R}}|+|\ell_{\bvec{P}}|
	+|\ell_{\bvec{R}}|}
	\right],\label{app_hyb_delta_jan}
}
\end{widetext}
where we use the following relation:
\eqsa{
	&\displaystyle T^{2}\sum_{i\omega_{\ell}}
	\sum_{i\omega_{n}}
	\frac{1}{i\omega_{\ell}-i\omega_{n}-\lambda_{1}}
	\cdot
	\frac{1}{i\omega_{\ell}-\lambda_{2}}
	\cdot
	\frac{1}{i\omega_{n}-\lambda_{3}}
	\nn
	&\displaystyle
	=
	\frac{
	\theta(\lambda_{1})\theta(-\lambda_{2})\theta(\lambda_{3})
	+
	\theta(-\lambda_{1})\theta(\lambda_{2})\theta(-\lambda_{3})
	}
	{\lambda_{1}-\lambda_{2}+\lambda_{3}}.
}
Equation (\ref{app_hyb_delta_jan}) will help us to estimate the amplitude of
the pseudogap in our theory. 
\bibliography{Yamaji_PRB_revtex_Sep_v4}
\end{document}